\documentclass[a4paper]{article}

\usepackage{tikz}
\usepackage{amssymb}
\usepackage{verbatim} 
\usetikzlibrary{matrix}
\usepackage{color} 
\usepackage{stmaryrd}
\usepackage{amsfonts} 
\usepackage{amsmath} 
\usepackage{amsthm}
\usepackage[all,cmtip]{xy}
\usepackage{catex} 			
\usepackage{fancybox}
\usepackage{relsize}
\usepackage{cmll} 
\usepackage{bussproofs}
\usepackage{standalone}
\usepackage[normalem]{ulem}
\usepackage{multicol}

\usepackage{mathdots}

\newenvironment{scprooftree}[1]%
  {\gdef\scalefactor{#1}\begin{center}\proofSkipAmount \leavevmode}%
  {\scalebox{\scalefactor}{\DisplayProof}\proofSkipAmount \end{center} }

\newtheorem{theorem}{Theorem}[section]
\newtheorem{corollary}[theorem]{Corollary}
\newtheorem{oss}{Remark} 
\newtheorem{proposition}[theorem]{Proposition}

\theoremstyle{definition}
\newtheorem{definition}{Definition}

\newcommand{\nota}{\noindent\textbf{Notation }}

\newcommand{\ls}{\mathcal{h}}						
\newcommand{\rs}{\mathcal{i}}						
\newcommand{\fr}{\rightarrow}						
\newcommand{\frr}{\Rightarrow}						

\newcommand{\R}{\mathcal R}

\newcommand{\N}{\mathbb N}
\newcommand{\s}{\Sigma}

\newcommand{\sig}{\mathcal S} 						
\newcommand{\pro}{\; \square \;}					

\newcommand{\inp}{\mathrm{in}} 						
\newcommand{\outp}{\mathrm{out}} 						

\newcommand{\catgen}[1]{\ls{#1}\rs} 						
\newcommand{\Sem}[1]{\mathcal {S}_{{#1}}} 					

\newcommand{\D}{\phi}

\newcommand{\id}{\textbf{id}}

\newcommand{\frrr}{\xymatrix@C-1pc{{} \ar@3[r]& {}}}

\newcommand{\cutelim}{\xymatrix@C-1pc{{} \ar@3[r]& {}}}
\newcommand{\vcutelim}{\xymatrix@R-0.5pc{~\ar@3[d] \\~ }}

\newcommand{\Mllt}{\tilde{\mathfrak{M}}}				
\newcommand{\Mellt}{\tilde{\mathfrak{E}}}				
\newcommand{\Mlltc}{{\tilde{\mathfrak{U}}}}			

\newcommand{\Mll}{\mathfrak{M}}                
\newcommand{\Mllc}{\mathfrak{U}}                

\newcommand{\fmll}{{\mathfrak{F}_{M\ell \ell}}}                
\newcommand{\smll}{{{\mathfrak{F}}^*_{M\ell \ell }}}                
\newcommand{\fmllc}{{\mathfrak{F}_{M\ell \ell _u}}}                
\newcommand{\smllc}{{{\mathfrak{F}}^*_{M\ell \ell _u}}}                
\newcommand{\fmell}{{\mathfrak{F}_{Me \ell \ell}}}                

\newcommand{\MLL}{\mathit{MLL}}				
\newcommand{\MLLc}{\mathit{MLL_u}}				

\deftwocell[crossing]{s : 2 -> 2}
\deftwocell[dots]{d : 2 -> 2}
\deftwocell[circle, black]{de : 0 -> 0}
\deftwocell[cap]{cap : 0 -> 2}
\deftwocell[cup]{cup : 2 -> 0}
\deftwocell[circle, black]{e1 : 1 -> 1}

\deftwocell[text=X,white]{g11 : 1 -> 1}				
\deftwocell[rectangle]{g : 2 -> 2} 				
\deftwocell[rectangle]{g35 :3 -> 5}				
\deftwocell[polygon]{g12:1 -> 2}				
\deftwocell[polygon]{g21 :2 -> 1}				
\deftwocell[text=\otimes,yellow]{net : 1 -> 2}			

\deftwocell[text=\alpha,white]{net1 : 2 -> 1}			
\deftwocell[text=\beta,white]{net2 : 2 -> 1}			
\deftwocell[text=\nu,white]{net3 : 4 -> 1}			
				
\deftwocell[black]{m : 2 -> 1}
\deftwocell[circle,black]{e : 0 -> 1}
\deftwocell[white]{penta : 4 -> 1}
\deftwocell[white]{tria : 2 -> 1}
\deftwocell[white,rectangle]{exa : 3 -> 1}
\deftwocell[white,circle]{exa2 : 3 -> 1}
\deftwocell[white, rectangle]{inv : 2 -> 1}
\deftwocell[white]{g : 3 -> 1}
\deftwocell[gray]{alpha : 3 -> 1}
\deftwocell[gray, rectangle]{gamma : 3 -> 1}
\deftwocell[gray, rectangle]{tau : 2 -> 1}
\deftwocell[gray,lefthalfcircle]{l : 1 -> 1}
\deftwocell[gray,righthalfcircle]{r : 1 -> 1}

\deftwocell[text=\phi, white]{phi : 2 -> 2} 			
\deftwocell[text=\phi', white]{phi1 : 2 -> 2} 			
\deftwocell[text=\psi, white]{psi : 2 -> 2} 			
\deftwocell[text=\psi', white]{psi1 : 2 -> 2} 			

\deftwocell[text=\phi, white]{phi3 : 3 -> 3} 			
\deftwocell[text=\phi, white]{phi23: 2 -> 3} 			
\deftwocell[text=\phi, white]{phi24: 2 -> 4} 			
\deftwocell[text=\psi, white]{psi24: 2 -> 4} 			
\deftwocell[text=\phi', white]{phi32 : 3 -> 2} 			
\deftwocell[text=\phi', white]{phi42 : 4 -> 2} 			
\deftwocell[text=\psi, white]{psi4 : 4 -> 4} 			
\deftwocell[text=\psi', white]{psip4 : 4 -> 4} 			
\deftwocell[text=\psi_1, white]{psi4b1 : 4 -> 4} 			
\deftwocell[text=\phi, white]{phi4 : 4 -> 4} 			
\deftwocell[text=\phi_1, white]{phi4b1 : 4 -> 4} 			
\deftwocell[text=\phi', white]{phip4 : 4 -> 4} 			

\deftwocell[text=\phi, white]{phi04 : 0 ->4} 			
\deftwocell[text=\phi, white]{phi05 : 0 -> 5} 			
\deftwocell[text=\phi, white]{phi06 : 0 -> 6} 			
\deftwocell[text=\phi, white]{phi07 : 0 -> 7} 			
\deftwocell[text=\phi, white]{phi08 : 0 -> 8} 			
\deftwocell[text = \phi' , white]{phi09 : 0 -> 9} 
\deftwocell[text = \phi' , white]{phi10 : 0 -> 10} 
\deftwocell[text = \psi , white]{psi07: 0 -> 7} 
\deftwocell[text = \psi' , white]{psi09 : 0 -> 9} 
\deftwocell[text = \psi' , white]{psi10 : 0 -> 10} 

\deftwocell[text=\phi_1, white]{phi104 : 0 -> 4} 			
\deftwocell[text=\phi_1, white]{phi105 : 0 -> 5} 			
\deftwocell[text=\phi_2, white]{phi205 : 0 -> 5} 			
\deftwocell[text=\phi_2, white]{phi204 : 0 -> 4} 			
\deftwocell[text=\phi_3, white]{phi303 : 0 -> 3} 			
\deftwocell[text=\psi_1, white]{psi105 : 0 -> 5} 			
\deftwocell[text=\psi_2, white]{psi205 : 0 -> 5} 			

\deftwocell[text=\phi']{G8: 0 ->8}
\deftwocell[text=\phi']{G6: 0 ->6}
\deftwocell[text=\phi'_l]{G51: 0 -> 5}
\deftwocell[text=\phi'_r]{G52: 0 -> 5}

\deftwocell[left=\dots]{midD : 1 ->1}	 			
\deftwocell[text=\mathfrak {h} , white]{hor :2  ->2} 			
\deftwocell[text=\mathfrak {h}' , white]{hor1 :2  ->2} 			
\deftwocell[text=\mathfrak {h}'' , white]{hori :2  ->2} 			
\deftwocell[text=\mathfrak {h}''' , white]{hori1 :2  ->2} 			
\deftwocell[text=\alpha , white]{blind1 : 0  ->2} 			

\deftwocell[text=x, lightgray]{gx1 : 1 -> 1} 			
\deftwocell[text=x, lightgray]{gx : 2 -> 2} 			
\deftwocell[text=g, lightgray]{g1 : 2 -> 2} 			
\deftwocell[text=g', lightgray]{g2 : 2 -> 2} 			
\deftwocell[text=N, white]{N : 2 ->1} 			
\deftwocell[text=N', white]{N1 : 2 ->1} 			
\deftwocell[text=N', white]{N31 : 3 ->1} 			
\deftwocell[text=\alpha, white]{split : 4 ->1} 			
\deftwocell[text=\beta, white]{split2 : 4 ->1} 			

\deftwocell[text=\alpha, white]{a : 2 -> 2} 			
\deftwocell[text=\sigma, white]{sigma : 2 -> 2} 		
\deftwocell[text=\mu, white]{sigma4 : 4 -> 4} 		
\deftwocell[text=\sigma, white]{sigma46 : 4 -> 6} 		
\deftwocell[text=\sigma, white]{sigma57 : 5 -> 7} 		
\deftwocell[text=\sigma, white]{sigma34 : 3 -> 4} 		
\deftwocell[text=\sigma, white]{sigma35 : 3 -> 5} 		
\deftwocell[text=\sigma, white]{sigma36 : 3 -> 6} 		
\deftwocell[text=\sigma, white]{sigma6 : 6 -> 6} 		
\deftwocell[text=\sigma, white]{sigma06 : 0 -> 6} 		
\deftwocell[text=\sigma', white]{sigma1 : 2 -> 2} 		

\deftwocell[text=\phi_1, white]{prem1 : 0 ->5} 			
\deftwocell[text=\phi_2, white]{prem2 : 0 -> 5} 			
\deftwocell[text=\phi_1', white]{phip : 5 -> 5} 			
\deftwocell[text=\phi_2', white]{phis : 5 -> 5} 			

\deftwocell[text=\chi_u, white]{up : 2 -> 6} 				
\deftwocell[text=\chi_d, white]{do : 6 -> 2} 				
\deftwocell[text=\chi_u, white]{up4 : 2 -> 4} 				
\deftwocell[text=\chi_d, white]{do4 : 4 -> 2} 				

\deftwocell[rectangle, green]{bigt2 : 2 -> 2} 			
\deftwocell[rectangle, green]{bigt : 8 -> 8} 			
\deftwocell[rectangle, green]{bigt10 : 10 -> 10} 			
\deftwocell[rectangle, green]{bigt12 : 12 -> 12} 				
\deftwocell[text=BT{(}W W' {)}, white]{btW : 8 -> 8} 			

\deftwocell[crossing2]{ln : 3 -> 3}				
\deftwocell[crossing1]{rn : 3 -> 3}				

\deftwocell[text=1, white]{dia1: 0 -> 2}
\deftwocell[text=2, white]{dia2: 0 -> 2}
\deftwocell[text=3 , white]{dia3: 0 -> 2}

\deftwocell[text=\phi, white]{diaphi02: 0 -> 2}
\deftwocell[text=\phi, white]{diaphi03: 0 -> 3}
\deftwocell[text=\phi, white]{diaphi04: 0 -> 4}
\deftwocell[text=\psi, white]{diapsi03: 0 -> 3}
\deftwocell[text=\psi, white]{diapsi04: 0 -> 4}
\deftwocell[text=\phi, white]{diaphi05: 0 -> 5}

\deftwocell[text=\chi_{ij}, white]{chigen : 2 -> 2}

\deftwocell[mid = \mathcal{C}]{backmC : 0 -> 0}				
\deftwocell[mid = \mathcal{D}]{backmD : 0 -> 0}				
\deftwocell[mid = \mathcal{E}]{backmE : 0 -> 0}				

\deftwocell[text=\circ, white]{circ : 2 -> 1} 			

\deftwocell[mid = A]{topA : 0 -> 1}						
\deftwocell[mid = {?}A]{topAw : 0 -> 1}					
\deftwocell[mid = {?}A^\bot]{topAbw : 0 -> 1}					
\deftwocell[mid = A^\bot]{topAb : 0 -> 1}					
\deftwocell[mid = B^\bot]{topBb : 0 -> 1}					
\deftwocell[mid = B]{topB : 0 -> 1}						
\deftwocell[mid = {?}B]{topBw : 0 -> 1}					
\deftwocell[mid = C]{topC : 0 -> 1}						
\deftwocell[mid = D]{topD : 0 -> 1}						

\deftwocell[mid = \Gamma]{topGam : 0 -> 2}					
\deftwocell[mid = \Gamma']{topGam1 : 0 -> 2}				
\deftwocell[mid = {?}\Gamma]{topGamw : 0 -> 2}				
\deftwocell[mid = {?}\Gamma']{topGam1w : 0 -> 2}				
\deftwocell[mid = {?}\Delta]{topDelw : 0 -> 2}				
\deftwocell[mid = {?}\Delta']{topDel1w : 0 -> 2}				
\deftwocell[mid = \sigma{(} {\Gamma } {)}]{topsiG : 0 -> 2}		
\deftwocell[mid = \Sigma]{topSig : 0 -> 2}					
\deftwocell[mid = \Delta]{topDel : 0 -> 2}					
\deftwocell[mid = L]{topL : 0 -> 1}						
\deftwocell[mid = R]{topR : 0 -> 1}						
\deftwocell[mid = W]{topW : 0 -> 2}						
\deftwocell[mid = W_1]{topWb1 : 0 -> 2}						
\deftwocell[mid = W_2]{topWb2 : 0 -> 2}						
\deftwocell[mid = W']{topW1 : 0 -> 2}						
\deftwocell[mid = W'_1]{topW1b1 : 0 -> 2}						
\deftwocell[mid = W'_2]{topW1b2 : 0 -> 2}						
\deftwocell[mid = W'']{topW2 : 0 -> 2}						
\deftwocell[mid = out {(}\alpha{)} ]{outa : 2 -> 2}				
\deftwocell[mid = in {(}\alpha{)} ]{ina : 2 -> 2}				
\deftwocell[mid = out {(}x{)} ]{outx : 2 -> 0}				
\deftwocell[mid = in {(}x{)} ]{inx : 0 -> 2}				

\deftwocell[mid = A]{midA : 1 -> 1}						
\deftwocell[mid = B]{midB : 1 -> 1}						
\deftwocell[mid = \Gamma]{midGam : 2 -> 2}					
\deftwocell[mid = \Gamma']{midGam1 : 2 -> 2}				
\deftwocell[mid = \Delta]{midDel : 2 -> 2}					
\deftwocell[mid = \Delta']{midDel1 : 2 -> 2}					
\deftwocell[mid = \Sigma]{midSig : 2 -> 2}					
\deftwocell[mid = \Sigma']{midSig1 : 2 -> 2}					

\deftwocell[left = \Delta]{leftDel : 1 -> 1}					
\deftwocell[left = \Sigma]{leftSig : 1 -> 1}					
\deftwocell[left =\wn \Gamma]{leftGamw : 1 -> 1}				

\deftwocell[mid = ~]{topV : 0 -> 1}						
\deftwocell[mid = ~]{pitV : 1 -> 0}						
\deftwocell[mid = A]{pitA : 1 -> 0}						
\deftwocell[mid = {?}A]{pitAw : 1 -> 0}						
\deftwocell[mid = {!}A]{pitAo : 1 -> 0}						
\deftwocell[mid = A^\bot]{pitAb : 1 -> 0}					
\deftwocell[mid = A^{\bot\bot}]{pitAbb : 1 -> 0}					
\deftwocell[mid = B]{pitB : 1 -> 0}						
\deftwocell[mid = {?}B]{pitBw : 1 -> 0}						
\deftwocell[mid = {!}B]{pitBo : 1 -> 0}						
\deftwocell[mid = C]{pitC : 1 -> 0}						
\deftwocell[mid = D]{pitD : 1 -> 0}						
\deftwocell[mid = {!}C]{pitCo : 1 -> 0}						
\deftwocell[mid = \bot]{pitbot : 1 -> 0}						
\deftwocell[mid = 1]{pit1 : 1 -> 0}						
\deftwocell[mid = L]{pitL : 1 -> 0}						
\deftwocell[mid = R]{pitR : 1 -> 0}						
\deftwocell[mid = \alpha]{pitAl : 1 -> 0}						
\deftwocell[mid = \beta]{pitBe : 1 -> 0}						

\deftwocell[mid = A \otimes {B}]{pitAtenB : 1 -> 0}				
\deftwocell[mid = {(}A^\bot \otimes {B^\bot}{)}^\bot]{pitAbtenBbb : 1 -> 0}				

\deftwocell[mid = A \parr {B}]{pitAparB : 1 -> 0}				
\deftwocell[mid = B \otimes {C}]{pitBtenC : 1 -> 0}				
\deftwocell[mid = B \parr {C}]{pitBparC : 1 -> 0}				
\deftwocell[mid = \Gamma]{pitGam : 2 -> 0}					
\deftwocell[mid = \Gamma']{pitGam1 : 2 -> 0}					
\deftwocell[mid = \Gamma'']{pitGam2 : 2 -> 0}					
\deftwocell[mid = \Gamma''']{pitGam3 : 2 -> 0}					
\deftwocell[mid = {?}\Gamma]{pitGamw : 2 -> 0}				
\deftwocell[mid = {?}\Gamma']{pitGam1w : 2 -> 0}				
\deftwocell[mid = {?}\Gamma]{pitGamw0 : 0 -> 0}				
\deftwocell[mid = {?}\Delta]{pitDelw : 2 -> 0}					
\deftwocell[mid = {?}\Delta']{pitDel1w : 2 -> 0}				
\deftwocell[mid = \Delta]{pitDel  : 2 -> 0}					
\deftwocell[mid = \Delta']{pitDel1  : 2 -> 0}					
\deftwocell[mid = \Sigma]{pitSig : 2 -> 0}					
\deftwocell[mid = \Sigma']{pitSig1 : 2 -> 0}					
\deftwocell[mid = \sigma{(} {\Gamma } {)}]{pitsiG : 2 -> 0}			
\deftwocell[mid = W]{pitW : 2 -> 0}						
\deftwocell[mid = W_1]{pitWb1 : 2 -> 0}						
\deftwocell[mid = W_2]{pitWb2 : 2 -> 0}						
\deftwocell[mid = W']{pitW1 : 2 -> 0}						
\deftwocell[mid = W'_1]{pitW1b1 : 2 -> 0}						
\deftwocell[mid = W'_2]{pitW1b2 : 2 -> 0}						
\deftwocell[mid = W'']{pitW2 : 2 -> 0}						

\deftwocell[text=\phi, white]{phi11 : 1 -> 1} 					
\deftwocell[mid = F]{topF : 0 -> 1}						
\deftwocell[mid = G]{topG : 0 -> 1}						
\deftwocell[mid = G]{pitG : 1 -> 0}						
\deftwocell[mid = G\circ {F}]{pitGcircF : 1 -> 0}				

\deftwocell[text=\otimes,yellow]{ten : 2 -> 1}		
\deftwocell[rectangle]{ax : 0 -> 2}				
\deftwocell[rectangle]{cut : 2 -> 0}				
\deftwocell[text=\parr, orange]{par : 2 -> 1}			
\deftwocell[circle, white]{v : 0 -> 1}				
\deftwocell[circle, black]{bot : 0 -> 1}					

\deftwocell[text=A,lightgray]{axA : 0 -> 2}				
\deftwocell[text=A^\bot,lightgray]{axAb : 0 -> 2}			

\deftwocell[text=A,lightgray]{cutA : 2 -> 0}				
\deftwocell[text=A^\bot,lightgray]{cutAb : 2 -> 0}			

\deftwocell[text=?,white]{D : 1 -> 1}				
\deftwocell[text=?,white]{C : 2 -> 1}				
\deftwocell[text=?,white]{W : 0 -> 1}				
\deftwocell[text=!,white]{P : 3 -> 3}				
\deftwocell[text=!,white]{P4 : 4  -> 4}				
\deftwocell[text=!,white]{P5 :5 -> 5}				
\deftwocell[text=!,white]{P1 :1 -> 1}				
\deftwocell[text=!,white]{P6 : 6 -> 6}				
\deftwocell[text=!,white]{P7 : 7 -> 7}				
\deftwocell[text=!,white]{P8 : 8 -> 8}				
\deftwocell[text=!,white]{P9 : 9 -> 9}				

\deftwocell[rectangle]{axt : 2 -> 4}			
\deftwocell[rectangle]{cutt : 4 -> 2}			

\deftwocell[text=\otimes,yellow]{tenc : 4 -> 1}		
\deftwocell[rectangle]{axc : 0 -> 4}			
\deftwocell[rectangle]{cutc : 4 -> 0}			
\deftwocell[text=A {\parr}{B},lightgray]{cutcAparB : 4 -> 0}			
\deftwocell[rectangle,white]{vc : 0 -> 3}		

\deftwocell[text=A,lightgray]{axcA : 0 -> 4}					
\deftwocell[text=A^\bot,lightgray]{axcAb : 0 -> 4}				
\deftwocell[text=A^{\bot\bot},lightgray]{axcAbb : 0 -> 4}				
\deftwocell[text=A {\otimes} {B},lightgray]{axcAtenB : 0 -> 4}					
\deftwocell[text=B^\bot {\otimes} {A^\bot},lightgray]{axcBbtenAb : 0 -> 4}					

\deftwocell[lefthalfcircle,red]{L : 1 -> 1}			
\deftwocell[righthalfcircle,blue]{R : 1 -> 1}			

\deftwocell[text=AX, lightgray]{elax : 2 -> 2} 			
\deftwocell[text=Cut, lightgray]{elcut : 2 -> 2} 			
\deftwocell[text=c,brown]{tenpar : 2 -> 2} 				
\usepackage{graphicx}

\begin{document}
\title{Proof Diagrams for Multiplicative Linear Logic: Syntax and Semantics}
\author{Matteo Acclavio}

\maketitle

\begin{abstract}
Proof nets are a syntax for linear logic proofs which gives a coarser notion of proof equivalence with respect to  syntactic equality together with an intuitive geometrical representation of proofs.

In this paper we give an alternative $2$-dimensional syntax for multiplicative linear logic derivations. The syntax of string diagrams authorizes the definition of a framework where the sequentializability of a term, i.e.~ deciding whether the term corresponds to a correct derivation, can be verified in linear time. 

Furthermore, we can use this syntax to define a denotational semantics for multiplicative linear logic with units by means of equivalence classes of proof diagrams modulo a terminating rewriting.
\end{abstract}

\section{Introduction}
Proof nets are a geometrical representation of \emph{linear logic} proofs  introduced by J-Y.Girard \cite{girll}. The building blocks of this syntax are called \emph{proof structures}, later generalized by Y. Lafont~\cite{lafint} in the so-called \emph{interaction nets}. 
To recognize if a proof structure is a proof net one needs to verify its \emph{sequentializability} property, that is, verifying whether it corresponds to a correct linear logic proof derivation.
Following Girard's original correction criterion, others methods have been introduced: the method by Danos-Regnier \cite{dan-reg}, that ensures graph acyclicity by a notion of  \emph{switchings} on $\otimes$ cells, and the method by Guerrini \cite{guerrini}, that reformulates correction by means of graph contractability. Unfortunately the aforementioned criteria become ineffective in presence of the multiplicative unit $\bot$. In order to recover a sequentialization condition for the multiplicative fragment with units ($\MLLc$)  Girard has introduced the notion of \emph{jumps} \cite{girjump}. These are untyped edges which assign a $\bot$ to an axiom in order to represent a dependency relation of the respective rules in sequentialization. 

One peculiar feature of this syntax for proofs is that proof structures allow to recover the semantical equivalence of derivations under some inference rules permutations \cite{lafint}. In the case of the multiplicative fragment of linear logic ($\MLL$), proof nets perfectly capture this equivalence by giving a canonical representative for each class. On the other hand, in presence of multiplicative units, proof nets are not canonical \cite{MLLu} and have to  be identified up to jump re-assignation, ruling out a satisfactory notion of proof net \cite{noMLL}.

In this work give an alternative syntax for $\MLLc$ proofs. For this purpose, we replace the underlying interaction nets syntax with the one of string diagrams. We show that this syntax, which also presents an intuitive $2$-dimensional representation of proofs, is  able to capture some inference rule permutations in derivations.

String diagrams \cite{string} are a syntax with a rigid structure for $2$-arrows (or $2$-cells) of a $2$-category. Although the two syntaxes may graphically look similar, string diagrams' \emph{strings} do not just denote connections between cells but they represent morphisms.
Since crossing strings is not allowed without the introduction of \emph{twisting operators}, we introduce the notion of \emph{twisting relations} in order to equate diagrams by permitting cells to cross certain strings.

As soon as we consider a derivation of a proof as a sequence  of $n$-ary operators applications over lists of formulas, we are able to express it by means of string diagrams which  keep track of lists reordering. In a sense, string diagrams diagrams keep track of edge crossing in  pictorial representations of  proof nets.

We study several diagram rewriting systems given by \emph{twisting polygraphs}. In this particular class of polygraph \cite{Bur} string crossings are restrained to a specific family of strings, while some rewriting rules recover the graph representation equivalence.

The syntax of string diagrams allows us to define a polygraph where we introduce some \emph{control strings} in order to encode the correct parenthesization of operators. In particular, these strings prevent the representation of non-correct applications of inference rules,  resulting  into a sound framework where sequentializability, that is if a \emph{proof diagram} corresponds to a derivation, linearly depends on diagram inputs and outputs pattern only. 

Furthermore, this syntax induces an equivalence relation over linear logic derivations representable by the same proof diagrams. However, this equivalence does not capture all rule permutations required for the elimination of the so called \emph{commutative cuts}. In fact, these rule permutations require the permutation of  derivation tree branches as shown in the following case:

{
\begin{scprooftree}{0.7}
\AxiomC{$\overset 1 \vdots$}
\noLine
\UnaryInfC{$\vdash \Gamma , A,B$}
\AxiomC{$\overset 2 \vdots$}
\noLine
\UnaryInfC{$\vdash \Delta, C $}
\RightLabel{$ \otimes_1 $}
\BinaryInfC{$ \vdash \Gamma, \Delta , (B \otimes_1 C), A$}
\AxiomC{$\overset 3\vdots$}
\noLine
\UnaryInfC{$\vdash \s ,  D$}
			\RightLabel{$ \otimes_2 $}
\BinaryInfC{$ \vdash \Gamma,\Delta,\s, (A \otimes_1 D), (B\otimes_2 C)$}

\AxiomC{~}
\noLine
\UnaryInfC{$\sim$}
\noLine
\UnaryInfC{$~$}
\noLine
\UnaryInfC{$~$}

\AxiomC{$\overset 1 \vdots$}
\noLine
\UnaryInfC{$\vdash \Gamma , A,B$}

\AxiomC{$\overset 3 \vdots$}
\noLine
\UnaryInfC{$\vdash \s ,  D$}
\RightLabel{$ \otimes_2 $}
\BinaryInfC{$ \vdash \Gamma, \s , (A \otimes_1 D), B$}
\AxiomC{$\overset 2 \vdots$}
\noLine
\UnaryInfC{$\vdash \Delta, C $}
			\RightLabel{$ \otimes_1 $}
\BinaryInfC{$ \vdash \Gamma,\Delta,\s, (A \otimes_1 D), (B\otimes_2 C)$}

\noLine
\TrinaryInfC{}
\end{scprooftree}
}
If a syntax does not equate derivations differing for rule permutations, it is crucial for a \emph{cut-elimination} theorem to explicitly authorize them.
On the other hand, this syntax makes equivalent some proofs which are representable by proof nets differing in jumps assignation only. 

With the purpose of keeping this last nice feature and extend the equivalence to include the missed rules permutations, we here extend the results presented in \cite{AccMLL} by enriching our polygraph with some additional generators and rewriting rules. The equivalence induced by these rewriting rules induces an equivalence over derivation (seen as syntactical expressions) effective to identify all and only $\MLLc$ derivations which we use to consider equivalent (with respect of independent inference rules permutations).

Extending the polygraph with the rewriting rules for cut-elimination achieving a a relative cut-elimination theorem. We conclude by giving a denotational semantics  \cite{girard1991new} for $\MLLc$ proofs by means of equivalence classes of proof diagrams.

\section{String diagrams}

In this section we recall some basic notions in string diagram rewriting \cite{string}. For an introduction to this syntax see Selinger's survey \cite{Sel} and refer to  in  J.Baez's notes \cite{baez} for some  interesting observations on the motivation and applications of this formalism.

Given two lists $\Gamma=\Gamma_1 * \dots * \Gamma_n$ and $\Delta=\Delta_1 * \dots * \Delta_m$ of symbols in an alphabet $\s$, a \emph{string diagram} $\phi : \Gamma \frr \Delta$ with \emph{inputs} $\inp(\phi)=\Gamma$ and \emph{outputs} $\outp(\phi)=\Delta$ is pictured as follows:
\vspace{-.3cm}
$$
\twocell{topGam *1 d *1 phi *1 d *1 pitDel}
$$

A string diagram can be interpreted as a function with multiple inputs and outputs  of type  respectively $\Gamma_1, \dots \Gamma_n$ and $\Delta_1 , \dots , \Delta_m$. 
Diagrams may be composed in two different ways. If $\phi : \Gamma \frr \Delta$ and $\phi': \Gamma' \frr \Delta'$ are diagrams, we define:
\begin{itemize}

\item \emph{sequential} composition: if $\Delta=\Gamma'$, the diagram $\phi' \circ \phi : p\frr q'$ corresponds to usual composition of maps as the notation suggests.

This composition is associative with units $\id_\Gamma: \Gamma \frr \Gamma$ for each possible list of inputs $\Gamma$. In other words, we have $\phi \circ \id_{\inp (\phi)}= \phi= \id_{\outp (\phi)} \circ \phi$. The\emph{ identity diagram} $id_\Gamma$ is pictured as follows: $\twocell{topGam *1 d *1 pitGam }$

\item \emph{parallel}  composition: the diagram $\phi * \phi': \Gamma * \Gamma' \frr \Delta * \Delta'$ is always defined.
This composition is associative with unit $\id_0: \emptyset \frr \emptyset$. In other words, we have $\id_0 * \phi=\phi=\phi * \id_0$. This $\id_0$ is called the \emph{empty diagram}.
\end{itemize}

These two compositions are respectively represented as follows:
$$
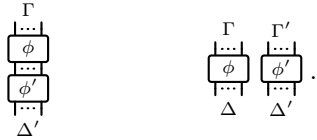
\twocell{topGam *1 d *1 phi *1 d *1 phi1 *1 d *1 pitDel1}
\qquad \qquad \qquad
\twocell{(topGam *0 topGam1)*1(d *0 d)*1( phi *0 phi1)*1(d *0 d)*1(pitDel *0 pitDel1)}\; .
$$
Our two compositions satisfy the \emph{interchange rule}: if $\phi:\Gamma \frr \Delta$ and $\phi' : \Gamma'\frr \Delta'$, then 
$$ (\id_\Delta * \phi' ) \circ (\phi * \id_{\Gamma'})= \phi * \phi' =(\phi * \id_{\Delta'})\circ (\id_\Gamma * \phi')$$ 
that corresponds to the following picture:
$$\twocell{(topGam *0 topGam1) *1 (d *0 2) *1 (phi *0 d) *1 (d *0 phi1) *1 (2 *0 d)*1 (pitDel *0 pitDel1)}=\twocell{(topGam *0 topGam1) *1 (d *0 d) *1 (phi *0 phi1) *1 (d *0 d)*1 (pitDel *0 pitDel1)}=\twocell{(topGam *0 topGam1) *1 (2 *0 d) *1  (d *0 phi1) *1 (phi *0 d) *1 (d *0 2)*1 (pitDel *0 pitDel1)}$$

String diagrams are a formalism for morphisms in a strict monoidal category with objects finite lists of symbols over an alphabet $\s$. The sequential composition $\circ$ denotes the usual morphisms composition while the product is the list concatenation and it is denoted by $*$.

\begin{definition}[Signature]
Fixed an alphabet $\s$ we denote by $\s^*$ the set of \emph{words} or \emph{lists} over $\s$. A \emph{signature} $\sig$ is a set of \emph{atomic diagrams} (or \emph{gates type}). Given a signature, a diagram $\phi : \Gamma \frr \Delta$ (with $\Gamma,\Delta \in \s^*$) represents a morphism in the  monoidal category $\sig ^*$ in which morphisms are freely generated by $\sig$, i.e.~by the two compositions $*$ and $\circ$ and identities. A \emph{gate} is an occurrence of an atomic diagram, we denote $g: \alpha$ or we say that $g$ is an $\alpha$-gate if $g$ is an occurrence of $\alpha \in \sig$.
\end{definition}

\begin{definition}
We say that $\phi$  is a \emph{subdiagram} of $\phi'$ if and only if there exist $ \psi_u,\psi_d\in \sig^*$ and $\Gamma, \Delta$ such that $\phi'=\chi_d \circ (\id_\Gamma *\phi*\id_{\Delta})\circ \chi_u $.
\end{definition}

\nota Given $\phi\in \sig^*$ and $\sig' \subseteq \sig$, we write $|\phi|_{\sig'}$ the number of gates in $\phi$ with gate type $\alpha \in \sig'$.

\begin{definition}
We call \emph{horizontal} a diagram $\phi$ generated by parallel composition (and identities) only in $\sig^*$. It is \emph{elementary} if $|\phi|_\sig=1$.
\end{definition}

\subsection{Diagram rewriting}
\begin{definition}[Diagram Rewriting System]
Fixed an alphabet $\s$, a \emph{diagram rewriting system} is  a couple $(\sig, \R)$ given by a signature $\sig$ and a set $\R$ of rewriting rules of the form
$$
\twocell{topGam *1 d *1 phi *1 d *1 pitDel}
\xymatrix{\ar@3[r]&}
\twocell{topGam *1 d *1 phi1 *1 d *1 pitDel}
$$
where $\D, \D': \Gamma \frr \Delta$ are diagrams in $\sig^*$ with same inputs and outputs. We call $\phi$ and $\phi'$ respectively \emph{source} and \emph{target} of the rewriting rules.
\end{definition}

\begin{definition}
We allow each rewriting  rule under any context, that is, if  $\xymatrix@C=1em{ \phi \ar@3[r] & \phi'}$ in $\R$ then, for every $\chi_u , \chi_d\in \sig^*$,
$$
\twocell{d *1 up *1 (2 *0 d *0 2) *1 (d *0 phi *0 d ) *1 ( 2 *0 d *0 2) *1 do *1 d}
\xymatrix{\ar@3[r]&}
\twocell{d *1 up *1 (2 *0 d *0 2) *1 (d *0 phi1 *0 d ) *1 ( 2 *0 d *0 2) *1 do *1 d} \; .$$
 We say that $\psi$ \emph{reduces}, or \emph{rewrites}, to $\psi'$  (denoted $\xymatrix@C=1em{\psi \ar@3[r]^{*} & \psi'}$) if there is a \emph{rewriting sequence} $P: \xymatrix@C=1em{\psi=\psi_0 \ar@3[r] & \psi_1 \ar@3[r] & \dots  \ar@3[r] & \psi_n=\psi'}$.
\end{definition}

We here recall some classical notions in rewriting:

\begin{itemize}
\item A diagram $\phi$ is \emph{irreducible} if there is no $\phi'$ such that $\xymatrix@C=1em{\phi \ar@3[r] & \phi'}$;

\item A rewriting system \emph{terminates} if there is  no infinite rewriting sequence;

\item A rewriting system is \emph{confluent} if for all $\phi_1,\phi_2$ and $\D$ such that $\xymatrix@C=1em{\phi \ar@3[r] & \phi_1}$ and $\xymatrix@C=1em{\phi \ar@3[r] & \phi_2 }$, there exists $\phi' $ such that $\xymatrix@C=1em{\phi_1 \ar@3[r]^{*} & \phi'} $  and $\xymatrix@C=1em{\phi_2 \ar@3[r]^{*} & \phi'}$;

\item A rewriting system is \emph{convergent} if both properties hold.

\end{itemize}

\section{Polygraphs}

In this section we formulate some basic notion in string diagram rewriting by using the language of \emph{polygraphs}. 
Introduced by Street \cite{Street} as \emph{computads}, later reformulated and extended by Burroni \cite{Bur}, polygraphs can be considered as the generalization for higher dimensional categories of the notion of monoid presentation and the construction of the free category generated by a quiver.

Here we study some diagram rewriting systems with labels on strings in terms of $3$-polygraphs, which are denoted $\s=(\s_0,\s_1,\s_2,\s_3)$.
In particular, we consider polygraphs with just one $0$-cell in $\s_0$ in order to avoid background labeling. The set of $1$-cells $\s_1$ represents string labels, the $2$-cells in $\s_2$ are the  signature $\sig_\s$ of our rewriting system with rules $\R_\s=\s_3$, the set of $3$-cells. We say that a polygraph $\s$ exhibits some computational properties when the relative diagram rewriting system does. 

\nota   We denote $\phi\in \s$ whenever $\phi $ is a diagram generated by the associated signature $\sig_\s$. If $\s$ is a $3$-polygraph with one $0$-cell, we denote by $\catgen{\s}$ the monoidal category with objects words in $\s_1$ and morphisms $[\phi]$ (we denote $[\phi]\in\catgen \s$) the equivalence classes of diagrams $\phi\in\sig^*$ modulo $\R_\s$. We say that $\s'$ \emph{extends} $\s$ if $\s'$ can be obtained by $\s$ by extending the sets of $i$-cells, that is $\s_i\subseteq \s'_i$ for all $i$.

\subsection{Twisting Polygraph}
In this section we introduce a notion of polygraph which generalizes polygraphic presentations of symmetric monoidal categories.

\begin{definition}[Symmetric polygraph]
We call the \emph{polygraph of permutation} the following monochrome $3$-polygraph:
$$\mathfrak S=\biggl(\s_0=\{\square\}, \s_1=\{\twocell{1}\} , \s_2=\{\twocell{s} \}, \s_3=\biggl\{ \xymatrix@C-1pc{ \twocell{s *1 s } \ar@3[r] & \twocell{2}}, \; 
\xymatrix@C-1pc{ {\twocell{(s *0 1)*1 (1 *0 s)*1 (s *0 1)}} \ar@3[r] & {\twocell{ (1 *0 s)*1 (s *0 1)*1 (1 *0 s) }}}\biggr\} \biggr)$$

We call \emph{symmetric} a  $3$-polygraph $\s$ with one $0$-cell, one $1$-cell (i.e. $\s_1=\{ \twocell{1}\}$), containing one $2$-cell $\twocell{s} \in \s_2$ and such that the following holds

$$\twocell{s *1 s } = \twocell{2} \quad \mbox{, }\quad \twocell{(d *0 1) *1 (a *0 1) *1 (d*0 1) *1 rn *1 (1*0 d)} =  \twocell{ (d*0 1) *1 rn *1 (1*0 d) *1 (1*0 a)*1 (1 *0 d)}
 \qquad \mbox{and} \qquad 
\twocell{(1 *0 d) *1 (1 *0 a) *1 (1*0 d) *1 ln *1 (d*0 1)} =  \twocell{ (1*0 d) *1 ln *1 (d*0 1) *1 (a*0 1)*1 (d *0 1)} \quad \mbox{ for all } \alpha \in \s_2 $$
in the $2$-category $\s^*$. In such $3$-polygraph to denote diagrams inputs and outputs it suffices to provide the respective numbers of their input and output strings.
\end{definition}

\begin{theorem}[Convergence of $\mathfrak S$]\label{permconf}
The polygraph $\mathfrak S$ is convergent.
\begin{proof}
As in \cite{LafBool}, in order to prove termination we interprete every diagram $\phi: n\frr m \in  \mathfrak S^*$ with a monotone function $[\phi]:\N^n\fr \N^m$. These have well-founded partial order induced by product order on $\N^p$ ($\bar x =(x_1, \dots,x_p)\leq(y_1, \dots y_p)=\bar y$ whenever $x_1\leq x_1 \land \dots \land x_p \leq y_p$):
$$f,g: \N^{*p}\fr \N^{*p} \mbox{ then } f< g \mbox { iff } f(\bar x)< g(\bar x) \mbox{ for all } \bar x \in \N^{*p}.$$
We interpret the gate $\twocell{s} $ by the function $[\;\twocell{s}\;] (x,y)\fr (y, x+y)$.
This allow us to associate to any $3$-cell $\xymatrix@C-1pc{ \phi \ar@3[r] & \psi}$ two monotone maps  $[\phi]$ and $[\psi]$ such that $[\phi]> [\psi]$:
{\small
$$\begin{gathered}
\Big[\; \twocell{s *1 s }\;\Big] (x,y)= (2x+y,x+y)> (x,y)=\Big[\;\twocell{2}\;\Big](x,y), \\
\Bigg[\;{\twocell{(s *0 1)*1 (1 *0 s)*1 (s *0 1)}}\;\Bigg](x,y,z)=(2x+y+z, x+y,x)> (x+y+z,x+y,x)=\Bigg[\;{\twocell{ (1 *0 s)*1 (s *0 1)*1 (1 *0 s) }}\;\Bigg](x,y,z)
\end{gathered}$$}
By the compatibility of the order with sequential and parallel composition, this suffices to prove that, for any couple of  diagrams, $[\phi] > [\psi]$ holds if $\xymatrix@C-1pc{\phi \ar@3^{*}[r]& \psi}$. Since this order on monotone maps on integers admits no infinite decreasing chain, infinite reduction paths can not exist.

In order to prove convergence, it suffices to check the confluence of the following critical peaks, that are the minimal critical branchings of the rewriting system (see \cite{AccCohe} for details):
$$
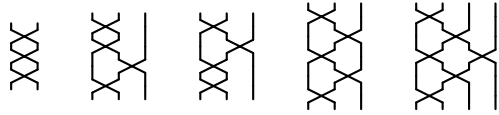
\twocell{s *1 s *1 s}								\quad \quad
\twocell{(s *0 1) *1 (s *0 1)*1 (1 *0 s)*1 (s *0 1)} 				\quad \quad
\twocell{(s *0 1)*1 (1 *0 s)*1 (s *0 1) *1 (s *0 1)} 				\quad \quad
\twocell{(s *0 1) *1 (1*0 s) *1 (s *0 1)*1 (1 *0 s)*1 (s *0 1)} 		\quad \quad
\twocell{(s *0 2) *1 (1*0 s *0 1) *1 (s *0 s)*1 (1 *0 s *0 1) *1 (s *0 2)}	
$$  
\end{proof}
\end{theorem}

Each diagram in $\mathfrak S$ can be interpreted as a permutation in the \emph{group of permutations over $n$ elements} $S_n$ with product $\circ$ defined as their function composition. On the other hand, each $\sigma\in S_n$ corresponds to some diagrams in $\mathfrak S$. In particular, we interpret the diagram $\id_{k-1} * \twocell{s} * \id_{n-(k+1)}: n \frr n$ as the transposition $(k,k+1)\in S_n$.  

\nota  We note $Lad^l_n= \twocell{ (1*0 d) *1 ln *1 (d*0 1) }: n\frr n $ and $Lad^r_n= \twocell{(d*0 1) *1 rn *1 (1*0 d) }:n\frr n$ the left and right \emph{ladder} diagrams corresponding respectively to the permutations $(1,n,n-1, \dots , 2)$ and $(n,1,2, \dots, n-1)$ in $S_n$.

\begin{proposition}\label{corrPer}
For any permutation $\sigma\in S_n$ there is a unique diagram in normal form $\hat \phi_\sigma: n\frr n \in \mathfrak S$ corresponding to $\sigma$. We call it the \emph{canonical diagram of $\sigma$}.
\begin{proof}
We define $\mathfrak S_1=\{\twocell{1}\}$ and $\mathfrak S_{n+1}$ the set of diagrams in $ \mathfrak S$ of the form:
$$\twocell{(1 *0 d) *1 (1 *0 sigma1) *1 (1 *0 d) *1 (g35) *1 (d *0 1 *0 d)}=\hat\phi_\sigma: n+1\frr n+1 $$
with $\twocell{ d *1 sigma1 *1 d} \in \mathfrak S_{n}$ and $\twocell{(1 *0 d)*1 g35 *1 (d *0 1 *0 d)} = \twocell{(1 *0 d *0 2)*1 (ln *0 d) *1( d *0 3)}= Lad^l_k * id_{(n+1-k)}$. 
We have $|\mathfrak S_n|=n!$ since $|\mathfrak S_1|=1$ and  $|\mathfrak S_{n+1}|=(n+1)|\mathfrak S_{n}| $ on account of $n+1=|\{Lad^l_k\}_{1\leq k\leq n+1}|=|\{Lad^l_k* \id_{(n+1-k)}\}_{1\leq k\leq n+1}|$.

To exhibit a one-to-one correspondence between $S_{n+1}$ and $\mathfrak S_{n+1}$, for any $\sigma \in S_{n+1}$ we define $Er(\sigma)\in S_n$ as the permutation
$$Er(\sigma)=\begin{cases}i \fr \sigma(i+1) & {if }\quad \sigma (i+1)<\sigma (1) \\ i \fr \sigma(i+1)+1 & {if }\quad \sigma (1)<\sigma (i+1)\end{cases}.$$
and $\hat \phi _\sigma=( Lad^l_k* \id_{(n+1-\sigma(1))}) \circ (\id_1 * \hat \phi _{Er(\sigma)})$.

No element in $\mathfrak S_n$ contains subdiagram of the form $\twocell{s *1 s}$ nor $\twocell{(s *0 1) *1 (1 *0 s) *1 (s *0 1)}$. This means that they are irreducible and so, by the confluence of $\mathfrak S$, in normal form.
\end{proof}
\end{proposition}

\begin{definition}[Twisting polygraph]
A \emph{twisting polygraph} is  a $3$-polygraph $\s$ with one $0$-cell equipped with a set $T_\s\subseteq \s_1 $ called \emph{twisting family} such that  for each $A,B \in T_\s$ there is a \emph{twisting operator} $\twocell{s}_{ A, B}: A *B \frr B*A \in \s_2$  and $\s_3$  includes the following families $T_\R$ of \emph{twisting relations}:
\begin{itemize}
\item For all $A,B,C \in T_\s$:

\begin{equation}\label{twist1}\xymatrix{ {\twocell{(topA *0 topB )*1(s *1 s) }} \ar@3[r] & {\twocell{(midA *0 midB)*1 2}}} \qquad \mbox{ and }\qquad 
\xymatrix{ \twocell{(topA *0 topB *0 topC)*1(s *0 1)*1 (1 *0 s)*1 (s *0 1)*1(pitC *0 pitB *0 pitA)} \ar@3[r] &{\twocell{ (topA *0 topB *0 topC)*1(1 *0 s)*1 (s *0 1)*1 (1 *0 s) *1(pitC *0 pitB *0 pitA) }}} 
\quad ; \end{equation}

\item For all $\alpha: \Gamma \fr \Gamma' \in \s_2$ with $\Gamma, \Gamma'\in T_\s^*$, $A\in T_\s$, at least one of the two possible orientation of the following rewriting rules is in $\s_3$:

\begin{equation}\label{twist2}
\xymatrix{ \twocell{(topGam *0 topA )*1 (d *0 1) *1 (a *0 1) *1 (d*0 1) *1 rn *1 (1*0 d) *1 (pitA *0 pitGam1)}  \ar@3@/^1pc/[r] &  
\twocell{ (topGam *0 topA)*1(d*0 1) *1 rn *1 (1*0 d) *1 (1*0 a)*1 (1 *0 d)*1 (pitA *0 pitGam1)} \ar@3@/^1pc/[l]}
\qquad \mbox{and} \qquad
\xymatrix{ \twocell{ (topA *0 topGam)*1 (1 *0 d) *1 (1 *0 a) *1 (1*0 d) *1 ln *1 (d*0 1) *1 (pitGam1 *0 pitA)} \ar@3@/^1pc/[r] & \twocell{(topA *0 topGam)*1 (1*0 d) *1 ln *1 (d*0 1) *1 (a*0 1)*1 (d *0 1)*1 (pitGam1 *0 pitA)}\ar@3@/^1pc/[l] }
\quad.
\end{equation}
\end{itemize}
Moreover, if $\phi, \psi$ are \emph{twisting diagrams} (i.e. diagrams made only of twisting operators) $\xymatrix@C=1em{\phi \ar@3[r]^{*}_{~\R_\s} & \psi}$ iff $\xymatrix@C=1em{\phi \ar@3[r]^{*}_{~\R_T} & \psi}$ where $\R_T$ is the set given by rewriting rules of (\ref{twist1}).
A \emph{total-twisting polygraphy} is a twisting polygraph with $T_\s=\s_1$.
\end{definition}

The idea  behind twisting polygraphs is to present diagram rewriting systems where, in equivalence classes modulo rewriting, the crossings of strings labeled by the twisting family are not taken into account. In fact, the family of relations (\ref{twist1}) says that these crossings are involutive and satisfy Yang-Baxter equation \cite{YBE} for braidings, while relations in (\ref{twist2}) allow gates to ``cross'' a string in case of fitting  labels. 

We interpret a twisting diagram $\phi_\sigma :\Gamma \frr \sigma(\Gamma)$ as the permutations  in $S_{|\Gamma|}$ acting over the order of occurrence of $1$-cells in the word $\Gamma \in T_\s^*$. For this reason, as in $\mathfrak S$, we define left ladders, right ladders and the standard diagrams $\hat \phi^\Gamma_\sigma : \Gamma \fr \sigma(\Gamma)$ (or simply $\hat \phi _\sigma$)  with source and target in $T_\s^*$. In conformity with the twisting polygraph restrictions over $\s_3$, we can prove the uniqueness of $\hat \phi_\sigma$ as in Proposition \ref{corrPer}.

\section{Multiplicative Linear Logic sequent calculus}\label{SecLL}

In this paper we focus on the multiplicative fragment of linear logic sequent calculus with units.
We here we recall the usual inference rules:

{\small\begin{tabular}
{p{2,5cm} |p{4cm}|p{4cm}}
&  Identity or Axiom  &  Cut \\
Structural &
{\begin{prooftree}
\AxiomC{}
\RightLabel{$Ax$}
\UnaryInfC{$ \vdash A, A^\bot$}
\end{prooftree}}
&
{\begin{prooftree}
\AxiomC{$\vdash \s , A$}
\AxiomC{$\vdash \Gamma, A^\bot$}
\RightLabel{$Cut$}
\BinaryInfC{$ \vdash \s , \Gamma$}
\end{prooftree}}\\
\hline
&  Tensor & Par\\
 Multiplicative &
\begin{prooftree}
\AxiomC{$\vdash \s , A$}
\AxiomC{$\vdash B , \Gamma $}
\RightLabel{$\otimes $}
\BinaryInfC{$ \vdash \s,  (A\otimes B), \Gamma $}
\end{prooftree}
&
\begin{prooftree}
\AxiomC{$\vdash \s, A,B $}
\RightLabel{$\parr $}
\UnaryInfC{$ \vdash \s , A\parr B$}
\end{prooftree}\\
\hline
& Bottom & 1\\
Units &
 \begin{prooftree}
\AxiomC{$\vdash \s $}
\RightLabel{$\bot $}
\UnaryInfC{$ \vdash \s ,\bot$}
\end{prooftree}
&
\begin{prooftree}
\AxiomC{$  $}
\RightLabel{$1 $}
\UnaryInfC{$ \vdash \mathit {1}$}
\end{prooftree}
\end{tabular}}

We also consider the usually omitted exchange rule: 
\begin{prooftree}
\AxiomC{$\vdash A_1, \dots , A_k $}
\RightLabel{$ \sigma \in S_k $}
\UnaryInfC{$\vdash A_{\sigma(1)}, \dots , A_{\sigma(k)}$}
\end{prooftree}

We call \emph{principal} a formula which occurs in the conclusion of a rule but does not occur in the premise(s) and \emph{active} a formula which occurs in the premise of a rule but not in the conclusion. In a derivation $d(\Gamma)$, we say that a $Cut$ rule is \emph{commutative} when one of its active formulas is not principal. Moreover, a commutative cut is \emph{pure} if the non-principal active formula is principal for a $\otimes$ rule.
A \emph{cut-free derivation} is a derivation with no occurrences of $Cut$ rules.

We finally recall that the \emph{multiplicative linear logic fragment with units} ($\MLLc$) is given by the aforementioned inference rules, while the \emph{multiplicative fragment} ($\MLL$) is the one given by the inference rules $Ax, Cut, \otimes , \parr$ (and exchange) only.

\begin{oss}[On Negation]\label{remNeg}
We assume negation to be involutive, i.e.~$A^{\bot \bot}=A$ and the De Morgan's laws to apply with respect to $\parr$ and $\otimes$, i.e. $(A \heartsuit B)^\bot= B^\bot \heartsuit^\bot A^\bot$ for any formulas $A,B$ where $\heartsuit=\parr$ and $\heartsuit^\bot=\otimes$ or vice versa $\heartsuit= \otimes $ and $\heartsuit^\bot= \parr$. Moreover $1^\bot=\bot$.
\end{oss}

\begin{oss}[On Rules]\label{arity}
In this work we interpret inference rules as operators with specific arities over the set of sequents: $Ax$ and $1$ are $0$-ary, $\parr$ and $\bot$ are unary and $\otimes$ and $Cut$ are binary.
\end{oss}

\nota We indicate with $\fmll$ and $\fmllc$ the sets of formulas respectively in $\MLL$ and $\MLLc$. Moreover we indicate with $\smll$ and $\smllc$ their respective sets of sequents.

In the formalism of sequent calculus, oftentimes two derivations are identified when they can be transformed one into the other by a sequence of permutations over inference rules. Indeed, this identification is crucial for a cut-elimination result whenever we face a commutative cuts.
In this paper we consider the equivalence among derivations only form a syntactical viewpoint: namely, two derivations are considered equal if and only if they display exactly the same sequents (multisets of formulas) and the same rules in the same order. We then formalize the equivalence relation $\sim$ over $\MLLc$ derivations given by the permutation of inference rules with disjoint sets of active formula occurrences:

\begin{definition}\label{standeqLL}
We define the \emph{standard equivalence over $\MLLc$ derivations} (denoted by $\sim$) as the equivalence derivations generated by the following equivalences for all $A,B,C,D\in \fmllc$, $\Gamma, \Delta, \s \in \smllc$:

\begin{itemize}
\item If  $\odot_1,\odot_2 \in \{\parr, \bot\}$:
\end{itemize}

\begin{scprooftree}{.7}
\AxiomC{$\vdots$}
\noLine
\UnaryInfC{$\vdash \Gamma , \Delta, \s$}
\RightLabel{$ \odot_1 $}
\UnaryInfC{$ \vdash \odot_1(\Gamma),  \Delta, \s$}
\RightLabel{$ \odot_2 $}
\UnaryInfC{$ \vdash  \odot_1(\Gamma), \odot_2(\Delta),\s$}

\AxiomC{~}
\noLine
\UnaryInfC{$\sim$}
\noLine
\UnaryInfC{$~$}
\noLine
\UnaryInfC{$~$}

\AxiomC{$\vdots$}
\noLine
\UnaryInfC{$\vdash \Gamma , \Delta,\s$}

\RightLabel{$ \odot_2 $}
\UnaryInfC{$ \vdash  \Gamma, \odot_2(\Delta),\s$}
\RightLabel{$ \odot_1 $}
\UnaryInfC{$ \vdash \odot_1(\Gamma),   \odot_2(\Delta),\s$}

\noLine
\TrinaryInfC{}
\end{scprooftree}

where

$$\odot_1(\Gamma)= \left\{\begin{array}{lll} 
\Gamma,\bot  & \mbox{ if } \odot_1=\bot \\ 
\Gamma', A\parr B & \mbox{ if } \odot_1=\parr \mbox { and } \Gamma=\Gamma',A,B
\end{array}\right. $$

and

$$\odot_2(\Delta)= \left\{\begin{array}{lll} 
\Delta,\bot  & \mbox{ if } \odot_2=\bot \\ 
\Delta', C\parr D & \mbox{ if } \odot_2=\parr \mbox { and } \Delta=\Delta',C,D
\end{array}\right. $$

\begin{itemize}
\item If  $\odot_1\in \{\otimes, Cut\}$, $\odot_2 \in \{\parr, \bot\}$:
\end{itemize}

\begin{scprooftree}{.7}
\AxiomC{$\vdots$}
\noLine
\UnaryInfC{$\vdash \Delta , A$}
\AxiomC{$\vdots$}
\noLine
\UnaryInfC{$\vdash  B, \Gamma ,\s$}
\RightLabel{$ \odot_1 $}
\BinaryInfC{$ \vdash \Delta, \odot_1 (A, B), \Gamma,\s$}
			\RightLabel{$ \odot_2 $}
\UnaryInfC{$ \vdash \Delta, \odot_1(A, B), \odot_2(\Gamma),\s$}

\AxiomC{~}
\noLine
\UnaryInfC{$\sim$}
\noLine
\UnaryInfC{$~$}
\noLine
\UnaryInfC{$~$}

\AxiomC{$\vdots$}
\noLine
\UnaryInfC{$\vdash \Delta , A$}
\AxiomC{$\vdots$}
\noLine
\UnaryInfC{$\vdash  B, \Gamma ,\s$}
			\RightLabel{$ \odot_2 $}
\UnaryInfC{$ \vdash \Delta,B, \odot_2(\Gamma),\s$}

\RightLabel{$ \odot_1 $}
\BinaryInfC{$ \vdash \Delta, \odot_1 (A , B), \odot_2(\Gamma),\s$}

\noLine
\TrinaryInfC{}

\end{scprooftree}

and

\begin{scprooftree}{.7}

\AxiomC{$\vdots$}
\noLine
\UnaryInfC{$\vdash \Gamma , A,\s$}
\AxiomC{$\vdots$}
\noLine
\UnaryInfC{$\vdash  B, \Delta $}
\RightLabel{$ \odot_1 $}
\BinaryInfC{$ \vdash \Gamma, \odot_1 (A , B), \Delta,\s$}
			\RightLabel{$ \odot_2 $}
\UnaryInfC{$ \vdash \odot_2(\Gamma),\odot_1 (A , B), \Delta,\s$}

\AxiomC{~}
\noLine
\UnaryInfC{$\sim$}
\noLine
\UnaryInfC{$~$}
\noLine
\UnaryInfC{$~$}

\AxiomC{$\vdots$}
\noLine
\UnaryInfC{$\vdash \Gamma , A,\s$}
			\RightLabel{$ \odot_2 $}
\UnaryInfC{$ \vdash \odot_2(\Gamma),A,\s$}
\AxiomC{$\vdots$}
\noLine
\UnaryInfC{$\vdash  B, \Delta$}

\RightLabel{$ \odot_1 $}
\BinaryInfC{$ \vdash \odot_2(\Gamma), \odot_1 (A , B), \Delta,\s$}

\noLine
\TrinaryInfC{}

\end{scprooftree}

where

$$\odot_1(A,B)= \left\{\begin{array}{lll} 
A \otimes B  & \mbox{ if } \odot_1=\otimes \\ 
\emptyset & \mbox{ if } \odot_1=Cut \mbox{ and } A=B^\bot
\end{array}\right. $$

$$\odot_2(\Gamma)= \left\{\begin{array}{lll} 
\Gamma,\bot  & \mbox{ if } \odot_2=\bot \\ 
\Gamma', A\parr B & \mbox{ if } \odot_2=\parr \mbox { and } \Gamma=\Gamma',A,B 
\end{array}\right. $$

\begin{itemize}
\item If  $\odot_1, \odot_2\in \{\otimes, Cut\}$:
\end{itemize}

\begin{scprooftree}{.7}
\AxiomC{$\vdots$}
\noLine
\UnaryInfC{$\vdash \Gamma , A$}
\AxiomC{$\vdots$}
\noLine
\UnaryInfC{$\vdash \s, B, C $}
\RightLabel{$ \odot_1 $}
\BinaryInfC{$ \vdash \Gamma, \s ,  \odot_1(A, B), C$}
\AxiomC{$\vdots$}
\noLine
\UnaryInfC{$\vdash \Delta ,  D$}
			\RightLabel{$ \odot_2 $}
\BinaryInfC{$ \vdash \Gamma,\s,\Delta,  \odot_1(A , B), (C\odot_2 D)$}

\AxiomC{~}
\noLine
\UnaryInfC{$\sim$}
\noLine
\UnaryInfC{$~$}
\noLine
\UnaryInfC{$~$}

\AxiomC{$\vdots$}
\noLine
\UnaryInfC{$\vdash \Gamma , A$}
               \AxiomC{$\vdots$}
                \noLine
               \UnaryInfC{$\vdash \s, B, C $}
			\AxiomC{$\vdots$}
			\noLine
			\UnaryInfC{$\vdash \Delta ,  D$}
			\RightLabel{$ \odot_2$}
		\BinaryInfC{$ \vdash \s, \Delta, B,  \odot_2(C, D)$}

\RightLabel{$ \odot_1 $}
\BinaryInfC{$ \vdash \Gamma, \s,\Delta ,  \odot_1(A , B),  \odot_2(C, D)$}

\noLine
\TrinaryInfC{}
\end{scprooftree}

\begin{scprooftree}{.7}
\AxiomC{$\vdots$}
\noLine
\UnaryInfC{$\vdash \Gamma , A,C$}
\AxiomC{$\vdots$}
\noLine
\UnaryInfC{$\vdash \s, B $}
\RightLabel{$ \odot_1 $}
\BinaryInfC{$ \vdash \Gamma, \s ,  \odot_1(A, B), C$}
\AxiomC{$\vdots$}
\noLine
\UnaryInfC{$\vdash \Delta ,  D$}
			\RightLabel{$ \odot_2 $}
\BinaryInfC{$ \vdash \Gamma,\s,\Delta,  \odot_1(A , B), (C\odot_2 D)$}

\AxiomC{~}
\noLine
\UnaryInfC{$\sim$}
\noLine
\UnaryInfC{$~$}
\noLine
\UnaryInfC{$~$}

\AxiomC{$\vdots$}
\noLine
\UnaryInfC{$\vdash \Gamma , A,C$}
\AxiomC{$\vdots$}
\noLine
\UnaryInfC{$\vdash \Delta, D $}
\RightLabel{$ \odot_2 $}
\BinaryInfC{$ \vdash \Gamma, \Delta ,  A, \odot_2( C,D)$}
\AxiomC{$\vdots$}
\noLine
\UnaryInfC{$\vdash \s ,  B$}
			\RightLabel{$ \odot_1 $}
\BinaryInfC{$ \vdash \Gamma,\s,\Delta,  \odot_1(A , B), (C\odot_2 D)$}

\noLine
\TrinaryInfC{}
\end{scprooftree}

\begin{scprooftree}{.7}
\AxiomC{$\vdots$}
\noLine
\UnaryInfC{$\vdash\s, C $}
               \AxiomC{$\vdots$}
                \noLine
               \UnaryInfC{$\vdash \Gamma , A $}
			\AxiomC{$\vdots$}
			\noLine
			\UnaryInfC{$\vdash \Delta ,  D,B$}
			\RightLabel{$ \odot_1$}
		\BinaryInfC{$ \vdash \Gamma, \Delta, D,  \odot_1(A, B)$}
\RightLabel{$ \odot_2 $}
\BinaryInfC{$ \vdash \Gamma, \s,\Delta ,  \odot_1(A , B),  \odot_2(C, D)$}

\AxiomC{~}
\noLine
\UnaryInfC{$\sim$}
\noLine
\UnaryInfC{$~$}
\noLine
\UnaryInfC{$~$}

\AxiomC{$\vdots$}
\noLine
\UnaryInfC{$\vdash \Gamma , A$}
               \AxiomC{$\vdots$}
                \noLine
               \UnaryInfC{$\vdash \s, C $}
			\AxiomC{$\vdots$}
			\noLine
			\UnaryInfC{$\vdash \Delta ,  D,B$}
			\RightLabel{$ \odot_2$}
		\BinaryInfC{$ \vdash \s, \Delta, B,  \odot_2(C, D)$}

\RightLabel{$ \odot_1 $}
\BinaryInfC{$ \vdash \Gamma, \s,\Delta ,  \odot_1(A , B),  \odot_2(C, D)$}

\noLine
\TrinaryInfC{}
\end{scprooftree}

$$\odot_1(A,B)= \left\{\begin{array}{lll} 
A \otimes B  & \mbox{ if } \odot_1=\otimes \\ 
\emptyset & \mbox{ if } \odot_1=Cut \mbox{ and } A=B^\bot
\end{array}\right. 
$$
$$
\odot_2(C,D)= \left\{\begin{array}{lll} 
C \otimes D  & \mbox{ if } \odot_2=\otimes \\ 
\emptyset & \mbox{ if } \odot_2=Cut \mbox { and } C=D^\bot
\end{array}\right. $$
\end{definition}

We define the \emph{cut-elimination procedure} by the following set of rewriting rules over derivations:

\begin{definition}[Cut-elimination procedure]
The {cut-elimination procedure} is the relation $\fr_{Cut}$ generated by the following (oriented) relations called \emph{cut-elimination steps}:

\begin{scprooftree}{.7}
\AxiomC{$\vdots$}
\noLine
\UnaryInfC{$\vdash \Gamma , A$}
\AxiomC{}
\RightLabel{Ax}
\UnaryInfC{$\vdash A^\bot, A $}
\RightLabel{$ Cut $}
\BinaryInfC{$ \vdash \Gamma, A$}

\AxiomC{~}
\noLine
\UnaryInfC{$\fr_{Cut}$}
\noLine
\UnaryInfC{$~$}
\noLine
\UnaryInfC{$~$}

\AxiomC{$\vdots$}
\noLine
\UnaryInfC{$\vdash \Gamma , A$}

\noLine
\TrinaryInfC{}

\AxiomC{~}

\AxiomC{}
\RightLabel{Ax}
\UnaryInfC{$\vdash A,  A^\bot $}

\AxiomC{$\vdots$}
\noLine
\UnaryInfC{$\vdash \Gamma , A$}
\RightLabel{$ Cut $}
\BinaryInfC{$ \vdash \Gamma, A$}

\AxiomC{~}
\noLine
\UnaryInfC{$\fr_{Cut}$}
\noLine
\UnaryInfC{$~$}
\noLine
\UnaryInfC{$~$}

\AxiomC{$\vdots$}
\noLine
\UnaryInfC{$\vdash \Gamma , A$}

\noLine
\TrinaryInfC{}

\noLine
\TrinaryInfC{}
\end{scprooftree}

\begin{scprooftree}{0.7}
\AxiomC{$\vdots$}
\noLine
\UnaryInfC{$\vdash \Gamma , A$}
\AxiomC{$\vdots$}
\noLine
\UnaryInfC{$\vdash B, \Delta $}
\RightLabel{$ \otimes $}
\BinaryInfC{$ \vdash \Gamma, \Delta,  A\otimes B$}
\AxiomC{$\vdots$}
\noLine
\UnaryInfC{$\vdash B^\bot, A^\bot, \s$}
			\RightLabel{$ \parr $}
\UnaryInfC{$\vdash B^\bot \parr A^\bot, \s$}
\RightLabel{Cut}
\BinaryInfC{$ \vdash \Gamma,\Delta,\s$}

\AxiomC{~}
\noLine
\UnaryInfC{{$\fr_{Cut}$}}
\noLine
\UnaryInfC{$~$}
\noLine
\UnaryInfC{$~$}

\AxiomC{$\vdots$}
\noLine
\UnaryInfC{$\vdash \Gamma , A$}
\AxiomC{$\vdots$}
\noLine
\UnaryInfC{$\vdash B, \Delta $}

\AxiomC{$\vdots$}
\noLine
\UnaryInfC{$\vdash B^\bot, A^\bot, \s$}
			\RightLabel{$ \parr $}
\RightLabel{$ Cut $}
\BinaryInfC{$ \vdash  \Delta,  A^\bot,\s$}
\RightLabel{$ Cut $}
\BinaryInfC{$ \vdash  \Gamma, \Delta,\s$}

\noLine
\TrinaryInfC{}
\end{scprooftree}

\begin{scprooftree}{0.7}
\AxiomC{$\vdots$}
\noLine
\UnaryInfC{$\vdash B^\bot, A^\bot, \s$}
			\RightLabel{$ \parr $}
\UnaryInfC{$\vdash B^\bot \parr A^\bot, \s$}
\AxiomC{$\vdots$}
\noLine
\UnaryInfC{$\vdash \Gamma , A$}
\AxiomC{$\vdots$}
\noLine
\UnaryInfC{$\vdash B, \Delta $}
\RightLabel{$ \otimes $}
\BinaryInfC{$ \vdash \Gamma, \Delta,  A\otimes B$}

\RightLabel{Cut}
\BinaryInfC{$ \vdash \Gamma,\Delta,\s$}

\AxiomC{~}
\noLine
\UnaryInfC{\large{$\fr_{Cut}$}}
\noLine
\UnaryInfC{$~$}
\noLine
\UnaryInfC{$~$}

\AxiomC{$\vdots$}
\noLine
\UnaryInfC{$\vdash B^\bot, A^\bot, \s$}

\AxiomC{$\vdots$}
\noLine
\UnaryInfC{$\vdash B, \Delta $}

\RightLabel{$ Cut $}
\BinaryInfC{$ \vdash  \Delta,  A^\bot,\s$}

\AxiomC{$\vdots$}
\noLine
\UnaryInfC{$\vdash \Gamma , A$}

\RightLabel{$ Cut $}
\BinaryInfC{$ \vdash  \Gamma, \Delta,\s$}

\noLine
\TrinaryInfC{}
\end{scprooftree}

\begin{scprooftree}{.7}
\AxiomC{$\vdots$}
\noLine
\UnaryInfC{$\vdash \Gamma $}
\RightLabel{$\bot$}
\UnaryInfC{$\vdash \Gamma, \bot $}

\AxiomC{}
\RightLabel{$1$}
\UnaryInfC{$\vdash 1 $}
\RightLabel{$ Cut $}
\BinaryInfC{$ \vdash \Gamma$}

\AxiomC{~}
\noLine
\UnaryInfC{$\fr_{Cut}$}
\noLine
\UnaryInfC{$~$}
\noLine
\UnaryInfC{$~$}

\AxiomC{$\vdots$}
\noLine
\UnaryInfC{$\vdash \Gamma$}

\noLine
\TrinaryInfC{}

\AxiomC{}

\AxiomC{}
\RightLabel{$1$}
\UnaryInfC{$\vdash 1$}

\AxiomC{$\vdots$}
\noLine
\UnaryInfC{$\vdash \Gamma $}
\RightLabel{$ \bot $}
\UnaryInfC{$\vdash \Gamma , \bot$}

\RightLabel{$ Cut $}
\BinaryInfC{$ \vdash \Gamma$}

\AxiomC{~}
\noLine
\UnaryInfC{$\fr_{Cut}$}
\noLine
\UnaryInfC{$~$}
\noLine
\UnaryInfC{$~$}

\AxiomC{$\vdots$}
\noLine
\UnaryInfC{$\vdash \Gamma $}

\noLine
\TrinaryInfC{}

\noLine
\TrinaryInfC{}
\end{scprooftree}

\end{definition}

The cut-elimination theorem for $\MLLc$ sequent calculus is proved  by showing the termination of the cut-elimination procedure \cite{girll}. This result requires the identification of derivations by the standard equivalence. Alternatively, the proof requires the definition of some additional rewriting rules which permute the commutative $Cut$ instances. 
We remark that even in non-commutative extensions of linear logics \cite{AbruNon} where permutations of formulas in a sequent are strongly restricted,  we (unexpectedly) require permutations over derivation branches for a proof of cut-elimination theorem. 

For this reason, any \emph{denotational semantic} of $\MLLc$ sequent calculus \cite{girss,Mel,Lollo} has to take into account the standard equivalence of derivations in order to capture the cut-elimination. It results that the equivalence relation over derivations  induced by any such semantics contains the equivalence relation $\approx$ over the derivations syntax generated by $(\fr_{Cut}\cup \sim)$.

\section{String diagram syntax for linear logic}

In this section we define some particular $3$-polygraphs which generate a family of string diagrams we call \emph{proof diagrams}. These  diagrams are a syntax for  linear logic sequent calculus with explicit exchange rules.

The first polygraph $\s_{\MLLc}$ we define generates a family of terms corresponding to the different representations of $\MLLc$ proof nets, with explicit notation for wire crossings but no jump assignations. 

We then improve this construction adding two non-twisting colors for strings and we adapt certain gate types in order to make them interact with these \emph{control strings}. Due to the more rigid structure of the diagrammatic syntax, in this polygraph $\Mlltc$ we are able to characterize diagrams corresponding to linear logic derivations by just checking their inputs and outputs patterns. On the other hand, the rewriting we define is able to capture all permutations of inference rules with exception of the ones between two binary rules ($\otimes$ or $Cut$), in particular the one needed to eliminate commutative cuts, crucial for the sequent calculus cut-elimination theorem.

We extend to $\Mllc$ the polygraphic presentation of this model  by extending $\Mlltc$ with two sets of generators and relations which allows us to perform some transformations corresponding to certain permutations of binary inference rules. We then show that the classes of equivalent diagrams modulo the rewriting of this polygraph are in one-to-one correspondence with the classes of $\sim$-equivalent $\MLLc$ proof.

We conclude with the polygraph $\Mllc_{Cut}$ which include the rewriting rules corresponding to cut-elimination steps of $\MLLc$ sequent calculus showing that the associated quotient over $\MLLc$ derivations captures the semantics equivalence of proof.

\nota From now on, in order to unify the notation $1$-cell composition with the one of sequents, we replace the symbol $*$ for string diagrams parallel composition with a comma.


\subsection{Proof diagrams and $\MLLc$ proof nets}
 
The first polygraph we introduce can be seen as a formal syntax for proof net representations.

\begin{definition}
The $3$-polygraph $ \s_{\MLLc}^{Cut}$ is the \emph{polygraph of multiplicative linear logic proof nets with units}. It is given by the following sets of cells:
\begin{multicols}{2}
\begin{itemize}
\item $\s^u_0=\{\pro\}$;
\item $\s^u_1=\fmllc $;
\end{itemize}
\end{multicols}
\begin{itemize}
\item $\s^u_2=\begin{Bmatrix}
\otimes_{A,B}: &  A , B &\frr & A\otimes B& = & \twocell{(topA *0 topB) *1 ten *1 (pitAtenB)}  \\
\parr_{A,B}: &  A , B &\frr & A\parr B  & = &      \twocell{(topA *0 topB) *1 par *1 (pitAparB)} \\
Ax_A: & \pro &\frr & A , A^\bot  & = & {\twocell{axA *1 (pitA *0 pitAb)}}\\
Cut_A: &  A , A^\bot &\frr & \pro  & = & {\twocell{(topA *0 topAb) *1 cutA}}\\
\twocell{s}_{A,B}: &  A , B &\frr & B , A & = & \twocell{(topA *0 topB)*1 (s)*1(pitB *0 pitA)}\\
1: & \pro &\frr & 1 & =& \twocell{v *1 pit1}\\
\bot: & \pro &\frr &\bot  &=& \twocell{bot *1 pitbot}
\end{Bmatrix}_{A,B\in \fmllc}$

If there is no ambiguity we note \twocell{ax} and \twocell{cut} instead of \twocell{axA} and \twocell{cutA}.

\item $\s^u_3=\s^M_{Twist} \cup \s^u_{Twist} \cup \s^{Ax}_{Cut} \cup \s^M_{Cut} \cup \s^u_{Cut}$ where:

\begin{itemize} 
\item $\s^M_{Twist}$ is given by the following twisting relations:
\end{itemize}

$$
\xymatrix@C=.3cm{ {\twocell{(topA *0 topB )*1(s *1 s)  *1 (pitA *0 pitB)}} \ar@3[r] & {\twocell{(midA *0 midB)}}} ,
\quad 
\xymatrix@C=.3cm{ \twocell{(topA *0 topB *0 topC)*1(s *0 1)*1 (1 *0 s)*1 (s *0 1)*1(pitC *0 pitB *0 pitA)} \ar@3[r] &{\twocell{ (topA *0 topB *0 topC)*1(1 *0 s)*1 (s *0 1)*1 (1 *0 s) *1(pitC *0 pitB *0 pitA) }}},
$$
\begin{center}\resizebox{10cm}{!}{
$$
\xymatrix@C=.3cm{ \twocell{topB *1 (ax *0 1) *1 (1 *0 s)*1 (s *0 1) *1 (pitB *0 pitA *0 pitAb)} \ar@3[r] & \twocell{topB *1 ( 1 *0 ax)*1 (pitB *0 pitA *0 pitAb) }},
\quad
\xymatrix@C=.3cm{  \twocell{topB *1 (1*0 ax )*1 (s*0 1) *1 (1*0 s) *1 (pitA *0 pitAb *0 pitB)} \ar@3[r] & \twocell{ topB *1 (ax *0 1 ) *1 (pitA *0 pitAb *0 pitB)}},
\quad 
\xymatrix@C=.3cm{\twocell{(topA *0 topAb *0 topB ) *1 ( 1 *0 s)*1 (s *0 1)*1 (1*0 cut)*1pitB} \ar@3[r] &  \twocell{(topA *0 topAb *0 topB ) *1( cut *0 1) *1 pitB }},
\quad
\xymatrix@C=.3cm{ \twocell{(topB *0 topA *0 topAb) *1 (s*0 1) *1 (1*0 s) *1 (cut *0 1)*1 pitB} \ar@3[r] &  \twocell{ (topB *0 topA *0 topAb) *1 (1 *0 cut) *1 pitB }},
$$
}
\end{center}
\begin{center}\resizebox{10cm}{!}{
$$ 
\xymatrix@C=.3cm{\twocell{(topA *0 topB *0 topC) *1 (ten *0 1) *1  s *1(pitC *0 pitAtenB) } \ar@3[r] & \twocell{(topA *0 topB *0 topC) *1 (1 *0 s)*1 (s *0 1)*1 (1 *0 ten)*1(pitC *0 pitAtenB)}},
\quad
\xymatrix@C=.3cm{ \twocell{(topA *0 topB *0 topC) *1 (1 *0 ten)*1 (s) *1 (pitBtenC *0 pitA)} \ar@3[r] & \twocell{(topA *0 topB *0 topC) *1 (s *0 1)*1 (1 *0 s)*1 (ten *0 1)*1 (pitBtenC *0 pitA) }},
\quad
\xymatrix@C=.3cm{\twocell{(topA *0 topB *0 topC) *1(par *0 1)*1 (s)*1 ( pitC *0 pitAparB)} \ar@3[r] & \twocell{(topA *0 topB *0 topC) *1 (1 *0 s)*1 (s *0 1)*1 (1 *0 par)*1 ( pitC *0 pitAparB) }},
\quad
\xymatrix@C=.3cm{\twocell{(topA *0 topB *0 topC) *1( 1 *0 par)*1 (s)*1(pitBparC *0 pitA)} \ar@3[r] & \twocell{(topA *0 topB *0 topC) *1 (s *0 1)*1 (1 *0 s)*1 (par *0 1)*1 (pitBparC *0 pitA) }};
$$
}
\end{center}
together with two rules representing the involution $A^{\bot\bot}=A$: 
$$
\xymatrix{ {\underset{~A\; ~A^\bot}{\twocell{ axA *1 s }}} \ar@3[r] & {\underset{A^\bot\;  A}{\twocell{ axAb }}}},
\;
\xymatrix{ {\overset{~A\; ~A^\bot}{\twocell{ s *1 cutA }}} \ar@3[r] & {\overset{A^\bot\;  A}{\twocell{ cutAb }}}};
$$

\begin{itemize} 
\item $\s^u_{Twist}$ is given by the following twisting relations:
\end{itemize}
\begin{center}\resizebox{10cm}{!}{
$$
\xymatrix{ \twocell{topA *1 (bot *0 1) *1  s *1( pitA *0 pitbot) } \ar@3[r] & \twocell{ topA *1 (1 *0 bot)  *1( pitA *0 pitbot) }},
\quad
\xymatrix{ \twocell{topA *1 (1 *0 bot)*1 (s) *1 (pitbot *0 pitA)} \ar@3[r] & \twocell{ topA *1 (bot *0 1) *1 (pitbot *0 pitA) }},
\quad
\xymatrix{ \twocell{topA *1 (v *0 1) *1  s *1( pitA *0 pit1) } \ar@3[r] & \twocell{topA *1  (1 *0 v) *1( pitA *0 pit1)}},
\quad
\xymatrix{ \twocell{topA *1 (1 *0 v)*1 (s) *1 (pit1 *0 pitA)} \ar@3[r] &\twocell{ topA *1 (v *0 1)*1 (pit1 *0 pitA) }};
$$
}\end{center}

\begin{itemize} 

\item $\s^{Ax}_{Cut}$ is following the set of rules for the cut elimination:
\begin{center}\resizebox{10cm}{!}{
$$
\xymatrix{ \twocell{(topGam *0 topA) *1 (ax *0 d *0 1) *1 (1 *0 ln *01) *1 (1*0 d*0 cut)*1 (pitA *0 pitGam) } \ar@3[r] & \twocell{(topGam *0 topA) *1 (d*0 1) *1 rn *1 (1*0 d)*1 (pitA *0 pitGam)}},
\qquad 
\xymatrix{  \twocell{(topA *0 topGam)*1 (1 *0 d *0 ax) *1 (1 *0 rn *01) *1 (cut*0 d*0 1) *1 (pitGam *0 pitA) } \ar@3[r] & \twocell{(topA *0 topGam) *1 (1*0 d) *1 ln*1 (d *0 1)*1 (pitGam *0 pitA)}}, \mbox{for any $\Gamma\in\ \fmll^*$}
$$
}\end{center}

\begin{center}\resizebox{10cm}{!}{
$$
\xymatrix{ \twocell{topA *1 (ax *0 1) *1 (1*0 s) *1 (cut *0 1) *1 pitA } \ar@3[r] & \twocell{midA}},
\quad 
\xymatrix{ \twocell{(topA  *0 topGam)  *1 (1 *0 d *0 ax) *1  (1 *0 rn *0 1)*1 (s *0 sigma *0 1)*1 (1*0 ln*0 1) *1 (1*0 d*0 cut) *1 (pitA *0 pitsiG)} \ar@3[r] & \twocell{(topA  *0 topGam)  *1 (1 *0 d)*1 (1 *0 sigma) *1 (1 *0 d)*1 (pitA *0 pitsiG) }} \; \mbox{, for any $\twocell{topGam *1 d *1 sigma *1 d *1 pitsiG}$ canonical diagram of $\sigma$;}
$$
}\end{center}

\item $\s^M_{Cut}$ is following the set of rules for the cut elimination:
\begin{center}\resizebox{10cm}{!}{
$$
\xymatrix{ \overset{~~A\; B\; ~B^\bot A^\bot}{\twocell{(par *0 ten)*1 cut }} \ar@3[r] & \overset{~A\; ~B\;~ B^\bot A^\bot}{\twocell{ (1 *0 cut *0 1) *1 ( cut)}}},
\quad
\xymatrix{ \overset{~~A\; B\; ~B^\bot A^\bot}{\twocell{(ten *0 par)*1 cut }} \ar@3[r] & \overset{~A\; ~B\;~ B^\bot A^\bot}{\twocell{ (1 *0 cut *0 1) *1 ( cut)}}},
$$
}\end{center}

\item  $\s^{u}_{Cut}$  the following set of rules for cut elimination:
$$\xymatrix{ {\twocell{(bot *0 v)*1 cut }} \ar@3[r] & \emptyset },
\;
\xymatrix{ {\twocell{(v *0 bot)*1 cut }} \ar@3[r] & \emptyset }
$$.

\end{itemize}
\end{itemize}

\end{definition}

\begin{oss}
The polygraph $ \s_{\MLLc}^{Cut}$ is twisting with twisting family $\fmllc$, i.e.~it is total twisting.
\end{oss}

\begin{theorem}[Interpretation of proofs in $\s_{\MLLc}^{Cut}$]\label{intmll}
For any derivation $d(\Gamma)$ of $\vdash \Gamma$ in $\MLLc$ there is a proof diagram $\phi_{d(\Gamma)}: \pro \frr \Gamma\in \s_{\MLLc}^{Cut}$.
\begin{proof}
Let $d(\Gamma)$ be a derivation in $\MLLc$ of $\vdash \Gamma$. First we observe that, if there is a diagram $\phi: \Delta \frr \Gamma$ so there also is a diagram $\phi^\sigma= \hat \phi_\sigma \circ \phi: \Delta \frr \sigma(\Gamma)$ for all permutation $\sigma\in S_{|\Gamma|}$. Thus,  we can proceed by induction on the number of inference rules appearing in $d(\Gamma)$:
\begin{itemize}
\item If just one inference rule occurs in $d(\Gamma)$, it must be an $Ax$ rule or  a $1$ rule. It follows that $\Gamma=A,A^\bot$ and $\phi_{d(\Gamma)}=Ax_A:\pro \frr A, A^\bot$ or that $\Gamma=1$ and $\phi_\Gamma= 1: \pro \frr 1$;

\item If $n+1$ inference rules occur in $d(\Gamma) $, then we consider the last one and we distinguish two cases in base of its arity (see Remark  \ref{arity}):
\begin{itemize}

\item If it is unary and $ \Gamma= \Gamma', A\parr B$, then, by inductive hypothesis, there is a diagram $\phi_{d(\Gamma',A,B)}: \pro \fr \Gamma',A,B$ of the derivation $d( \Gamma',A,B)$ with $n$ inference rules. Therefore
$$\phi_{d(\Gamma)}=(\id_{\Gamma'},\parr_{A,B}) \circ \phi_{d(\Gamma',A,B)}: \pro \frr \Gamma ;$$

\item If it  is an unary $\bot$ and $\Gamma= \Gamma', \bot$, then, by inductive  hypothesis, there is a  diagram $\phi_{\Gamma'}: \pro \frr \Gamma'$ and $\phi_\Gamma= \phi_{\Gamma'}, \bot$.

\item If it is binary and $ \Gamma=\Delta, A\otimes B, \Delta'$, then, by inductive hypothesis, there are two diagrams $\phi_{d(\Delta, A)}:\pro \frr \Delta, A$ and $\phi_{d(B, \Delta')}:\pro \frr B ,\Delta'$ relative to the two derivations $d(\Delta, A)$ and $d(B, \Delta')$ with at most $n$ inference rules. Therefore
$$\phi_{d(\Gamma)}=(\id_{\Delta},\otimes_{A,B}, \id_{\Delta'}) \circ (\phi_{d(\Delta, A)} , \phi_{d(B, \Delta')}):  \pro \frr \Gamma ;$$

\item Similarly, if it is binary and $\Gamma= \Delta, Cut(A, A^\bot), \Delta'$, then
$$\phi_{d(\Gamma)}=(\id_{\Delta},cut_{A}, \id_{\Delta'}) \circ (\phi_{d(\Delta, A)} , \phi_{d(A^\bot, \Delta')}):  \pro \frr \Gamma . $$
\end{itemize}
\end{itemize}
\end{proof}
\end{theorem}


The $2$-cells of this syntax reminds $\MLLc$ proof structure representations. We remark two important differences: cells are always top-to-bottom orientated, that is with the active port on the bottom, and wire crossing are part of this syntax by means of twisting operators.  
This intuition leads to the following:

\begin{proposition}[Proof structure interpretation]\label{psint}
We can associate to any proof diagram $\phi$ in $\s_{\MLLc}^{Cut}$ a $\MLLc$ proof structures $P_\phi$. 
\begin{proof}
It suffices to consider a proof diagram as a specific representation of a proof structure with no specific jumps assignation: strings, $Ax$-gates and $Cut$-gates are interpreted as wires ($\twocell{cap}\leftrightsquigarrow \twocell{ax}$ and $\twocell{cup}\leftrightsquigarrow \twocell{cut}$),  twisting operators as wire crossing and  gates of type $\otimes, \parr, \bot$ and $1$ as the corresponding cells of the proof structure with a  coherent  labeling with respect of gate types. Then, since proof diagrams in $\s_{\MLLc}^{Cut}$ keep no records about jump assignations, for each $\bot$-gate we assign arbitrary  jump.
\end{proof}
\end{proposition}

However, the converse is not true. In fact even if we interpret down-to-down and up-to-up wire turn-backs as $Ax$ and $Cut$ gates respectively (i.e. $\twocell{cap}\rightsquigarrow \twocell{ax}$ and $\twocell{cup}\rightsquigarrow \twocell{cut}$)  and  wire crossing  as occurrences of twisting operators, in the syntax of proof diagrams we are not able to represent some (incorrect)  proof structures because of the type of inputs and outputs of $Ax$ and $Cut$ gates. By means of example, consider  the proof structure $\twocell{(cap *0 1 ) *1 (1 *0 par) *1 (cup) }$ whose translation in proof diagram syntax requires the existence of $A,B\in \fmllc$ with $A^\bot=A^\bot \parr B$ in order to be well defined.


\subsection{Proof diagram with control for $\MLLc$}
In order to have an analogous of the proof net correctness criterion formalized inside a syntax of $\MLLc$ proof diagrams, we enrich the set of string labels with two new non-twisting colors $L=\twocell{L}$ (left) and $R=\twocell{R}$ (right) that we call \emph{control strings}.

The idea is to use these strings to reproduce a $2$-dimensional notation for parenthesization, in order to internalize a notion of well-paranthesization in a setting where a proof derivation can be seen as a sequence of operations over lists of sequents. Thus,  unary derivation rules act on single sequents (as in the case of $\parr$ and $\bot$), binary ones act on two sequent (as in the case of $\otimes$ and $Cut$) and the 0-ary one, that are $Ax$ and $1$, generates a new sequent. For this purpose we re-define the $2$-cells for $0$-ary and binary rules in order to make them interact with control strings.

\nota In order to help reader, $L$ and $R$ control strings are represented in diagrams by strings decorated by a certain number of $\twocell{L}$ and $\twocell{R}$ respectively. These have to be considered as string labels and not gates.


\begin{definition}
The \emph{control polygraph of multiplicative linear logic with units} $\Mlltc$ is given by the following sets of cells:
\begin{multicols}{2}
\begin{itemize}
\item $\Mlltc_0=\{\pro\}$;
\item $\Mlltc_1= \fmllc \cup \{L=\twocell{L},R=\twocell{R}\}$;
\end{itemize}
\end{multicols}
\begin{itemize}
\item 
{$\Mlltc_2=\begin{Bmatrix}
\otimes_{A,B}: &  A ,R,L, B &\frr & A\otimes B &=& \twocell{(topA *0 R *0 L *0 topB)*1 tenc *1 pitAtenB}\\
\parr_{A,B}: &  A , B &\frr & A\parr B  &=& \twocell{(topA *0 topB )*1 par *1 pitAparB}\\
Ax_A: & \pro &\frr &L, A , A^\bot,R &=& \twocell{axcA*1 (L*0 pitA *0 pitAb*0 R)} \\
Cut_A: &  A ,R,L, A^\bot &\frr & \pro  &=& \twocell{(topA *0 R *0 L *0 topAb)*1 cutc}\\ 
\twocell{s}_{A,B}: &  A ,B &\frr & B , A &=& \twocell{(topA *0 topB) *1 s *1 (pitB *0 pitA)}\\
1: & \pro &\frr &L, 1,R  &=&\twocell{vc *1 (L *0 pit1 *0 R)}\\
\bot: & \pro &\frr &\bot  &=& \twocell{bot *1 pitbot} \\
\end{Bmatrix}_{A,B\in \fmllc}$
}
\item $\Mlltc_3= \Mllt_{Twist}\cup \Mlltc_{Twist}$ where:

\begin{itemize}
\item $\Mllt_{Twist}$  is given by the  following twisting relations:
\end{itemize}
\begin{center}\resizebox{10cm}{!}{
$$
\xymatrix{ {\twocell{(topA *0 topB )*1(s *1 s) *1 (pitA *0 pitB) }} \ar@3[r] & {\twocell{(midA *0 midB)}}}, 
\; 
\xymatrix{ \twocell{(topA *0 topB *0 topC)*1(s *0 1)*1 (1 *0 s)*1 (s *0 1)*1(pitC *0 pitB *0 pitA)} \ar@3[r] &{\twocell{ (topA *0 topB *0 topC)*1(1 *0 s)*1 (s *0 1)*1 (1 *0 s) *1(pitC *0 pitB *0 pitA) }}} ,
\; 
\xymatrix{\twocell{(topA *0 topB *0 topC) *1(par *0 1)*1 (s)*1 ( pitC *0 pitAparB)} \ar@3[r] & \twocell{(topA *0 topB *0 topC) *1 (1 *0 s)*1 (s *0 1)*1 (1 *0 par)*1 ( pitC *0 pitAparB) }},
\;
\xymatrix{\twocell{(topA *0 topB *0 topC) *1( 1 *0 par)*1 (s)*1(pitBparC *0 pitA)} \ar@3[r] & \twocell{(topA *0 topB *0 topC) *1 (s *0 1)*1 (1 *0 s)*1 (par *0 1)*1 (pitBparC *0 pitA) }};
$$
}\end{center}

together with one rule representing the involution $A^{\bot\bot}=A$: 
$$\xymatrix{ \twocell{ axcA *1 (L *0  s *0 R) *1 (pitV *0 pitAb *0 pitA *0 pitV) } \ar@3[r] & \twocell{ axcAb *1 (L *0 2 *0 R) *1 (pitV *0 pitAb *0 pitA *0 pitV) }}$$

\begin{itemize}
\item $\Mlltc_{Twist}$  is given by the  following twisting relations:
\end{itemize}
$$
\xymatrix{ \twocell{topA *1 (bot *0 1) *1  s *1( pitA *0 pitbot) } \ar@3[r] & \twocell{ topA *1 (1 *0 bot)  *1( pitA *0 pitbot) }},
\qquad
\xymatrix{ \twocell{topA *1 (1 *0 bot)*1 (s) *1 (pitbot *0 pitA)} \ar@3[r] & \twocell{ topA *1 (bot *0 1) *1 (pitbot *0 pitA) }}.
$$
\end{itemize}
\end{definition}


\begin{oss}
The polygraph $\Mlltc$ is twisting with twisting family $\fmllc$. This means that we can represent any crossing of strings labeled by $\MLLc$ formulas and these crossings  interact as we attend with $2$-cells which are not  connected to control strings.
\end{oss}

\begin{oss}[$Cut$-gates shape]\label{CutDef}
We assume the De Morgan's laws in the definition of $Cut$-gate inputs, as given in Remark \ref{remNeg}. That is, for any $A,B\in \fmllc$:
\begin{itemize}
\item $Cut_{A\parr B}: (A\parr B) , R, L, (B^\bot \otimes A^\bot)  \frr \pro$;
\item $Cut_{A\otimes B}: (A\otimes B) , R, L,( B^\bot \parr A^\bot)  \frr \pro$;
\item $Cut_{\bot}: \bot , R, L, 1 \parr A^\bot  \frr \pro$;
\item $Cut_{1}: 1 , R, L, \bot  \frr \pro$;
\end{itemize}
\end{oss}

In this setting we are able to prove that the sequentializability of a diagram depends only on its inputs and outputs. Moreover, we are able to characterize $\MLLc$ provable sequents in terms of existence of proof diagrams with a specific type.

\begin{theorem}[Controlled proof diagram correspondence in $\Mlltc$]\label{corrMLLc}
$$\vdash_{MLLu} \Gamma \Leftrightarrow \exists \phi \in \Mlltc \mbox{ such that } \phi: \pro \frr L,\Gamma, R.$$

\begin{proof}
To prove the left-to-right implication $\Rightarrow$, as in Teor. \ref{intmll}, we remark that, if there is a diagram $\phi: \pro \frr L, \Gamma, R$ with $\Gamma$ sequent in $\MLLc$, so there is a diagram 
$$\phi^\sigma= (\id_L ,\hat \phi_\sigma, \id_R) \circ \phi: \pro \frr L,\sigma(\Gamma), R $$
 for any permutation $\sigma\in S_{|\Gamma|}$. Then we proceed by induction on the number of inference rules in a derivation $d(\Gamma)$ in $\MLLc$:
\begin{itemize}

\item If just one inference rule occurs $d(\Gamma)$, then it is an $Ax$ or a $1$, then $\Gamma=A,A^\bot$ and $\phi_{d(\Gamma)}=Ax_A: \pro \frr L,A, A^\bot , R$ or $\Gamma= 1$ and $\phi_{d(\Gamma)}= 1: \pro \frr L, 1,R$;

\item If $n+1$ inference rules appear, then we consider the last one and we distinguish two cases in base of its arity:

\begin{itemize}

\item If it is an unary $\parr$ and $ \Gamma= \Gamma', A\parr B$, then, by inductive hypothesis, there is a diagram $\phi_{d(\Gamma',A,B)}: \pro \frr L, \Gamma',A,B, R$ of the derivation $d( \Gamma',A,B)$ and 
$$\phi_{d(\Gamma)}=(\id_{L,\Gamma'},\parr_{A,B},\id_R) \circ \phi_{d(\Gamma',A,B)}: \pro \frr L, \Gamma, R ;$$

\item Similarly, if it is a unary $\bot$ and $\Gamma= \Gamma', \bot$, then, by inductive  hypothesis, there is a  diagram $\phi_{\Gamma'}: \pro \frr L, \Gamma', R$ and $\phi_\Gamma= (L,\bot, \id_{\Gamma'},  R) \circ \phi_{\Gamma'}$;

\item If it is a binary $\otimes$ and $ \Gamma=\Delta, A\otimes B, \Delta'$, then, by inductive hypothesis, there are two diagrams $\phi_{d(\Delta, A)}:\pro \frr L, \Delta, A, R$ and $\phi_{d(B, \Delta')}:\pro \frr L, B ,\Delta', R$ relative to the two derivations $d(\Delta, A)$ and $d(B, \Delta')$ with at most $n$ inference rules. Therefore
$$\phi_{d(\Gamma)}=(\id_{L,\Delta},\otimes_{A,B}, \id_{\Delta',R}) \circ (\phi_{d(\Delta, A)} , \phi_{d(B, \Delta')}): \pro \frr L, \Gamma, R $$

\item Similarly, if it is a binary $Cut$ and $\Gamma= \Delta, Cut(A, A^\bot), \Delta'$, then
$$\phi_{d(\Gamma)}=(\id_{L,\Delta},Cut_{A^\bot}, \id_{\Delta',R}) \circ (\phi_{d(\Delta, A)} , \phi_{d(A^\bot, \Delta')}): \pro \frr L, \Gamma, R. $$

\end{itemize}
\end{itemize}

In order to prove sequentialization, i.e. the right-to-left implication $\Leftarrow$, we proceed by induction on the number $|\phi|_\sig$ of gates in $\phi$:
\begin{itemize}

\item If $|\phi|_{ \Mllt}=0$ so $\phi: \id_\Gamma: \Gamma \frr \Gamma$. By hypothesis $\phi$ has no input (i.e. $s_2(\phi)=\pro$) so it is the identity  diagram over the empty string, this is the empty diagram $\id_0: \pro\frr \pro$ which it is not sequentializable since $t_2(\phi)=\pro \neq L,R$;

\item If  $|\phi|_{ \Mlltc}=1$ then $\phi$ is an elementary diagram. The elementary diagrams with source $\pro$ and target $L,\Gamma, R$ with $\Gamma \in \smllc$ are atomic made of a unique $2$-cell of type $Ax_A: \pro \fr L, A, A^\bot, R$ for some  $A\in \fmllc$ or $1: 0 \fr L,1,R$. The associated sequent $\vdash A, A^\bot$ or $\vdash 1$ is derivable in $\MLLc$;

\item Otherwise there is  $2$-cell of type  $\alpha: \Gamma' \frr \alpha ( \Gamma' ) \in \Mllt_2$ and  $\Gamma= \Delta, \alpha (\Gamma' ), \Delta'$. In this case $\phi= (\id_{L, \Delta} , \alpha , \id_{\Delta, R}) \circ \phi' $ where $\phi': \pro \frr L,\Delta, \Gamma', \Delta',R$. We have the following cases:

\begin{itemize}
\item If $\alpha= \twocell{s}_{A, B} $, $\Gamma' = A,B$ and $\alpha(\Gamma')= B,A$. The diagram $\phi'$ is sequentializable by inductive hypothesis since $|\phi|_{\Mlltc}= |\phi' |_{\Mllt} +1$:
$$
{\raisebox{-12.50pt}
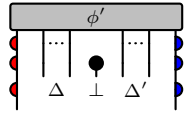
{\begin{tikzpicture} \begin{scope} [ x = 10pt, y = 10pt, join = round, cap = round, thick, black, solid, -]
\draw [rounded corners = 1pt, fill = lightgray] (-.25,2.50) rectangle (6.25, 3.5) ; \node at (3,3) {$\scriptstyle \phi'$};
 \draw (0.00,2.50)--(0.00,2.25) (1.00,2.50)--(1.00,2.00) (2.00,2.50)--(2.00,2.00) (4.00,2.50)--(4.00,2.00) (5.00,2.50)--(5.00,2.00) (6.00,2.50)--(6.00,2.25) ; \draw (0.00,0.00)--(0.00,-0.50) (6.00,0.00)--(6.00,-0.50) ; \draw (0.00,2.25)--(0.00,2.00) ; \draw (0.00,1.75)--(0.00,2.00) ; \draw [fill = red] (0.00,2.25) arc (90:270:0.25) ; \draw (0.00,1.75)--(0.00,2.25) ; \draw [dash pattern = on 0.25pt off 2pt] (1.25,2.00)--(1.75,2.00) ; \draw [dash pattern = on 0.25pt off 2pt] (4.25,2.00)--(4.75,2.00) ; \draw (6.00,2.25)--(6.00,2.00) ; \draw (6.00,1.75)--(6.00,2.00) ; \draw [fill = blue] (6.00,1.75) arc (-90:90:0.25) ; \draw (6.00,1.75)--(6.00,2.25) ; \draw (0.00,1.75)--(0.00,1.50) (1.00,2.00)--(1.00,1.25) (2.00,2.00)--(2.00,1.25) (4.00,2.00)--(4.00,1.25) (5.00,2.00)--(5.00,1.25) (6.00,1.75)--(6.00,1.50) ; \draw (0.00,1.50)--(0.00,1.25) ; \draw (0.00,1.00)--(0.00,1.25) ; \draw [fill = red] (0.00,1.50) arc (90:270:0.25) ; \draw (0.00,1.00)--(0.00,1.50) ;  \draw  ; \draw (3.00,1.00)--(3.00,1.25) ; \draw [fill = black] (3.00,1.25) circle (0.25) ;  \draw (6.00,1.50)--(6.00,1.25) ; \draw (6.00,1.00)--(6.00,1.25) ; \draw [fill = blue] (6.00,1.00) arc (-90:90:0.25) ; \draw (6.00,1.00)--(6.00,1.50) ; \draw (0.00,1.00)--(0.00,0.50) (1.00,1.25)--(1.00,0.75) (2.00,1.25)--(2.00,0.75) (3.00,1.00)--(3.00,0.75) (4.00,1.25)--(4.00,0.75) (5.00,1.25)--(5.00,0.75) (6.00,1.00)--(6.00,0.50) ; \draw (0.00,0.50)--(0.00,0.25) ; \draw (0.00,0.00)--(0.00,0.25) ; \draw [fill = red] (0.00,0.50) arc (90:270:0.25) ; \draw (0.00,0.00)--(0.00,0.50) ; \node at (1.50,0.25) {$\scriptstyle \Delta$} ; \node at (3.00,0.25) {$\scriptstyle \bot$} ; \node at (4.50,0.25) {$\scriptstyle \Delta'$} ; \draw (6.00,0.50)--(6.00,0.25) ; \draw (6.00,0.00)--(6.00,0.25) ; \draw [fill = blue] (6.00,0.00) arc (-90:90:0.25) ; \draw (6.00,0.00)--(6.00,0.50) ; \end{scope} \end{tikzpicture}}}$$
 
\item Similarly if  $\alpha= \parr_{A,B}$, $\Gamma'= A,B$ and $\alpha(\Gamma')= A\parr B$ or if  $\alpha= \bot$, $\Gamma'=\emptyset$ and $\alpha(\Gamma')=\bot$:
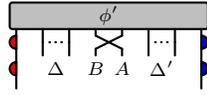
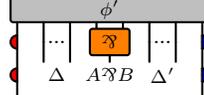
$$
\twocell{(G8) *1 (L *0 d *0 s *0 d *0 R) *1 (L *0 pitDel *0 pitB *0 pitA *0 pitDel1 *0 R)}
\qquad \qquad\qquad
\twocell{(G8) *1 (L *0 d *0 par *0 d *0 R) *1 (L *0 pitDel *0 pitAparB  *0 pitDel1 *0 R)}
$$

\item If $\alpha= \otimes_{A,B}$ so $\Gamma'= A, R, L, B$,  $\alpha(\Gamma')= A\otimes B$ and 
$$\phi': \pro \frr L,\Delta, A, R, L, B , \Delta',R. $$
This diagram is a parallel composition $\phi=\phi'_l ,\phi'_r$ with 
$$\phi'_l : \pro \frr L,\Delta, A, R\quad \mbox{ and } \quad  \phi'_r : \pro \frr L,B, \Delta',  R$$
of two diagrams which satisfy
 inductive hypothesis since $|\phi|_{\Mllt}=|\phi'_l|_{\Mllt} +|\phi'_r|_{ \Mllt}+1$:
$$ \twocell{(G51 *0 G52) *1 (L *0 3 *0 R *0 L *0 3 *0 R) *1 (L *0 d *0 tenc *0 d *0 R) *1 (L *0 pitDel *0 pitAtenB *0 pitDel1 *0 R)  }$$

\item Similarly if $\alpha= Cut_{A}$ with $B=A^\bot$ we have $\Gamma'=A,R, L,  A^\bot$ and $\alpha(\Gamma')=\emptyset$:
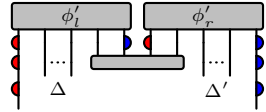
$$
\twocell{(G51 *0 G52) *1 (L *0 3 *0 R *0 L *0 3 *0 R) *1 (L *0 d *0 cutc *0 d *0 R) *1 (L *0 pitDel  *0 pitDel1 *0 R)  }
$$
\end{itemize}
\end{itemize}
\end{proof}
\end{theorem}
In particular, this theorem gives and \emph{representation procedure} to associate a diagram to a derivation and a \emph{sequentialization procedure} to associate a derivation to a proof diagram.

\begin{definition}[Representation]
We say that a proof diagram  $\phi\in\Mllc$ with $\phi: \pro \frr L, \Gamma, R$ \emph{represents} a derivation $d(\Gamma)$ if it can be sequentialized into the derivation $d(\Gamma)$, and  that a derivation $d(\Gamma)$ is represented by  $\phi$ or that $\phi$ is a \emph{diagrammatic representation of $d(\Gamma)$} if the derivation $d(\Gamma)$ can be imitated by $\phi$.
\end{definition}

\begin{definition}[Proof diagram branch]
We says that $\psi$ is a \emph{branch} of a sequentializable proof diagram $\phi$ if it is a subdiagram of the form $\psi: \pro \frr L, \Gamma, R$. 
\end{definition}
A branch $\psi\subseteq \psi$ represents to a sub-derivation of the derivation represented by $\phi$, in other words it is a branch of the relative derivation tree.

We  prove the termination of the polygraph $\Mlltc$ in order to give a definition of \emph{irreducible proof diagram}.

\begin{theorem}[Termination of $\Mlltc$]\label{termlltc}
The polygraph $\Mlltc$ is terminating.
\begin{proof}
We define a \emph{termination order} \cite{GuirTer} by associating to 
any proof diagram $\phi: \Gamma \frr \Delta$  a function $[ \phi] : \N ^{|\Gamma|} \frr \N ^{|\Delta|}$ according to the following interpretations:
$$
[\;\twocell{axc *1 (L *0 2 *0 R)}\;] : \emptyset \fr (1,1,1,1) \; , \qquad 
[\;\twocell{(1 *0 R *0 L *0 1) *1 cutc}\;]: (z_1,x,y,z_2)\fr \emptyset\; ,
$$
$$[\; \twocell{par}\; ] : (x,y)\fr x+y+1 \; ,\qquad 
[\twocell{(1 *0 R *0 L *0 1) *1 tenc}\; ] (x,z_1, z_2,y)\fr x+y+1 \; ,
$$
$$
[\;\twocell{s}\;]: (x,y)\fr (y, x+y) \; ,\qquad 
[\twocell{bot}]:(\emptyset )\fr 1 \; ,\qquad 
[\twocell{vc *1 ( L *0 1 *0 R)}]:(\emptyset )\fr (1,1,1) \; ,
$$
In particular, for any rule $\xymatrix@C=1em{\phi \ar@3[r]& \phi'}\in \Mlltc_3$ we have such that $[ \phi] > [{\phi'}]$:
{\small
$$
\Big[\; \twocell{s *1 s }\;\Big] (x,y)= (2x+y,x+y)> (x,y)=\Big[\;\twocell{2}\;\Big](x,y), 
$$
$$
\Bigg[\;{\twocell{(s *0 1)*1 (1 *0 s)*1 (s *0 1)}}\;\Bigg](x,y,z)=(2x+y+z, x+y,x)> (x+y+z,x+y,x)=\Bigg[\;{\twocell{ (1 *0 s)*1 (s *0 1)*1 (1 *0 s) }}\;\Bigg](x,y,z),
$$
$$
\Big[\; \twocell{axc *1 (L *0 s *0 R) }\;\Big] \emptyset= (0,2,1,0)> (0,1,1,0)=\Big[\;\twocell{axc *1 (L *0 2 *0 R)}\;\Big]\emptyset, 
$$
$$
\Bigg[\; \twocell{(par *0 1)*1 s}\;\Bigg](x,y,z)= (x+y+z+1, x+y+1) > (x+y+z,x+y+1)=\Bigg[\; \twocell{ (1 *0 s)*1 (s *0 1)*1 (1 *0 par) }\;\Bigg](x,y,z),
$$
$$
\Bigg[\; \twocell{(1 *0 par)*1 s}\;\Bigg](x,y,z)= (y+z+1, x+y+z+1) > (y+z+1,x+y)=\Bigg[\; \twocell{  (s *0 1)*1 (1 *0 s)*1(par *0 1) }\;\Bigg](x,y,z),
$$
$$
\Big[\; \twocell{(bot *0 1) *1 s }\;\Big](x)= (x+2,1) > (x,1)=\Big[\;\twocell{(1 *0 bot) *1 2}\;\Big] (x),
$$
$$
\Big[\;\twocell{(1 *0 bot) *1 s}\;\Big](x)= (x+2,x)> (1,x)=\Big[\; \twocell{(bot *0 1) *1 2}\;\Big](x),
$$
}
The compatibility of the order with sequential and parallel composition suffices to conclude that for any couple of  diagrams $[ \phi] > [ {\phi'}]$ holds whenever $\xymatrix@C=1em{ \phi \ar@3[r]^{*} & \psi}$. This rules out the existence of an infinite reduction path by the same argumentations given inTheorem \ref{permconf} proof.
\end{proof}
\end{theorem}

In the next section we study the quotient over derivations  induced by the morphisms in $\catgen{\Mlltc}$.

\subsection{The quotient over derivations induced by $\Mlltc$}\label{secQ}

The polygraph $\Mlltc$ generates a monoidal category $\catgen{ \Mlltc}$ where morphisms are the equivalence classes of proof diagrams generated by the signature $\Mlltc_2$ modulo the rewriting rules in $\Mlltc_3$. The representability of a derivation by means of a proof diagram gives raise to  an important question about the correlation between two derivations represented by the same proof diagram.

In this section we study the equivalence relation between derivations which can be represented by the same proof diagrams and by proof diagrams belonging to the same equivalence class in $\catgen{ \Mlltc }$. We compare it with the standard equivalence relation $\sim$ and the equivalence relation induced over derivation by the proof net syntax \cite{MLLu}.

If we denote $N_{d(\Gamma)}$ the proof net representing the derivation $d(\Gamma)$ and  $\phi_{d(\Gamma)}\in \Mlltc$ a proof diagram representing a derivation $d(\Gamma)$, we can define the following equivalence relations over $\MLLc$ derivations:
\begin{itemize}

\item we denote $\sim_N$ the equivalence relation over derivations induced by proof nets syntax. It is defined as follows:
$$d'(\Gamma)\sim_N d''(\Gamma) \mbox{ iff } N_{d'(\Gamma)}=N_{d''(\Gamma)}.$$
In other words, $d'(\Gamma)\sim_N d''(\Gamma)$ if and only if they can be represented by the same proof net.

\item we denote $\simeq_{D}$ the equivalence relation over derivations induced by proof diagram syntax. It is defined as follows:
$$d'(\Gamma)\simeq_{ D} d''(\Gamma) \mbox{ iff } \exists \phi \in \Mlltc \mbox{ such that }\phi_{d'(\Gamma)}=\phi=\phi_{d''(\Gamma)}.$$
In other words, $d'(\Gamma)\simeq_{ D} d''(\Gamma)$ if and only if they can be represented by the same proof diagram in  $\Mlltc_3$.

\item we denote $\sim_{\tilde D}$ the equivalence relation over derivations induced by $\catgen {\Mlltc}$. It is defined as follows:
$$d'(\Gamma)\sim_{\tilde D} d''(\Gamma) \mbox{ iff } \exists \phi_{d'(\Gamma)},\phi_{d''(\Gamma)}\in \Mlltc \mbox{ s.t.  } [\phi_{d'(\Gamma)}]_{\Mlltc}=[\phi_{d''(\Gamma)}]_{\Mlltc}.$$
In other words, $d'(\Gamma)\sim_{\tilde D} d''(\Gamma)$ if and only if they can be represented by two proof diagrams which are equivalent modulo $\Mlltc_3$. 
\end{itemize}

It is well-known that $\sim_N$ captures all permutation of multiplicative inference rules except the ones changing the jump assignation for a $\bot$ cell. This implies that $\sim_N=\sim$ over the pure multiplicative fragment of linear logic but that $\sim_N$ is finer than $\sim$ in presence of multiplicative units \cite{noMLL}.

We remark that $\sim_{\tilde D}$ captures all commutations of unary inference rules ($\bot $, $\parr$ and exchange) with disjoint sets of principal and active formula occurrences (by the interchange rule) together with permutations between $\bot$ or $\parr$ rules and exchange rules (by twisting relations). 

At the same time, in $\MLLc$ sequent calculus we usually consider sequents as multisets; thus, the equivalence relation $\simeq_{ D}$ does not really take into account the geometry of twisting operators in proof diagrams. In fact, we can always re-arrange the order of occurrences of formulas in the sequents inside a  derivation before represent it by a proof diagram. This allows to  shape at will the geometry of twisting operators of the representation of the derivation. For this reason,  unexpectedly (but not that much) it emerges that the two equivalence relations $\simeq_D$ and $\sim_{\tilde D}$ are equivalent.

However, this equivalence relation $\simeq_D$ is not able to capture all permutations of binary inference rules ($\otimes$ and $Cut$): let $\alpha, \beta \in \{\otimes, Cut\}$, then $\sim$ equates only permutations of the kind that follows:

\begin{scprooftree}{0.7}
\AxiomC{$\overset  1 \vdots$}
\noLine
\UnaryInfC{$\vdash \s, A $}

\AxiomC{$\overset  2 \vdots$}
\noLine
\UnaryInfC{$\vdash B, \Gamma, C$}

\RightLabel{$ \alpha $}
\BinaryInfC{$\vdash \s, \alpha(A, B),  \Gamma , C$}
\AxiomC{$\overset  3 \vdots$}
\noLine
\UnaryInfC{$\vdash D, \Delta $}
\RightLabel{$ \beta $}

\BinaryInfC{$ \vdash \s , \alpha (A, B) , \Gamma, \beta(C, D), \Delta$}

\AxiomC{~}
\noLine
\UnaryInfC{$~$}
\noLine
\UnaryInfC{$~$}
\noLine
\UnaryInfC{$~$}
\noLine
\UnaryInfC{$~$}
\noLine
\UnaryInfC{$~$}
\noLine
\UnaryInfC{$\sim$}

\AxiomC{$\overset  1 \vdots$}
\noLine
\UnaryInfC{$\vdash \s, A $}

\AxiomC{$\overset  2 \vdots$}
\noLine
\UnaryInfC{$\vdash B, \Gamma, C$}

\AxiomC{$\overset  3 \vdots$}
\noLine
\UnaryInfC{$\vdash D, \Delta $}

\RightLabel{$ \beta $}
\BinaryInfC{$\vdash A , \Gamma,   \beta(C, D), \Delta$}
\RightLabel{$ \alpha $}
\BinaryInfC{$ \vdash \s , \alpha (A , B) ,\Gamma, \beta(C ,D) , \Delta$}

\noLine
\TrinaryInfC{}

\end{scprooftree}

that is, permutations of $\otimes$ or $Cut$  rules that do not change the order of the branching in a derivation tree. 

For an actual example of these particular cases, consider the linear logic sequent $B\otimes C, A\otimes D$. This exhibits two different $\sim$-equivalent (but also $\sim_N$-equivalent) derivations  which are not $\simeq_{ D}$-equivalent:
\begin{scprooftree}{.7}
\AxiomC{$\overset  1 \vdots$}
\noLine
\UnaryInfC{$\vdash A,B $}

\AxiomC{$\overset  2 \vdots$}
\noLine
\UnaryInfC{$\vdash C$}

\RightLabel{$ \otimes $}
\BinaryInfC{$\vdash A, B\otimes C$}
\AxiomC{$\overset  3 \vdots$}
\noLine
\UnaryInfC{$\vdash D $}
\RightLabel{$ \otimes $}

\BinaryInfC{$ \vdash A\otimes D, B\otimes C$}

\AxiomC{~}
\noLine
\UnaryInfC{$\sim$}
\noLine
\UnaryInfC{$~$}
\noLine
\UnaryInfC{$~$}

\AxiomC{$\overset  1 \vdots$}
\noLine
\UnaryInfC{$\vdash  A,B $}

\AxiomC{$\overset  3 \vdots$}
\noLine
\UnaryInfC{$\vdash D$}

\RightLabel{$ \otimes $}
\BinaryInfC{$\vdash A\otimes D, B$}
\AxiomC{$\overset  2 \vdots$}
\noLine
\UnaryInfC{$\vdash C $}
\RightLabel{$ \otimes $}

\BinaryInfC{$ \vdash A \otimes D , B\otimes D$}

\noLine
\TrinaryInfC{}

\end{scprooftree}
in fact, their diagrammatic representations belong to two different equivalence classes in $\ls \Mlltc \rs$:
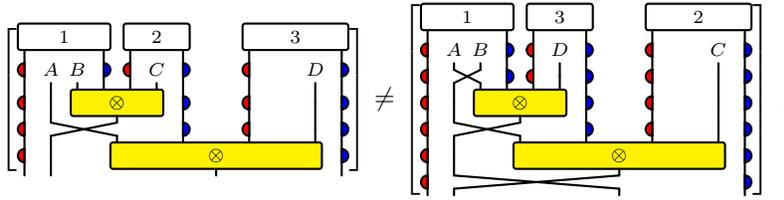
$$\begin{bmatrix}{\twocell{(dia1 *0 dia2 *0 dia3) *1 (L *0 topA *0 topB *0 R *0 L *0 topC *0 R *0 L *0 topD *0 R)*1 (L *0 1 *0 tenc *0 R *0 L *0 1 *0 R) *1 (L *0 s *0 R *0 L *0 1 *0 R) *1 (L *0 1 *0 tenc *0 R)}}\end{bmatrix} \neq
\begin{bmatrix}\twocell{(dia1 *0 dia3 *0 dia2)*1 (L *0 topA *0 topB *0 R *0 L *0 topD *0 R*0 L *0 topC *0 R)*1 (L *0 s *0 R *0 L *0 1 *0 R *0 L *0 1 *0 R) *1 (L *0 1 *0 tenc *0 R *0 L *0 1 *0 R) *1 (L *0 s *0 R *0 L *0 1 *0 R) *1 (L *0 1 *0 tenc *0 R)*1 (L *0 s *0 R)}\end{bmatrix} \;.$$ 

 It follows that $\simeq_{ D}$ equates less than $\sim$. Most of all, the equivalence relation $\simeq_{D}$ does not capture the part of the semantical equivalence which is required in order to take into account the elimination of commutative cuts and, consequently, to have an equivalence relation compatible  with the cut-elimination result.

In the next section, we extend our polygraph in order to make compatible with cut-elimination the induced equivalence relation over derivations.

\begin{oss}
The two equivalences $\simeq_D$ and $\sim_N$ are not comparable. In fact, we have that $\simeq_{ D}$ captures $\bot$ rules  permutations which change jump assignations that are not captured by $\sim_N $, but $\simeq_{D}$ does not capture permutations of binary inference rules which are perfectly captured by $\sim_N$.
\end{oss}
\subsection{The polygraph of $\MLLc$ proof diagrams}

In this section  we extend $\Mlltc$ to a polygraph $\Mllc$ in order to induce an equivalence over  proof diagrams which captures the standard equivalence over derivations. To this end,  we extend $\Mlltc$ with  generators and rewriting rules  in order to  enable some permutations of proof diagram branches. In effect, these transformations are forbidden in $\Mlltc$ by the presence of control strings which  impeach the definition of several twisting operators.

As remarked in the previous section, proof diagram syntax is inefficient to capture the standard proof equivalence in presence of some configurations including the ones of pure commutative cuts. This is because we keep records of how we manage occurrences of formulas in derivations (by means of twisting operators) revealing an  hidden ``tangle'' structure. 

\begin{definition}[Crossing split]
If $\phi\in \Mellt$ is an irreducible proof diagram, we says that $\phi$ has a \emph{crossing split} if it contains a subdiagram of the form

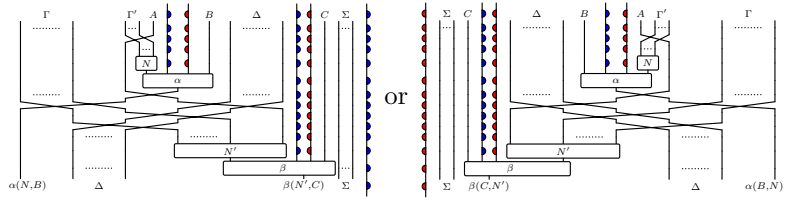
\begin{center}\resizebox{5cm}{!}{
$
{\raisebox{-66.25pt}{\begin{tikzpicture} \begin{scope} [ x = 10pt, y = 10pt, join = round, cap = round, thick, black, solid, -] \draw (12.50,13.50)--(12.50,13.00) (14.00,13.50)--(14.00,13.00) (21.75,13.50)--(21.75,13.00) (22.75,13.50)--(22.75,13.00) (26.75,13.50)--(26.75,13.00) ; \draw (26.75,0.25)--(26.75,-0.25)  ; \node at (3.88,12.75) {$\scriptstyle \Gamma$} ; \node at (10.00,12.75) {$\scriptstyle \Gamma'$} ; \node at (11.50,12.75) {$\scriptstyle A$} ; \draw (12.50,13.00)--(12.50,12.75) ; \draw (12.50,12.50)--(12.50,12.75) ; \draw [fill = blue] (12.50,12.50) arc (-90:90:0.25) ; \draw (12.50,12.50)--(12.50,13.00) ; \draw (14.00,13.00)--(14.00,12.75) ; \draw (14.00,12.50)--(14.00,12.75) ; \draw [fill = red] (14.00,13.00) arc (90:270:0.25) ; \draw (14.00,12.50)--(14.00,13.00) ; \node at (15.50,12.75) {$\scriptstyle B$} ; \node at (18.88,12.75) {$\scriptstyle \Delta$} ; \draw (21.75,13.00)--(21.75,12.75) ; \draw (21.75,12.50)--(21.75,12.75) ; \draw [fill = blue] (21.75,12.50) arc (-90:90:0.25) ; \draw (21.75,12.50)--(21.75,13.00) ; \draw (22.75,13.00)--(22.75,12.75) ; \draw (22.75,12.50)--(22.75,12.75) ; \draw [fill = red] (22.75,13.00) arc (90:270:0.25) ; \draw (22.75,12.50)--(22.75,13.00) ; \node at (23.75,12.75) {$\scriptstyle C$} ; \node at (25.25,12.75) {$\scriptstyle \Sigma$} ; \draw (26.75,13.00)--(26.75,12.75) ; \draw (26.75,12.50)--(26.75,12.75) ; \draw [fill = blue] (26.75,12.50) arc (-90:90:0.25) ; \draw (26.75,12.50)--(26.75,13.00) ; \draw  (2.00,12.25)--(2.00,11.75) (5.75,12.25)--(5.75,11.75) (9.50,12.25)--(9.50,11.75) (10.50,12.25)--(10.50,11.75) (11.50,12.25)--(11.50,11.75) (12.50,12.50)--(12.50,12.00) (14.00,12.50)--(14.00,12.00) (15.50,12.25)--(15.50,11.75) (17.00,12.25)--(17.00,11.75) (20.75,12.25)--(20.75,11.75) (21.75,12.50)--(21.75,12.00) (22.75,12.50)--(22.75,12.00) (23.75,12.25)--(23.75,11.75) (24.75,12.25)--(24.75,11.75) (25.75,12.25)--(25.75,11.75) (26.75,12.50)--(26.75,12.00) ; \draw [dash pattern = on 0.25pt off 2pt] (2.94,11.75)--(4.81,11.75) ; \draw [dash pattern = on 0.25pt off 2pt] (9.75,11.75)--(10.25,11.75) ;  \draw (12.50,12.00)--(12.50,11.75) ; \draw (12.50,11.50)--(12.50,11.75) ; \draw [fill = blue] (12.50,11.50) arc (-90:90:0.25) ; \draw (12.50,11.50)--(12.50,12.00) ; \draw (14.00,12.00)--(14.00,11.75) ; \draw (14.00,11.50)--(14.00,11.75) ; \draw [fill = red] (14.00,12.00) arc (90:270:0.25) ; \draw (14.00,11.50)--(14.00,12.00) ;  \draw [dash pattern = on 0.25pt off 2pt] (17.94,11.75)--(19.81,11.75) ; \draw (21.75,12.00)--(21.75,11.75) ; \draw (21.75,11.50)--(21.75,11.75) ; \draw [fill = blue] (21.75,11.50) arc (-90:90:0.25) ; \draw (21.75,11.50)--(21.75,12.00) ; \draw (22.75,12.00)--(22.75,11.75) ; \draw (22.75,11.50)--(22.75,11.75) ; \draw [fill = red] (22.75,12.00) arc (90:270:0.25) ; \draw (22.75,11.50)--(22.75,12.00) ;  \draw [dash pattern = on 0.25pt off 2pt] (25.00,11.75)--(25.50,11.75) ; \draw (26.75,12.00)--(26.75,11.75) ; \draw (26.75,11.50)--(26.75,11.75) ; \draw [fill = blue] (26.75,11.50) arc (-90:90:0.25) ; \draw (26.75,11.50)--(26.75,12.00) ; \draw  (2.00,11.75)--(2.00,11.00) (5.75,11.75)--(5.75,11.00) (9.50,11.75)--(9.50,11.25) (10.50,11.75)--(10.50,11.25) (11.50,11.75)--(11.50,11.25) (12.50,11.50)--(12.50,11.25) (14.00,11.50)--(14.00,11.25) (15.50,11.75)--(15.50,11.00) (17.00,11.75)--(17.00,11.00) (20.75,11.75)--(20.75,11.00) (21.75,11.50)--(21.75,11.25) (22.75,11.50)--(22.75,11.25) (23.75,11.75)--(23.75,11.00) (24.75,11.75)--(24.75,11.00) (25.75,11.75)--(25.75,11.00) (26.75,11.50)--(26.75,11.25) ;   \draw (9.50,11.25)--(10.50,10.75) (10.50,11.25)--(11.50,10.75) (11.50,11.25)--(9.50,10.75) ; \draw (12.50,11.25)--(12.50,11.00) ; \draw (12.50,10.75)--(12.50,11.00) ; \draw [fill = blue] (12.50,10.75) arc (-90:90:0.25) ; \draw (12.50,10.75)--(12.50,11.25) ; \draw (14.00,11.25)--(14.00,11.00) ; \draw (14.00,10.75)--(14.00,11.00) ; \draw [fill = red] (14.00,11.25) arc (90:270:0.25) ; \draw (14.00,10.75)--(14.00,11.25) ;   \draw (21.75,11.25)--(21.75,11.00) ; \draw (21.75,10.75)--(21.75,11.00) ; \draw [fill = blue] (21.75,10.75) arc (-90:90:0.25) ; \draw (21.75,10.75)--(21.75,11.25) ; \draw (22.75,11.25)--(22.75,11.00) ; \draw (22.75,10.75)--(22.75,11.00) ; \draw [fill = red] (22.75,11.25) arc (90:270:0.25) ; \draw (22.75,10.75)--(22.75,11.25) ;   \draw (26.75,11.25)--(26.75,11.00) ; \draw (26.75,10.75)--(26.75,11.00) ; \draw [fill = blue] (26.75,10.75) arc (-90:90:0.25) ; \draw (26.75,10.75)--(26.75,11.25) ; \draw  (2.00,11.00)--(2.00,10.25) (5.75,11.00)--(5.75,10.25) (9.50,10.75)--(9.50,10.25) (10.50,10.75)--(10.50,10.25) (11.50,10.75)--(11.50,10.25) (12.50,10.75)--(12.50,10.50) (14.00,10.75)--(14.00,10.50) (15.50,11.00)--(15.50,10.25) (17.00,11.00)--(17.00,10.25) (20.75,11.00)--(20.75,10.25) (21.75,10.75)--(21.75,10.50) (22.75,10.75)--(22.75,10.50) (23.75,11.00)--(23.75,10.25) (24.75,11.00)--(24.75,10.25) (25.75,11.00)--(25.75,10.25) (26.75,10.75)--(26.75,10.50) ;    \draw [dash pattern = on 0.25pt off 2pt] (10.75,10.25)--(11.25,10.25) ; \draw (12.50,10.50)--(12.50,10.25) ; \draw (12.50,10.00)--(12.50,10.25) ; \draw [fill = blue] (12.50,10.00) arc (-90:90:0.25) ; \draw (12.50,10.00)--(12.50,10.50) ; \draw (14.00,10.50)--(14.00,10.25) ; \draw (14.00,10.00)--(14.00,10.25) ; \draw [fill = red] (14.00,10.50) arc (90:270:0.25) ; \draw (14.00,10.00)--(14.00,10.50) ;   \draw (21.75,10.50)--(21.75,10.25) ; \draw (21.75,10.00)--(21.75,10.25) ; \draw [fill = blue] (21.75,10.00) arc (-90:90:0.25) ; \draw (21.75,10.00)--(21.75,10.50) ; \draw (22.75,10.50)--(22.75,10.25) ; \draw (22.75,10.00)--(22.75,10.25) ; \draw [fill = red] (22.75,10.50) arc (90:270:0.25) ; \draw (22.75,10.00)--(22.75,10.50) ;   \draw (26.75,10.50)--(26.75,10.25) ; \draw (26.75,10.00)--(26.75,10.25) ; \draw [fill = blue] (26.75,10.00) arc (-90:90:0.25) ; \draw (26.75,10.00)--(26.75,10.50) ; \draw  (2.00,10.25)--(2.00,9.25) (5.75,10.25)--(5.75,9.25) (9.50,10.25)--(9.50,9.25) (10.50,10.25)--(10.50,9.75) (11.50,10.25)--(11.50,9.75) (12.50,10.00)--(12.50,9.50) (14.00,10.00)--(14.00,9.50) (15.50,10.25)--(15.50,9.25) (17.00,10.25)--(17.00,9.25) (20.75,10.25)--(20.75,9.25) (21.75,10.00)--(21.75,9.50) (22.75,10.00)--(22.75,9.50) (23.75,10.25)--(23.75,9.25) (24.75,10.25)--(24.75,9.25) (25.75,10.25)--(25.75,9.25) (26.75,10.00)--(26.75,9.50) ;    \draw [rounded corners = 1pt, fill = white] (10.25,8.75) rectangle (11.75,9.75) ; \node at (11.00,9.25) {$\scriptstyle N$} ; \draw (12.50,9.50)--(12.50,9.25) ; \draw (12.50,9.00)--(12.50,9.25) ; \draw [fill = blue] (12.50,9.00) arc (-90:90:0.25) ; \draw (12.50,9.00)--(12.50,9.50) ; \draw (14.00,9.50)--(14.00,9.25) ; \draw (14.00,9.00)--(14.00,9.25) ; \draw [fill = red] (14.00,9.50) arc (90:270:0.25) ; \draw (14.00,9.00)--(14.00,9.50) ;   \draw (21.75,9.50)--(21.75,9.25) ; \draw (21.75,9.00)--(21.75,9.25) ; \draw [fill = blue] (21.75,9.00) arc (-90:90:0.25) ; \draw (21.75,9.00)--(21.75,9.50) ; \draw (22.75,9.50)--(22.75,9.25) ; \draw (22.75,9.00)--(22.75,9.25) ; \draw [fill = red] (22.75,9.50) arc (90:270:0.25) ; \draw (22.75,9.00)--(22.75,9.50) ;   \draw (26.75,9.50)--(26.75,9.25) ; \draw (26.75,9.00)--(26.75,9.25) ; \draw [fill = blue] (26.75,9.00) arc (-90:90:0.25) ; \draw (26.75,9.00)--(26.75,9.50) ; \draw  (2.00,9.25)--(2.00,8.00) (5.75,9.25)--(5.75,8.00) (9.50,9.25)--(9.50,8.00) (11.00,8.75)--(11.00,8.50) (12.50,9.00)--(12.50,8.50) (14.00,9.00)--(14.00,8.50) (15.50,9.25)--(15.50,8.50) (17.00,9.25)--(17.00,8.00) (20.75,9.25)--(20.75,8.00) (21.75,9.00)--(21.75,8.25) (22.75,9.00)--(22.75,8.25) (23.75,9.25)--(23.75,8.00) (24.75,9.25)--(24.75,8.00) (25.75,9.25)--(25.75,8.00) (26.75,9.00)--(26.75,8.25) ;    \draw [rounded corners = 1pt, fill = white] (10.75,7.50) rectangle (15.75,8.50) ; \node at (13.25,8.00) {$\scriptstyle \alpha$} ;  \draw (21.75,8.25)--(21.75,8.00) ; \draw (21.75,7.75)--(21.75,8.00) ; \draw [fill = blue] (21.75,7.75) arc (-90:90:0.25) ; \draw (21.75,7.75)--(21.75,8.25) ; \draw (22.75,8.25)--(22.75,8.00) ; \draw (22.75,7.75)--(22.75,8.00) ; \draw [fill = red] (22.75,8.25) arc (90:270:0.25) ; \draw (22.75,7.75)--(22.75,8.25) ;   \draw (26.75,8.25)--(26.75,8.00) ; \draw (26.75,7.75)--(26.75,8.00) ; \draw [fill = blue] (26.75,7.75) arc (-90:90:0.25) ; \draw (26.75,7.75)--(26.75,8.25) ; \draw (2.00,8.00)--(2.00,7.00) (5.75,8.00)--(5.75,7.00) (9.50,8.00)--(9.50,7.25) (13.25,7.50)--(13.25,7.25) (17.00,8.00)--(17.00,7.00) (20.75,8.00)--(20.75,7.00) (21.75,7.75)--(21.75,7.25) (22.75,7.75)--(22.75,7.25) (23.75,8.00)--(23.75,7.00) (24.75,8.00)--(24.75,7.00) (25.75,8.00)--(25.75,7.00) (26.75,7.75)--(26.75,7.25) ;  \draw [dash pattern = on 0.25pt off 2pt] (2.94,7.00)--(4.81,7.00) ; \draw (9.50,7.25)--(13.25,6.75) (13.25,7.25)--(9.50,6.75) ; \draw [dash pattern = on 0.25pt off 2pt] (17.94,7.00)--(19.81,7.00) ; \draw (21.75,7.25)--(21.75,7.00) ; \draw (21.75,6.75)--(21.75,7.00) ; \draw [fill = blue] (21.75,6.75) arc (-90:90:0.25) ; \draw (21.75,6.75)--(21.75,7.25) ; \draw (22.75,7.25)--(22.75,7.00) ; \draw (22.75,6.75)--(22.75,7.00) ; \draw [fill = red] (22.75,7.25) arc (90:270:0.25) ; \draw (22.75,6.75)--(22.75,7.25) ;   \draw (26.75,7.25)--(26.75,7.00) ; \draw (26.75,6.75)--(26.75,7.00) ; \draw [fill = blue] (26.75,6.75) arc (-90:90:0.25) ; \draw (26.75,6.75)--(26.75,7.25) ; \draw  (2.00,7.00)--(2.00,6.50) (5.75,7.00)--(5.75,6.50) (9.50,6.75)--(9.50,6.50) (13.25,6.75)--(13.25,6.50) (17.00,7.00)--(17.00,6.50) (20.75,7.00)--(20.75,6.50) (21.75,6.75)--(21.75,6.50) (22.75,6.75)--(22.75,6.50) (23.75,7.00)--(23.75,6.25) (24.75,7.00)--(24.75,6.25) (25.75,7.00)--(25.75,6.25) (26.75,6.75)--(26.75,6.50) ;  \draw (2.00,6.50)--(5.75,6.00) (5.75,6.50)--(9.50,6.00) (9.50,6.50)--(2.00,6.00) ; \draw (17.00,6.50)--(13.25,6.00) (20.75,6.50)--(17.00,6.00) (13.25,6.50)--(20.75,6.00) ; \draw (21.75,6.50)--(21.75,6.25) ; \draw (21.75,6.00)--(21.75,6.25) ; \draw [fill = blue] (21.75,6.00) arc (-90:90:0.25) ; \draw (21.75,6.00)--(21.75,6.50) ; \draw (22.75,6.50)--(22.75,6.25) ; \draw (22.75,6.00)--(22.75,6.25) ; \draw [fill = red] (22.75,6.50) arc (90:270:0.25) ; \draw (22.75,6.00)--(22.75,6.50) ;   \draw (26.75,6.50)--(26.75,6.25) ; \draw (26.75,6.00)--(26.75,6.25) ; \draw [fill = blue] (26.75,6.00) arc (-90:90:0.25) ; \draw (26.75,6.00)--(26.75,6.50) ; \draw  (2.00,6.00)--(2.00,5.50) (5.75,6.00)--(5.75,5.50) (9.50,6.00)--(9.50,5.75) (13.25,6.00)--(13.25,5.75) (17.00,6.00)--(17.00,5.75) (20.75,6.00)--(20.75,5.50) (21.75,6.00)--(21.75,5.75) (22.75,6.00)--(22.75,5.75) (23.75,6.25)--(23.75,5.50) (24.75,6.25)--(24.75,5.50) (25.75,6.25)--(25.75,5.50) (26.75,6.00)--(26.75,5.75) ;   \draw (13.25,5.75)--(9.50,5.25) (17.00,5.75)--(13.25,5.25) (9.50,5.75)--(17.00,5.25) ;  \draw (21.75,5.75)--(21.75,5.50) ; \draw (21.75,5.25)--(21.75,5.50) ; \draw [fill = blue] (21.75,5.25) arc (-90:90:0.25) ; \draw (21.75,5.25)--(21.75,5.75) ; \draw (22.75,5.75)--(22.75,5.50) ; \draw (22.75,5.25)--(22.75,5.50) ; \draw [fill = red] (22.75,5.75) arc (90:270:0.25) ; \draw (22.75,5.25)--(22.75,5.75) ;   \draw (26.75,5.75)--(26.75,5.50) ; \draw (26.75,5.25)--(26.75,5.50) ; \draw [fill = blue] (26.75,5.25) arc (-90:90:0.25) ; \draw (26.75,5.25)--(26.75,5.75) ; \draw (2.00,5.50)--(2.00,4.75) (5.75,5.50)--(5.75,5.00) (9.50,5.25)--(9.50,5.00) (13.25,5.25)--(13.25,5.00) (17.00,5.25)--(17.00,4.75) (20.75,5.50)--(20.75,4.75) (21.75,5.25)--(21.75,5.00) (22.75,5.25)--(22.75,5.00) (23.75,5.50)--(23.75,4.75) (24.75,5.50)--(24.75,4.75) (25.75,5.50)--(25.75,4.75) (26.75,5.25)--(26.75,5.00) ;   \draw (9.50,5.00)--(5.75,4.50) (13.25,5.00)--(9.50,4.50) (5.75,5.00)--(13.25,4.50) ;  \draw (21.75,5.00)--(21.75,4.75) ; \draw (21.75,4.50)--(21.75,4.75) ; \draw [fill = blue] (21.75,4.50) arc (-90:90:0.25) ; \draw (21.75,4.50)--(21.75,5.00) ; \draw (22.75,5.00)--(22.75,4.75) ; \draw (22.75,4.50)--(22.75,4.75) ; \draw [fill = red] (22.75,5.00) arc (90:270:0.25) ; \draw (22.75,4.50)--(22.75,5.00) ;   \draw (26.75,5.00)--(26.75,4.75) ; \draw (26.75,4.50)--(26.75,4.75) ; \draw [fill = blue] (26.75,4.50) arc (-90:90:0.25) ; \draw (26.75,4.50)--(26.75,5.00) ; \draw (2.00,4.75)--(2.00,4.00) (5.75,4.50)--(5.75,4.00) (9.50,4.50)--(9.50,4.00) (13.25,4.50)--(13.25,4.00) (17.00,4.75)--(17.00,4.00) (20.75,4.75)--(20.75,4.00) (21.75,4.50)--(21.75,4.25) (22.75,4.50)--(22.75,4.25) (23.75,4.75)--(23.75,4.00) (24.75,4.75)--(24.75,4.00) (25.75,4.75)--(25.75,4.00) (26.75,4.50)--(26.75,4.25) ;   \draw [dash pattern = on 0.25pt off 2pt] (6.69,4.00)--(8.56,4.00) ; \draw [dash pattern = on 0.25pt off 2pt] (14.19,4.00)--(16.06,4.00) ;  \draw (21.75,4.25)--(21.75,4.00) ; \draw (21.75,3.75)--(21.75,4.00) ; \draw [fill = blue] (21.75,3.75) arc (-90:90:0.25) ; \draw (21.75,3.75)--(21.75,4.25) ; \draw (22.75,4.25)--(22.75,4.00) ; \draw (22.75,3.75)--(22.75,4.00) ; \draw [fill = red] (22.75,4.25) arc (90:270:0.25) ; \draw (22.75,3.75)--(22.75,4.25) ;   \draw (26.75,4.25)--(26.75,4.00) ; \draw (26.75,3.75)--(26.75,4.00) ; \draw [fill = blue] (26.75,3.75) arc (-90:90:0.25) ; \draw (26.75,3.75)--(26.75,4.25) ; \draw  (2.00,4.00)--(2.00,3.00) (5.75,4.00)--(5.75,3.00) (9.50,4.00)--(9.50,3.00) (13.25,4.00)--(13.25,3.50) (17.00,4.00)--(17.00,3.50) (20.75,4.00)--(20.75,3.50) (21.75,3.75)--(21.75,3.25) (22.75,3.75)--(22.75,3.25) (23.75,4.00)--(23.75,3.00) (24.75,4.00)--(24.75,3.00) (25.75,4.00)--(25.75,3.00) (26.75,3.75)--(26.75,3.25) ;    \draw [rounded corners = 1pt, fill = white] (13.00,2.50) rectangle (21.00,3.50) ; \node at (17.00,3.00) {$\scriptstyle N'$} ; \draw (21.75,3.25)--(21.75,3.00) ; \draw (21.75,2.75)--(21.75,3.00) ; \draw [fill = blue] (21.75,2.75) arc (-90:90:0.25) ; \draw (21.75,2.75)--(21.75,3.25) ; \draw (22.75,3.25)--(22.75,3.00) ; \draw (22.75,2.75)--(22.75,3.00) ; \draw [fill = red] (22.75,3.25) arc (90:270:0.25) ; \draw (22.75,2.75)--(22.75,3.25) ;   \draw (26.75,3.25)--(26.75,3.00) ; \draw (26.75,2.75)--(26.75,3.00) ; \draw [fill = blue] (26.75,2.75) arc (-90:90:0.25) ; \draw (26.75,2.75)--(26.75,3.25) ; \draw (2.00,3.00)--(2.00,1.75) (5.75,3.00)--(5.75,1.75) (9.50,3.00)--(9.50,1.75) (17.00,2.50)--(17.00,2.25) (21.75,2.75)--(21.75,2.25) (22.75,2.75)--(22.75,2.25) (23.75,3.00)--(23.75,2.25) (24.75,3.00)--(24.75,1.75) (25.75,3.00)--(25.75,1.75) (26.75,2.75)--(26.75,2.00) ;  \draw [dash pattern = on 0.25pt off 2pt] (6.69,1.75)--(8.56,1.75) ; \node at (17.00,1.75) {$\scriptstyle ~$} ; \draw [rounded corners = 1pt, fill = white] (16.50,1.25) rectangle (24.50,2.25) ; \node at (20.75,1.75) {$\scriptstyle \beta$} ; \node at (23.75,1.75) {$\scriptstyle ~$} ; \draw [dash pattern = on 0.25pt off 2pt] (25.00,1.75)--(25.50,1.75) ; \draw (26.75,2.00)--(26.75,1.75) ; \draw (26.75,1.50)--(26.75,1.75) ; \draw [fill = blue] (26.75,1.50) arc (-90:90:0.25) ; \draw (26.75,1.50)--(26.75,2.00) ; \draw (2.00,1.75)--(2.00,1.00) (5.75,1.75)--(5.75,1.00) (9.50,1.75)--(9.50,1.00) (22.25,1.25)--(22.25,1.00) (24.75,1.75)--(24.75,1.00) (25.75,1.75)--(25.75,1.00) (26.75,1.50)--(26.75,0.75); \node at (2.50,0.50) {$\scriptstyle \alpha(N, B)$} ; \node at (7.63,0.50) {$\scriptstyle \Delta$} ; \node at (22.25,0.50) {$\scriptstyle \beta(N',C)$} ; \node at (25.25,0.50) {$\scriptstyle \Sigma$} ; \draw (26.75,0.75)--(26.75,0.50) ; \draw (26.75,0.25)--(26.75,0.50) ; \draw [fill = blue] (26.75,0.25) arc (-90:90:0.25) ; \draw (26.75,0.25)--(26.75,0.75) ; \end{scope} \end{tikzpicture}}}$
}
 or \resizebox{5cm}{!}{
$
{\raisebox{-66.25pt}{\begin{tikzpicture} \begin{scope} [ x = 10pt, y = 10pt, join = round, cap = round, thick, black, solid, -] \draw (0.00,13.25)--(0.00,12.75) (4.00,13.25)--(4.00,12.75) (5.00,13.25)--(5.00,12.75) (12.75,13.25)--(12.75,12.75) (14.25,13.25)--(14.25,12.75) ; \draw (0.00,0.00)--(0.00,-0.50) ; \draw (0.00,12.75)--(0.00,12.50) ; \draw (0.00,12.25)--(0.00,12.50) ; \draw [fill = red] (0.00,12.75) arc (90:270:0.25) ; \draw (0.00,12.25)--(0.00,12.75) ; \node at (1.50,12.50) {$\scriptstyle \Sigma$} ; \node at (3.00,12.50) {$\scriptstyle C$} ; \draw (4.00,12.75)--(4.00,12.50) ; \draw (4.00,12.25)--(4.00,12.50) ; \draw [fill = blue] (4.00,12.25) arc (-90:90:0.25) ; \draw (4.00,12.25)--(4.00,12.75) ; \draw (5.00,12.75)--(5.00,12.50) ; \draw (5.00,12.25)--(5.00,12.50) ; \draw [fill = red] (5.00,12.75) arc (90:270:0.25) ; \draw (5.00,12.25)--(5.00,12.75) ; \node at (7.88,12.50) {$\scriptstyle \Delta$} ; \node at (11.25,12.50) {$\scriptstyle B$} ; \draw (12.75,12.75)--(12.75,12.50) ; \draw (12.75,12.25)--(12.75,12.50) ; \draw [fill = blue] (12.75,12.25) arc (-90:90:0.25) ; \draw (12.75,12.25)--(12.75,12.75) ; \draw (14.25,12.75)--(14.25,12.50) ; \draw (14.25,12.25)--(14.25,12.50) ; \draw [fill = red] (14.25,12.75) arc (90:270:0.25) ; \draw (14.25,12.25)--(14.25,12.75) ; \node at (15.25,12.50) {$\scriptstyle A$} ; \node at (16.75,12.50) {$\scriptstyle \Gamma'$} ; \node at (22.88,12.50) {$\scriptstyle \Gamma$}  ; \draw (0.00,12.25)--(0.00,11.75) (1.00,12.00)--(1.00,11.50) (2.00,12.00)--(2.00,11.50) (3.00,12.00)--(3.00,11.50) (4.00,12.25)--(4.00,11.75) (5.00,12.25)--(5.00,11.75) (6.00,12.00)--(6.00,11.50) (9.75,12.00)--(9.75,11.50) (11.25,12.00)--(11.25,11.50) (12.75,12.25)--(12.75,11.75) (14.25,12.25)--(14.25,11.75) (15.25,12.00)--(15.25,11.50) (16.25,12.00)--(16.25,11.50) (17.25,12.00)--(17.25,11.50) (21.00,12.00)--(21.00,11.50) (24.75,12.00)--(24.75,11.50) ; \draw (0.00,11.75)--(0.00,11.50) ; \draw (0.00,11.25)--(0.00,11.50) ; \draw [fill = red] (0.00,11.75) arc (90:270:0.25) ; \draw (0.00,11.25)--(0.00,11.75) ; \draw [dash pattern = on 0.25pt off 2pt] (1.25,11.50)--(1.75,11.50) ;  \draw (4.00,11.75)--(4.00,11.50) ; \draw (4.00,11.25)--(4.00,11.50) ; \draw [fill = blue] (4.00,11.25) arc (-90:90:0.25) ; \draw (4.00,11.25)--(4.00,11.75) ; \draw (5.00,11.75)--(5.00,11.50) ; \draw (5.00,11.25)--(5.00,11.50) ; \draw [fill = red] (5.00,11.75) arc (90:270:0.25) ; \draw (5.00,11.25)--(5.00,11.75) ; \draw [dash pattern = on 0.25pt off 2pt] (6.94,11.50)--(8.81,11.50) ;  \draw (12.75,11.75)--(12.75,11.50) ; \draw (12.75,11.25)--(12.75,11.50) ; \draw [fill = blue] (12.75,11.25) arc (-90:90:0.25) ; \draw (12.75,11.25)--(12.75,11.75) ; \draw (14.25,11.75)--(14.25,11.50) ; \draw (14.25,11.25)--(14.25,11.50) ; \draw [fill = red] (14.25,11.75) arc (90:270:0.25) ; \draw (14.25,11.25)--(14.25,11.75) ;  \draw [dash pattern = on 0.25pt off 2pt] (16.50,11.50)--(17.00,11.50) ; \draw [dash pattern = on 0.25pt off 2pt] (21.94,11.50)--(23.81,11.50) ; \draw (0.00,11.25)--(0.00,11.00) (1.00,11.50)--(1.00,10.75) (2.00,11.50)--(2.00,10.75) (3.00,11.50)--(3.00,10.75) (4.00,11.25)--(4.00,11.00) (5.00,11.25)--(5.00,11.00) (6.00,11.50)--(6.00,10.50) (9.75,11.50)--(9.75,10.50) (11.25,11.50)--(11.25,11.00) (12.75,11.25)--(12.75,11.00) (14.25,11.25)--(14.25,11.00) (15.25,11.50)--(15.25,11.00) (16.25,11.50)--(16.25,11.00) (17.25,11.50)--(17.25,11.00) (21.00,11.50)--(21.00,10.75) (24.75,11.50)--(24.75,10.75)  ; \draw (0.00,11.00)--(0.00,10.75) ; \draw (0.00,10.50)--(0.00,10.75) ; \draw [fill = red] (0.00,11.00) arc (90:270:0.25) ; \draw (0.00,10.50)--(0.00,11.00) ;  \draw (4.00,11.00)--(4.00,10.75) ; \draw (4.00,10.50)--(4.00,10.75) ; \draw [fill = blue] (4.00,10.50) arc (-90:90:0.25) ; \draw (4.00,10.50)--(4.00,11.00) ; \draw (5.00,11.00)--(5.00,10.75) ; \draw (5.00,10.50)--(5.00,10.75) ; \draw [fill = red] (5.00,11.00) arc (90:270:0.25) ; \draw (5.00,10.50)--(5.00,11.00) ;  \draw (12.75,11.00)--(12.75,10.75) ; \draw (12.75,10.50)--(12.75,10.75) ; \draw [fill = blue] (12.75,10.50) arc (-90:90:0.25) ; \draw (12.75,10.50)--(12.75,11.00) ; \draw (14.25,11.00)--(14.25,10.75) ; \draw (14.25,10.50)--(14.25,10.75) ; \draw [fill = red] (14.25,11.00) arc (90:270:0.25) ; \draw (14.25,10.50)--(14.25,11.00) ; \draw (16.25,11.00)--(15.25,10.50) (17.25,11.00)--(16.25,10.50) (15.25,11.00)--(17.25,10.50) ;  \draw (0.00,10.50)--(0.00,10.25) (1.00,10.75)--(1.00,10.00) (2.00,10.75)--(2.00,10.00) (3.00,10.75)--(3.00,10.00) (4.00,10.50)--(4.00,10.25) (5.00,10.50)--(5.00,10.25) (6.00,10.50)--(6.00,9.75) (9.75,10.50)--(9.75,9.75) (11.25,11.00)--(11.25,10.25) (12.75,10.50)--(12.75,10.25) (14.25,10.50)--(14.25,10.25) (15.25,10.50)--(15.25,10.00) (16.25,10.50)--(16.25,10.00) (17.25,10.50)--(17.25,10.25) (21.00,10.75)--(21.00,10.00) (24.75,10.75)--(24.75,10.00)  ; \draw (0.00,10.25)--(0.00,10.00) ; \draw (0.00,9.75)--(0.00,10.00) ; \draw [fill = red] (0.00,10.25) arc (90:270:0.25) ; \draw (0.00,9.75)--(0.00,10.25) ;  \draw (4.00,10.25)--(4.00,10.00) ; \draw (4.00,9.75)--(4.00,10.00) ; \draw [fill = blue] (4.00,9.75) arc (-90:90:0.25) ; \draw (4.00,9.75)--(4.00,10.25) ; \draw (5.00,10.25)--(5.00,10.00) ; \draw (5.00,9.75)--(5.00,10.00) ; \draw [fill = red] (5.00,10.25) arc (90:270:0.25) ; \draw (5.00,9.75)--(5.00,10.25) ;  \draw (12.75,10.25)--(12.75,10.00) ; \draw (12.75,9.75)--(12.75,10.00) ; \draw [fill = blue] (12.75,9.75) arc (-90:90:0.25) ; \draw (12.75,9.75)--(12.75,10.25) ; \draw (14.25,10.25)--(14.25,10.00) ; \draw (14.25,9.75)--(14.25,10.00) ; \draw [fill = red] (14.25,10.25) arc (90:270:0.25) ; \draw (14.25,9.75)--(14.25,10.25) ; \draw [dash pattern = on 0.25pt off 2pt] (15.50,10.00)--(16.00,10.00) ;  \draw (0.00,9.75)--(0.00,9.25) (1.00,10.00)--(1.00,9.00) (2.00,10.00)--(2.00,9.00) (3.00,10.00)--(3.00,9.00) (4.00,9.75)--(4.00,9.25) (5.00,9.75)--(5.00,9.25) (6.00,9.75)--(6.00,8.75) (9.75,9.75)--(9.75,8.75) (11.25,10.25)--(11.25,9.25) (12.75,9.75)--(12.75,9.25) (14.25,9.75)--(14.25,9.25) (15.25,10.00)--(15.25,9.50) (16.25,10.00)--(16.25,9.50) (17.25,10.25)--(17.25,9.25) (21.00,10.00)--(21.00,9.00) (24.75,10.00)--(24.75,9.00)  ; \draw (0.00,9.25)--(0.00,9.00) ; \draw (0.00,8.75)--(0.00,9.00) ; \draw [fill = red] (0.00,9.25) arc (90:270:0.25) ; \draw (0.00,8.75)--(0.00,9.25) ;  \draw (4.00,9.25)--(4.00,9.00) ; \draw (4.00,8.75)--(4.00,9.00) ; \draw [fill = blue] (4.00,8.75) arc (-90:90:0.25) ; \draw (4.00,8.75)--(4.00,9.25) ; \draw (5.00,9.25)--(5.00,9.00) ; \draw (5.00,8.75)--(5.00,9.00) ; \draw [fill = red] (5.00,9.25) arc (90:270:0.25) ; \draw (5.00,8.75)--(5.00,9.25) ;  \draw (12.75,9.25)--(12.75,9.00) ; \draw (12.75,8.75)--(12.75,9.00) ; \draw [fill = blue] (12.75,8.75) arc (-90:90:0.25) ; \draw (12.75,8.75)--(12.75,9.25) ; \draw (14.25,9.25)--(14.25,9.00) ; \draw (14.25,8.75)--(14.25,9.00) ; \draw [fill = red] (14.25,9.25) arc (90:270:0.25) ; \draw (14.25,8.75)--(14.25,9.25) ; \draw [rounded corners = 1pt, fill = white] (15.00,8.50) rectangle (16.50,9.50) ; \node at (15.75,9.00) {$\scriptstyle N$} ;  \draw (0.00,8.75)--(0.00,8.00) (1.00,9.00)--(1.00,7.75) (2.00,9.00)--(2.00,7.75) (3.00,9.00)--(3.00,7.75) (4.00,8.75)--(4.00,8.00) (5.00,8.75)--(5.00,8.00) (6.00,8.75)--(6.00,7.75) (9.75,8.75)--(9.75,7.75) (11.25,9.25)--(11.25,8.25) (12.75,8.75)--(12.75,8.25) (14.25,8.75)--(14.25,8.25) (15.75,8.50)--(15.75,8.25) (17.25,9.25)--(17.25,8.00) (21.00,9.00)--(21.00,7.75) (24.75,9.00)--(24.75,7.75) ; \draw (0.00,8.00)--(0.00,7.75) ; \draw (0.00,7.50)--(0.00,7.75) ; \draw [fill = red] (0.00,8.00) arc (90:270:0.25) ; \draw (0.00,7.50)--(0.00,8.00) ;  \draw (4.00,8.00)--(4.00,7.75) ; \draw (4.00,7.50)--(4.00,7.75) ; \draw [fill = blue] (4.00,7.50) arc (-90:90:0.25) ; \draw (4.00,7.50)--(4.00,8.00) ; \draw (5.00,8.00)--(5.00,7.75) ; \draw (5.00,7.50)--(5.00,7.75) ; \draw [fill = red] (5.00,8.00) arc (90:270:0.25) ; \draw (5.00,7.50)--(5.00,8.00) ;  \draw [rounded corners = 1pt, fill = white] (11.00,7.25) rectangle (16.00,8.25) ; \node at (13.50,7.75) {$\scriptstyle \alpha$} ;  \draw (0.00,7.50)--(0.00,7.00) (1.00,7.75)--(1.00,6.75) (2.00,7.75)--(2.00,6.75) (3.00,7.75)--(3.00,6.75) (4.00,7.50)--(4.00,7.00) (5.00,7.50)--(5.00,7.00) (6.00,7.75)--(6.00,6.75) (9.75,7.75)--(9.75,6.75) (13.50,7.25)--(13.50,7.00) (17.25,8.00)--(17.25,7.00) (21.00,7.75)--(21.00,6.75) (24.75,7.75)--(24.75,6.75)  ; \draw (0.00,7.00)--(0.00,6.75) ; \draw (0.00,6.50)--(0.00,6.75) ; \draw [fill = red] (0.00,7.00) arc (90:270:0.25) ; \draw (0.00,6.50)--(0.00,7.00) ;  \draw (4.00,7.00)--(4.00,6.75) ; \draw (4.00,6.50)--(4.00,6.75) ; \draw [fill = blue] (4.00,6.50) arc (-90:90:0.25) ; \draw (4.00,6.50)--(4.00,7.00) ; \draw (5.00,7.00)--(5.00,6.75) ; \draw (5.00,6.50)--(5.00,6.75) ; \draw [fill = red] (5.00,7.00) arc (90:270:0.25) ; \draw (5.00,6.50)--(5.00,7.00) ; \draw [dash pattern = on 0.25pt off 2pt] (6.94,6.75)--(8.81,6.75) ; \draw (13.50,7.00)--(17.25,6.50) (17.25,7.00)--(13.50,6.50) ; \draw [dash pattern = on 0.25pt off 2pt] (21.94,6.75)--(23.81,6.75) ;  \draw (0.00,6.50)--(0.00,6.25) (1.00,6.75)--(1.00,6.00) (2.00,6.75)--(2.00,6.00) (3.00,6.75)--(3.00,6.00) (4.00,6.50)--(4.00,6.25) (5.00,6.50)--(5.00,6.25) (6.00,6.75)--(6.00,6.25) (9.75,6.75)--(9.75,6.25) (13.50,6.50)--(13.50,6.25) (17.25,6.50)--(17.25,6.25) (21.00,6.75)--(21.00,6.25) (24.75,6.75)--(24.75,6.25)  ; \draw (0.00,6.25)--(0.00,6.00) ; \draw (0.00,5.75)--(0.00,6.00) ; \draw [fill = red] (0.00,6.25) arc (90:270:0.25) ; \draw (0.00,5.75)--(0.00,6.25) ;  \draw (4.00,6.25)--(4.00,6.00) ; \draw (4.00,5.75)--(4.00,6.00) ; \draw [fill = blue] (4.00,5.75) arc (-90:90:0.25) ; \draw (4.00,5.75)--(4.00,6.25) ; \draw (5.00,6.25)--(5.00,6.00) ; \draw (5.00,5.75)--(5.00,6.00) ; \draw [fill = red] (5.00,6.25) arc (90:270:0.25) ; \draw (5.00,5.75)--(5.00,6.25) ; \draw (6.00,6.25)--(9.75,5.75) (9.75,6.25)--(13.50,5.75) (13.50,6.25)--(6.00,5.75) ; \draw (21.00,6.25)--(17.25,5.75) (24.75,6.25)--(21.00,5.75) (17.25,6.25)--(24.75,5.75) ;  \draw (0.00,5.75)--(0.00,5.50) (1.00,6.00)--(1.00,5.25) (2.00,6.00)--(2.00,5.25) (3.00,6.00)--(3.00,5.25) (4.00,5.75)--(4.00,5.50) (5.00,5.75)--(5.00,5.50) (6.00,5.75)--(6.00,5.25) (9.75,5.75)--(9.75,5.50) (13.50,5.75)--(13.50,5.50) (17.25,5.75)--(17.25,5.50) (21.00,5.75)--(21.00,5.50) (24.75,5.75)--(24.75,5.25) ; \draw (0.00,5.50)--(0.00,5.25) ; \draw (0.00,5.00)--(0.00,5.25) ; \draw [fill = red] (0.00,5.50) arc (90:270:0.25) ; \draw (0.00,5.00)--(0.00,5.50) ;  \draw (4.00,5.50)--(4.00,5.25) ; \draw (4.00,5.00)--(4.00,5.25) ; \draw [fill = blue] (4.00,5.00) arc (-90:90:0.25) ; \draw (4.00,5.00)--(4.00,5.50) ; \draw (5.00,5.50)--(5.00,5.25) ; \draw (5.00,5.00)--(5.00,5.25) ; \draw [fill = red] (5.00,5.50) arc (90:270:0.25) ; \draw (5.00,5.00)--(5.00,5.50) ;  \draw (9.75,5.50)--(13.50,5.00) (13.50,5.50)--(17.25,5.00) (17.25,5.50)--(9.75,5.00) ;  \draw (0.00,5.00)--(0.00,4.75) (1.00,5.25)--(1.00,4.50) (2.00,5.25)--(2.00,4.50) (3.00,5.25)--(3.00,4.50) (4.00,5.00)--(4.00,4.75) (5.00,5.00)--(5.00,4.75) (6.00,5.25)--(6.00,4.50) (9.75,5.00)--(9.75,4.50) (13.50,5.00)--(13.50,4.75) (17.25,5.00)--(17.25,4.75) (21.00,5.50)--(21.00,4.75) (24.75,5.25)--(24.75,4.50)  ; \draw (0.00,4.75)--(0.00,4.50) ; \draw (0.00,4.25)--(0.00,4.50) ; \draw [fill = red] (0.00,4.75) arc (90:270:0.25) ; \draw (0.00,4.25)--(0.00,4.75) ;  \draw (4.00,4.75)--(4.00,4.50) ; \draw (4.00,4.25)--(4.00,4.50) ; \draw [fill = blue] (4.00,4.25) arc (-90:90:0.25) ; \draw (4.00,4.25)--(4.00,4.75) ; \draw (5.00,4.75)--(5.00,4.50) ; \draw (5.00,4.25)--(5.00,4.50) ; \draw [fill = red] (5.00,4.75) arc (90:270:0.25) ; \draw (5.00,4.25)--(5.00,4.75) ;  \draw (13.50,4.75)--(17.25,4.25) (17.25,4.75)--(21.00,4.25) (21.00,4.75)--(13.50,4.25) ;  \draw (0.00,4.25)--(0.00,4.00) (1.00,4.50)--(1.00,3.75) (2.00,4.50)--(2.00,3.75) (3.00,4.50)--(3.00,3.75) (4.00,4.25)--(4.00,4.00) (5.00,4.25)--(5.00,4.00) (6.00,4.50)--(6.00,3.75) (9.75,4.50)--(9.75,3.75) (13.50,4.25)--(13.50,3.75) (17.25,4.25)--(17.25,3.75) (21.00,4.25)--(21.00,3.75) (24.75,4.50)--(24.75,3.75) ; \draw (0.00,4.00)--(0.00,3.75) ; \draw (0.00,3.50)--(0.00,3.75) ; \draw [fill = red] (0.00,4.00) arc (90:270:0.25) ; \draw (0.00,3.50)--(0.00,4.00) ;  \draw (4.00,4.00)--(4.00,3.75) ; \draw (4.00,3.50)--(4.00,3.75) ; \draw [fill = blue] (4.00,3.50) arc (-90:90:0.25) ; \draw (4.00,3.50)--(4.00,4.00) ; \draw (5.00,4.00)--(5.00,3.75) ; \draw (5.00,3.50)--(5.00,3.75) ; \draw [fill = red] (5.00,4.00) arc (90:270:0.25) ; \draw (5.00,3.50)--(5.00,4.00) ;  \draw [dash pattern = on 0.25pt off 2pt] (10.69,3.75)--(12.56,3.75) ; \draw [dash pattern = on 0.25pt off 2pt] (18.19,3.75)--(20.06,3.75)  ; \draw (0.00,3.50)--(0.00,3.00) (1.00,3.75)--(1.00,2.75) (2.00,3.75)--(2.00,2.75) (3.00,3.75)--(3.00,2.75) (4.00,3.50)--(4.00,3.00) (5.00,3.50)--(5.00,3.00) (6.00,3.75)--(6.00,3.25) (9.75,3.75)--(9.75,3.25) (13.50,3.75)--(13.50,3.25) (17.25,3.75)--(17.25,2.75) (21.00,3.75)--(21.00,2.75) (24.75,3.75)--(24.75,2.75) ; \draw (0.00,3.00)--(0.00,2.75) ; \draw (0.00,2.50)--(0.00,2.75) ; \draw [fill = red] (0.00,3.00) arc (90:270:0.25) ; \draw (0.00,2.50)--(0.00,3.00) ;  \draw (4.00,3.00)--(4.00,2.75) ; \draw (4.00,2.50)--(4.00,2.75) ; \draw [fill = blue] (4.00,2.50) arc (-90:90:0.25) ; \draw (4.00,2.50)--(4.00,3.00) ; \draw (5.00,3.00)--(5.00,2.75) ; \draw (5.00,2.50)--(5.00,2.75) ; \draw [fill = red] (5.00,3.00) arc (90:270:0.25) ; \draw (5.00,2.50)--(5.00,3.00) ; \draw [rounded corners = 1pt, fill = white] (5.75,2.25) rectangle (13.75,3.25) ; \node at (9.75,2.75) {$\scriptstyle N'$} ;  \draw (0.00,2.50)--(0.00,1.75) (1.00,2.75)--(1.00,1.50) (2.00,2.75)--(2.00,1.50) (3.00,2.75)--(3.00,2.00) (4.00,2.50)--(4.00,2.00) (5.00,2.50)--(5.00,2.00) (9.75,2.25)--(9.75,2.00) (17.25,2.75)--(17.25,1.50) (21.00,2.75)--(21.00,1.50) (24.75,2.75)--(24.75,1.50) ; \draw (0.00,1.75)--(0.00,1.50) ; \draw (0.00,1.25)--(0.00,1.50) ; \draw [fill = red] (0.00,1.75) arc (90:270:0.25) ; \draw (0.00,1.25)--(0.00,1.75) ;  \node at (3.00,1.50) {$\scriptstyle ~$} ; \draw [rounded corners = 1pt, fill = white] (2.75,1.00) rectangle (10.25,2.00) ; \node at (6.00,1.50) {$\scriptstyle \beta$} ; \node at (9.75,1.50) {$\scriptstyle ~$} ; \draw [dash pattern = on 0.25pt off 2pt] (18.19,1.50)--(20.06,1.50) ;   \draw (0.00,1.25)--(0.00,0.50) (1.00,1.50)--(1.00,0.75) (2.00,1.50)--(2.00,0.75) (4.50,1.00)--(4.50,0.75) (17.25,1.50)--(17.25,0.75) (21.00,1.50)--(21.00,0.75) (24.75,1.50)--(24.75,0.75) ; \draw (0.00,0.50)--(0.00,0.25) ; \draw (0.00,0.00)--(0.00,0.25) ; \draw [fill = red] (0.00,0.50) arc (90:270:0.25) ; \draw (0.00,0.00)--(0.00,0.50) ; \node at (1.50,0.25) {$\scriptstyle \Sigma$} ; \node at (4.50,0.25) {$\scriptstyle \beta(C,N')$} ; \node at (19.13,0.25) {$\scriptstyle \Delta$} ; \node at (24.00,0.25) {$\scriptstyle \alpha(B, N)$} ;  \end{scope} \end{tikzpicture}}}
$
}\end{center}
where $\alpha, \beta$ are \emph{splitting gates}, that are gates of type $\otimes$ or $Cut$. 
\end{definition}

In other words, we  have a crossing split in a proof diagram whenever the corresponding derivation exhibits two  binary inference rules $\alpha$ after $\beta$ such that the left (resp. right) active formula of $\alpha$ derives by the rightmost (resp. leftmost) sub-derivation branch of the left (resp. right) branch of  $\beta$. For example, consider the two following configurations with $A$ active formula of $\alpha$:

{
\begin{scprooftree}{0.6}
\AxiomC{$  \overset{\Gamma_1\quad  \Gamma_B\quad \Gamma_2\quad  \fbox{ $\Gamma_A$}}{\ddots \quad \vdots \quad \iddots}$}

\noLine
\UnaryInfC{$\vdash \Gamma, \fbox{$\Gamma'_A$}, B$}

\AxiomC{$\vdots$}
\noLine
\UnaryInfC{$\vdash C, \Delta$}

\RightLabel{$ \beta (B,C)$}
\BinaryInfC{$\vdash \Gamma,  \Delta , \fbox{$\Gamma'_A$}, \beta(B,C)$}
\doubleLine
\UnaryInfC{$\vdash \Gamma', \Delta', \fbox{$A$}, \beta(B,C)$}
\AxiomC{$ \vdots$}
\noLine
\UnaryInfC{$\vdash D, \s $}
\RightLabel{$ \alpha (A,D)$}

\BinaryInfC{$ \vdash \Gamma', \Delta',\beta (B,C) , \alpha(\fbox{$A$}, D), \s$}

\AxiomC{~}
\noLine
\UnaryInfC{{\LARGE or}}
\noLine
\UnaryInfC{$~$}
\noLine
\UnaryInfC{$~$}

\AxiomC{$ \vdots$}
\noLine
\UnaryInfC{$\vdash \s, D$}

\AxiomC{$ \vdots$}
\noLine
\UnaryInfC{$\vdash \Delta , C$}

\AxiomC{$\overset {\fbox{$\Gamma_A$} \quad \Gamma_1\quad  \Gamma_B\quad  \Gamma_2}{\ddots\quad \vdots\quad \iddots}$}
\noLine
\UnaryInfC{$\vdash \Gamma, \fbox{$\Gamma'_A$}, B$}
\RightLabel{$\beta(C,B)$}
\BinaryInfC{$\vdash  \Delta , \beta(B,C), \fbox{$\Gamma'_A$}, \Gamma$}
\doubleLine
\UnaryInfC{$\vdash  \beta(B,C), \fbox{$A$}, \Gamma',\Delta'$}

\RightLabel{$ \alpha(D,A)$}
\BinaryInfC{$ \vdash \s   , \beta(B,C), \alpha(D,\fbox{$A$}), \Gamma'\Delta'$}

\noLine
\TrinaryInfC{}
\end{scprooftree}
} 
\noindent where $\Gamma_A$ and $\Gamma'_A$ are sequents made of subformulas of $A$ only (similarly for the formula $B$ and $\Gamma_B$).

These configurations can be avoided in a proof diagram by giving a specific order to $Ax$ and $1$-gates, in the same way we permute  branches in derivation trees by $\sim$.

We call \emph{untangle procedure} the method of remove crossing split from a proof diagram. This requires to perform some rewritings which permute  proof diagram  branches.  For this purpose, we define some  gates type with the following shape:
$$
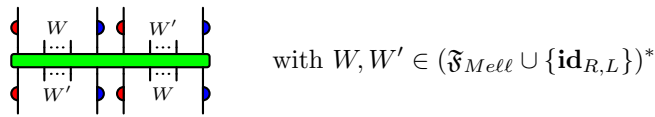
\twocell{(L *0 topW *0 R *0 L *0 topW1 *0 R) *1 (1 *0 d *0 2 *0 d *0 1) *1 bigt *1 (1 *0 d *0 2 *0 d *0 1) *1 (L *0 pitW1 *0 R *0 L *0 pitW *0 R)} \qquad \mbox{ with } W,W' \in (\fmell \cup\{ \id_{R,L}\})^*
$$
These gates can be seen as some ``big twisting operators'' able to cross a two sheafs of strings labeled by $L,W,R$ and $L,W',R$ where $W,W'$ are lists made not only by formulas but also by $ L$ and $R$. 

\begin{definition}[Polygraph of $\MLLc$]
The \emph{polygraph of multiplicative proof diagrams} is the polygraph $\Mllc$ obtained extended the polygraph $\Mlltc$ as follows:
\begin{itemize}
\begin{multicols}{2}
\item $\Mllc_0=\Mlltc_0$;
\item $\Mllc_1= \Mlltc_1$;
\end{multicols}
\item $\Mllc_2=\Mlltc_2 \cup \mathfrak Big$ where $$\mathfrak Big =\begin{Bmatrix}
B_{W, W}= \twocell{(L *0 topW *0 R *0 L *0 topW1 *0 R) *1 (1 *0 d *0 2 *0 d *0 1) *1 bigt *1 (1 *0 d *0 2 *0 d *0 1) *1 (L *0 pitW1 *0 R *0 L *0 pitW *0 R)}
\end{Bmatrix}_{W, W' \in (\fmllc\cup \{\id_{R,L}\})^*};$$

\item $\Mllc_3=\Mlltc_3 \cup \Mllc_{Big}$  where $\Mllc_{Big}$ is made of the following sets of $3$-cells: 

\begin{itemize} 
\item $\mathfrak B$-introduction: for any $\alpha, \beta \in \{Cut, \otimes\}$  and $\phi,\phi_1,\phi_2,\psi, \psi_1, \psi_2, N,N'$ irreducible in $\Mlltc$, with $N,N'\in \{\twocell{s}, \parr, \bot \}^*$,
we define:

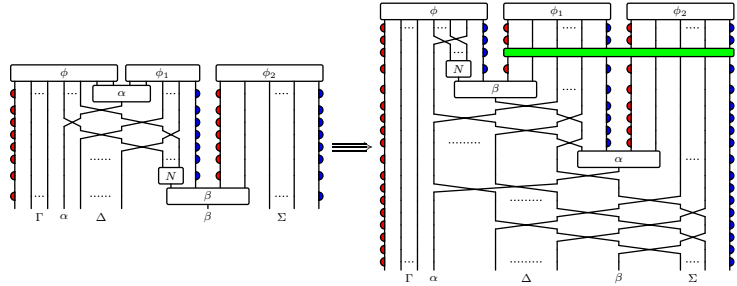
\begin{center}\resizebox{10cm}{!}{
$\xymatrix{ 
\twocell{
(phi07 *0 phi105 *0 phi205 ) *1
(L*0  d *0 d *0 split  *0 d *0  R *0 L *0 1 *0 d *0 R)*1
(L*0 3 *0 s *0 2 *0  R *0 L *0 1 *0 2 *0 R) *1
(L*0 2 *0 s *0 s *0 1 *0  R *0 L *0 1 *0 2 *0 R) *1
(L*0 3 *0 s *0 s *0  R *0 L *0 1 *0 2 *0 R) *1
(L*0 2 *0 2 *0 s *0 1 *0  R *0 L *0 1 *0 2 *0 R) *1
(L*0 3 *0 d *0 d *0  R *0 L *0 1 *0 2 *0 R) *1
(L*0 2 *0 2 *0 1 *0 N *0  R *0 L *0 1 *0 2 *0 R) *1
(L*0 d *0 1 *0 d *0 split2 *0 d *0 R) *1
(pitV*0 pitGam *0 pitAl *0 pitDel  *0 pitBe  *0 pitSig *0 pitV )
}
\ar@3[r] &
\twocell{
(phi07 *0 phi105 *0 phi205 ) *1
(L*0 d *0 d *0 1 *0 R *0 L *0 1  *0 d *0  R *0 L *0 1 *0 d *0 R)*1
(L*0 2 *0 rn *0 R *0 L *0 1  *0 2 *0  R *0 L *0 1 *0 2 *0 R)*1
(L*0 2 *0 1 *0 d *0 R *0 bigt10)*1
(L*0 3 *0 N *0 R *0 L *0 3 *0  R *0 L *0 1 *0 2 *0 R)*1
(L*0 3 *0 split2  *0 d *0  R *0 L *0 1 *0 2 *0 R)*1
(L*0 2 *0 1 *0 s *0 1 *0  R *0 L *0 1 *0 2 *0 R) *1
(L*0 2 *0 s *0 s *0  R *0 L *0 1 *0 2 *0 R) *1
(L*0 2 *0 1 *0 s *0 1 *0  R *0 L *0 1 *0 2 *0 R) *1
(L*0 2 *0 d *0 s  *0  R *0 L *0 1 *0 2 *0 R) *1
(L*0 2 *0 2 *0 1 *0 split *0 d *0 R) *1
(L*0 2 *0 2 *0 s *0 2 *0 R) *1
(L*0 2 *0 rn *0 1 *0 2 *0 R) *1
(L*0 2 *0 1 *0 d  *0 s *0 1 *0 R )*1
(L*0 2 *0 1 *0 1 *0 s *0 s *0 R )*1
(L*0 2 *0 1 *0 s  *0 s *0 1 *0 R )*1
(L*0 2 *0 1 *0 1 *0 s *0 s *0 R )*1
(L*0 2 *0 1 *0 2 *0 s *0 1 *0 R )*1
(L*0 d *0 1 *0 d  *0 1 *0 d *0 R )*1
(pitV*0 pitGam *0 pitAl *0 pitDel  *0 pitBe  *0 pitSig *0 pitV )
}
}$
}
\end{center}
and
\begin{center}\resizebox{10cm}{!}{
$
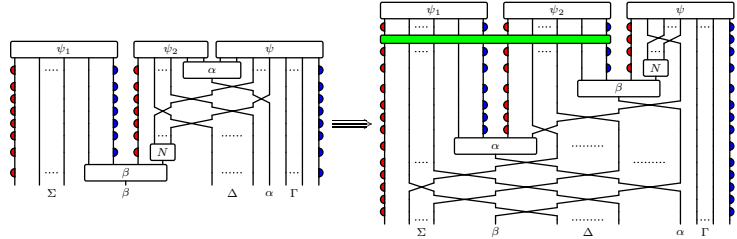
\xymatrix{ 
\twocell{
(psi105 *0 psi205 *0 psi07 ) *1
(L *0 d *0 1 *0 R *0 L *0 d *0  split *0 d *0 d *0 R) *1
(L *0 2 *0 1 *0 R *0 L *0 2 *0  s *0 1 *0 2 *0 R) *1
(L *0 2 *0 1 *0 R *0 L *0 1 *0  s *0 s *0 2 *0 R) *1
(L *0 2 *0 1 *0 R *0 L *0   s *0 s *0 3 *0 R) *1
(L *0 2 *0 1 *0 R *0 L *0  1 *0  s *0 4 *0 R) *1
(L *0 2 *0 1 *0 R *0 L *0  d *0 d *0 3 *0 R) *1
(L *0 2 *0 1 *0 R *0 L *0  N *0 2 *0 3 *0 R) *1
(L *0 d *0 split2 *0 d *0 1 *0 d *0 R) *1
(pitV *0 pitSig *0 pitBe *0 pitDel *0 pitAl *0 pitGam *0 pitV)
}
\ar@3[r] &
\twocell{
(psi105 *0 psi205 *0 psi07 ) *1
(L *0 d *0 1 *0 R *0 L *0 d *0  1 *0 R *0 L *0 1 *0 d *0 d *0 R) *1
(bigt10 *0 L *0 ln *0 2 *0 R) *1
(L *0 d *0 1 *0 R *0 L *0 d *0  1 *0 R *0 L *0 d *0 1 *0 2 *0 R) *1
(L *0 2 *0 1 *0 R *0 L *0 2 *0  1 *0 R *0 L *0 N *0 3 *0 R) *1
(L *0 2 *0 1 *0 R *0 L *0 2 *0  split2 *0 3 *0 R) *1
(L *0 2 *0 1 *0 R *0 L *0 2 *0  s *0 2 *0 R)*1
(L *0 2 *0 1 *0 R *0 L *0 1 *0  s *0 3 *0 R)*1
(L *0 2 *0 1 *0 R *0 L *0  s *0 4 *0 R)*1
(L *0 2 *0 split *0  d *0 3 *0 R)*1
(L *0 d *0 s *0  d *0 2 *0 R)*1
(L *0 1 *0 s  *0  s *0 3 *0 R)*1
(L *0 s *0 s *0  s *0 2 *0 R)*1
(L *0 1 *0 s *0  s *0 3 *0 R)*1
(L *0 2 *0 s *0 1 *0 3 *0 R)*1
(1 *0 d *0 1 *0 d *0 1 *0 d *0 1)*1
(pitV *0 pitSig *0 pitBe *0 pitDel *0 pitAl *0 pitGam *0 pitV)
}
}$
}
\end{center}

where $\phi$ and $\psi$ are respectively of the form:
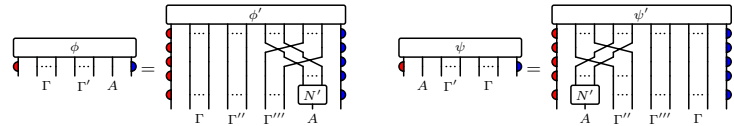
\begin{center}\resizebox{10cm}{!}{
$$
\twocell{phi07 *1 (L *0 d *0 d *0 1 *0 R)*1
 (pitV *0 pitGam *0 pitGam1 *0 pitA*0 pitV)
}= 
\twocell{
phi10 *1 
(L *0 d *0 d *0 d *0 d *0 R)*1
(L *0 2 *0 2 *0 rn *0 1 *0 R)*1
(L *0 2 *0 3 *0 rn *0 R)*1
(L *0 2 *0 2 *0 2 *0 d *0 R)*1
(L *0 d *0 d *0 d *0 N1 *0 R)*1
(pitV *0 pitGam *0 pitGam2 *0 pitGam3 *0 pitA *0 pitV)
} 
\qquad
\twocell{psi07 *1 (L *0 1 *0 d *0 d *0 R)*1
 (pitV *0 pitA *0 pitGam1 *0 pitGam *0 pitV)
}= 
\twocell{
psi10 *1 
(L *0 d *0 d *0 d *0 d *0 R)*1
(L *0 1 *0 ln *0 2 *0 2 *0 R)*1
(L *0 ln *0 3 *0 2 *0 R)*1
(L *0 d *0 d *0 2 *0 2 *0 R)*1
(L *0 N1 *0 d *0 d *0 d *0 R)*1
(pitV *0 pitA *0 pitGam2 *0 pitGam3 *0 pitGam *0 pitV)
}
$$
}\end{center}
with $\Gamma'=\Gamma'',\Gamma'''$;

\item The untangle relations: for any $\twocell{inx *1 d *1 gx *1 d *1 outx}\in \Mellt_2$, $B_{W,W'}\in \mathfrak Big$
\end{itemize}
\end{itemize}
\begin{center}\resizebox{12cm}{!}{
$\xymatrix{ 
\twocell{(1 *0  d  *0  2  *0 1) *1 (L*0 topWb1 *0 gx *0 topWb2 *0 R *0 L *0 topW1 *0 R) *1 (1 *0 d *0 d *0 d *0 2 *0 d *0 1 ) *1 bigt12 *1 (1 *0 d *0 2 *0 d *0 d *0 d *0 1) *1(L *0 pitW1 *0 R *0 L *0 pitWb1 *0 outx *0 pitWb2 *0 R)} 
\ar@3[r] & 
\twocell{ (L*0 topWb1 *0 inx *0 topWb2 *0 R *0 L *0 topW1 *0 R) *1 (1 *0 d *0 d *0 d *0 2 *0 d *0 1 ) *1 bigt12 *1 (1 *0 d *0 2 *0 d *0 d *0 d *0 1) *1(L *0 pitW1 *0 R *0 L *0 pitWb1 *0 gx *0 pitWb2 *0 R)*1 (1 *0 2 *0 d *0 1) }
},
\qquad
\xymatrix{ 
\twocell{(1  *0 2  *0  d *0 1) *1 (L*0 topW *0 R *0 L *0 topW1b1 *0 gx *0 topW1b2  *0 R) *1 (1 *0 d *0 2 *0 d *0 d *0 d *0 1 ) *1 bigt12 *1 (1 *0 d *0 d *0 d *0 2 *0 d *0 1) *1 (L *0 pitW1b1 *0 outx *0 pitW1b2 *0 R *0 L *0 pitW1  *0 R)} 
\ar@3[r] & 
\twocell{ (L*0 topW *0 R *0 L *0 topW1b1 *0 inx *0 topW1b2  *0 R) *1 (1 *0 d *0 2 *0 d *0 d *0 d *0 1 ) *1 bigt12 *1 (1 *0 d *0 d *0 d *0 2 *0 d *0 1) *1 (L *0 pitW1b1 *0 gx *0 pitW1b2 *0 R *0 L *0 pitW *0 R)*1 (1 *0 d *0 2 *0  1) } 
} \; .$
}
\end{center}
\end{definition}

To have an intuition, a $B$-gate can be visualized as follows (but we remind the reader that  such diagrams can not be defined in our syntax since twisting operators are not defined for control strings):
$$
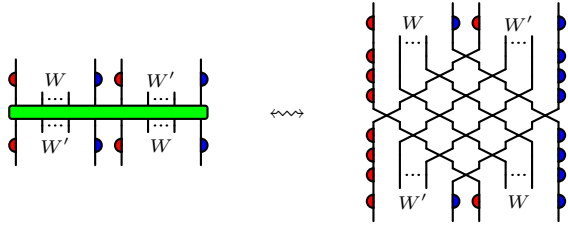
\twocell{(L *0 topW *0 R *0 L *0 topW1 *0 R) *1 (1 *0 d *0 2 *0 d *0 1) *1 bigt *1 (1 *0 d *0 2 *0 d *0 1) *1 (L *0 pitW1 *0 R *0 L *0 pitW *0 R)}
\qquad
\leftrightsquigarrow
\qquad
\twocell{(L *0 topW *0 R *0 L *0 topW1 *0 R) *1 (1 *0 d *0 2 *0 d *0 1) *1 
(L *0 2 *0 s *0 2 *0 R) *1
(L *0 1 *0 s *0 s *0 1 *0 R) *1
(L *0 s *0 s *0 s *0 R) *1
(s *0 s *0 s *0 s) *1
(L *0 s *0 s *0 s *0 R) *1
(L *0 1 *0 s *0 s *0 1 *0 R) *1
(L *0 d *0 s *0 d *0 R) *1
(L *0 pitW1 *0 R *0 L *0 pitW *0 R)}$$

A $\mathfrak B$-introduction rule eliminates from a $\Mlltc_3$-irreducible proof diagram a crossing split: it exchanges the order of splitting gate, it  modifies some twisting operators and it triggers the crossing of two proof diagram branches by the introduction of a $B$-gate. 

At the same time, the untangle relations move gates from the top to the bottom of a $B$-gate according with our intuition:  when a gate ``crosses'' a $B$-gate, it slides on the sheaf of strings passing from the  left to the right and vice versa. These rules untangle, step-by-step, two crossed branches of a diagram:

\begin{proposition}\label{Bigera}[$B$-gate elimination]
If $\phi,\psi \in \Mlltc$ are proof diagrams of type $\phi: \pro \frr L, W_1, R$ and $\psi: \pro \frr L, W_2, R$ with $W_1,W_2' \in (\fmllc \cup \{\id_{R,L}\})^*$, then there are rewritings path made only of untangle relations of the following forms for gates of type $B_{W_1,W_2}$:
$$
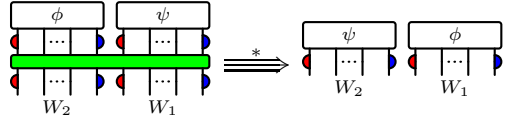
\xymatrix{ 
\twocell{
(diaphi04 *0 diapsi04) *1
(L *0 d *0 R *0 L *0 d *0 R)*1
(bigt)*1
(L *0 d *0 R *0 L *0 d *0 R)*1
(pitV *0 pitWb2 *0 pitV *0 pitV *0 pitWb1 *0 pitV)
}
\ar@3^{*}[r] & 
\twocell{
(diapsi04*0 diaphi04)*1
(L *0 d *0 R *0 L *0 d *0 R)*1
(pitV *0 pitWb2 *0 pitV *0 pitV *0 pitWb1 *0 pitV)
}
}
$$
We call this rewriting path a \emph{$B$-gate elimination}. Moreover, if $\phi',\psi'\in \Mlltc$ are proof diagrams of type $\phi': W_1 \frr L, W'_1, R$ and $\psi': W_2 \frr L, W_2', R$ with $W_1, W_1',W_2, W_2' \in (\fmllc \cup \{\id_{R,L}\})^*$, then  there are rewritings path made only of untangle relations of the following forms:
$$
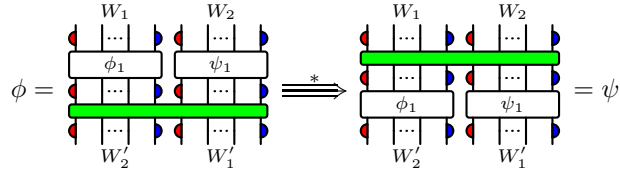
\xymatrix{ 
\phi=
\twocell{
(topV *0 topWb1 *0 topV *0 topV *0 topWb2 *0 topV) *1
(L *0 d *0 R *0 L *0 d *0 R)*1
(phi4b1 *0 psi4b1) *1
(L *0 d *0 R *0 L *0 d *0 R)*1
(bigt)*1
(L *0 d *0 R *0 L *0 d *0 R)*1
(pitV *0 pitW1b2 *0 pitV *0 pitV *0 pitW1b1 *0 pitV)
}
\ar@3^{*}[r] & 
\twocell{
(topV *0 topWb1 *0 topV *0 topV *0 topWb2 *0 topV) *1
(L *0 d *0 R *0 L *0 d *0 R)*1
(bigt)*1
(L *0 d *0 R *0 L *0 d *0 R)*1
(phi4b1 *0 psi4b1) *1
(L *0 d *0 R *0 L *0 d *0 R)*1
(pitV *0 pitW1b2 *0 pitV *0 pitV *0 pitW1b1 *0 pitV)
}
=\psi
}
$$
We call this rewriting path a \emph{$B$-gate reduction}.
\begin{proof}
By induction over the number of gates in the diagram $\phi,\psi$: each untangle relation decrease it.
\end{proof}
\end{proposition}

We call \emph{untangle sequence} a rewriting path made of one $\mathfrak B$-introduction rule followed by its relative $B$-gate elimination rewriting path.
Each untangle sequence corresponds to the elimination of a crossing split and it terminates after a finite number of steps depending on the number of gates in the diagram. 

We have a maximal $B$-gate reduction $\xymatrix{\phi \ar@3^{*}[r] &  \phi'}$ when $\phi'$ is of the form:
$$\phi'=\chi_d \circ ( \id_W , (\psi' \circ B) , \id_{W'}) =
\twocell{(topW *0 d *0 topW1) *1 (d *0 psi1 *0 d)*1 ( 2 *0 d *0 2) *1  (2 *0 bigt2 *0 2 ) *1 ( 2 *0 d *0 2) *1 do *1 d}
\mbox{ with }\psi'\in \mathfrak Big ^*$$
We call any such rewriting path a \emph{$B$-deactivation}.

We assume that any of this sequence generate no new crossing split. In fact, the elimination of a crossing split generate a new one if and only if there is a gate corresponding to a binary inference rule in parallel with respect of the lower splitting gate, for example:
$$\twocell{
(phi04 *0 phi104 *0 phi204  *0 phi303) *1
(L *0 1 *0 tenc  *0 1 *0  R *0 L *0 2 *0 R*0 L *0 1*0 R)*1
(L*0 s *0 1 *0  R *0 L *0 2 *0 R*0 L *0 1 *0 R) *1
(L*0 1 *0 s *0  R *0 L *0 2 *0 R *0 L *0 1 *0 R) *1
(L*0 2 *0 tenc *0 tenc *0 R) 
}$$

In these cases it is possible to verify that either we apply the $\mathfrak B$-introduction rule in such a way as to maintain these two gates in the same branching of the diagram, or we perform a second untangle sequence we are able to recover a configuration where they are in parallel again. In the previous example we have:

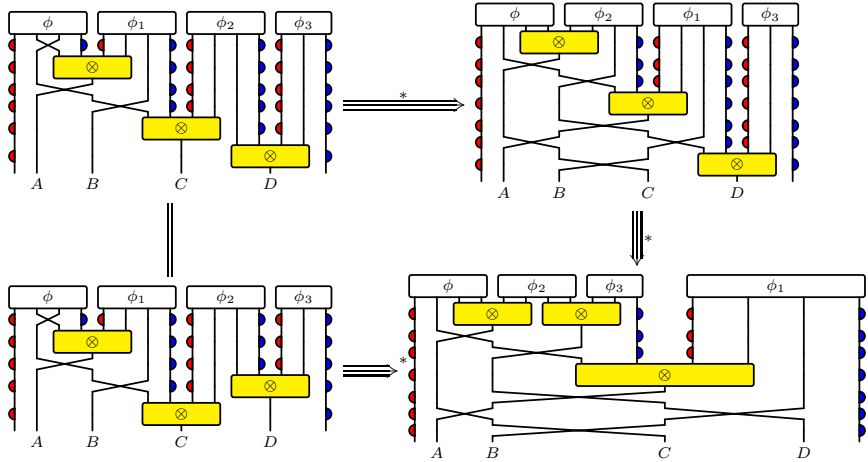
\begin{center}\resizebox{12cm}{!}{
$$
\xymatrix{
\twocell{
(phi04 *0 phi104 *0 phi204  *0 phi303) *1
(L *0 s *0 R *0 L *0 1  *0 1 *0  R *0 L *0 2 *0 R*0 L *0 1*0 R)*1
(L *0 1 *0 tenc  *0 1 *0  R *0 L *0 2 *0 R*0 L *0 1*0 R)*1
(L*0 s *0 1 *0  R *0 L *0 2 *0 R*0 L *0 1 *0 R) *1
(L*0 1 *0 s *0  R *0 L *0 2 *0 R *0 L *0 1 *0 R) *1
(L*0 2 *0 tenc *0 1 *0 R *0 L *0 1 *0 R) *1
(L*0 2 *0 1 *0 tenc *0 R) *1
(pitV *0 pitA *0 pitB *0 pitC *0 pitD *0 pitV)
}
\ar@3[r]^{*} \ar@{=}[d]&
\twocell{
(phi04 *0 phi204 *0 phi104  *0 phi303) *1
(L *0 1 *0 tenc  *0 1 *0  R *0 L *0 2 *0 R*0 L *0 1*0 R)*1
(L*0 s *0 1 *0  R *0 L *0 2 *0 R*0 L *0 1 *0 R) *1
(L*0 1 *0 s *0  R *0 L *0 2 *0 R *0 L *0 1 *0 R) *1
(L*0 2 *0 tenc *0 1 *0 R *0 L *0 1 *0 R) *1
(L*0 1 *0 s *0 1 *0 R *0 L *0 1 *0 R) *1
(L*0 s *0 s *0  R *0 L *0 1 *0 R) *1
(L*0 1 *0 s *0 tenc *0 R)*1
(pitV *0 pitA *0 pitB *0 pitC *0 pitD *0 pitV)
}
\ar@3[d]^{*}
\\
\twocell{
(phi04 *0 phi104 *0 phi204  *0 phi303) *1
(L *0 s *0 R *0 L *0 1  *0 1 *0  R *0 L *0 2 *0 R*0 L *0 1*0 R)*1
(L *0 1 *0 tenc  *0 1 *0  R *0 L *0 2 *0 R*0 L *0 1*0 R)*1
(L*0 s *0 1 *0  R *0 L *0 2 *0 R*0 L *0 1 *0 R) *1
(L*0 1 *0 s *0  R *0 L *0 1 *0 tenc *0 R) *1
(L*0 2 *0 tenc *0 1 *0 R) *1
(pitV *0 pitA *0 pitB *0 pitC *0 pitD *0 pitV)
}
\ar@3[r]^{*} &
\twocell{
(phi04 *0  phi204  *0 phi303 *0 phi104 ) *1
(L *0 1 *0 tenc  *0 tenc *0 R*0 L *0 2*0 R)*1
(L*0 s *0 1 *0 R*0 L *0 2 *0 R) *1
(L*0 1 *0 s *0  R *0 L *0 2 *0 R) *1
(L*0 2 *0 tenc *0 1 *0 R) *1
(L*0 1 *0 s *0 1 *0 R) *1
(L*0 s *0 s  *0 R) *1
(L*0 1 *0 s *0 1 *0 R) *1
(pitV *0 pitA *0 pitB *0 pitC *0 pitD *0 pitV)
}
}
$$
}\end{center}

The choice of define $\mathfrak B$-introduction rules with premises $\Mlltc_3$-irreducible diagrams with no $B$-gates leads the following result:

\begin{corollary}\label{Bigcom}
Conflicts between a  $\mathfrak B$-introduction rule and an untangle relation and conflicts between a rule in $\Mllc_{Big}$ and a rule in  $\Mlltc_3$ are trivially solvable. Then we can assume the corresponding rewritings paths commute.
\begin{proof}
The subdiagram rewritten by a $\mathfrak B$-introduction rules is $\Mlltc_3$-irreducible and contains no $B$-gates then all possible non-trivial conflicts are the ones between two  $\mathfrak B$-introduction rules  discussed above.
The confluence of non-trivial critical pairs between untangle relations and  rules in  $\Mlltc_3$ follows by argumentations similar to the ones given in the Proposition \ref{Bigera}.
\end{proof}
\end{corollary}

This lead the following theorem about the termination of rewriting in $\Mllc$.

\begin{theorem}[Termination in $\Mllc$]\label{TerMllu}
The polygraph $\Mllc$ is terminating.
\begin{proof}
Corollary \ref{Bigcom} implies that a rewriting path in $\Mllc$ can be written as an alternate sequence of rewriting paths in $\Mlltc_3$, untangle sequences and $B$-deactivations. We know that the length of untangle sequences and $B$-deactivations are finite and linearly depend on the number of gates in a diagram. Moreover, Theorem \ref{termlltc} proves that there are not infinite rewriting paths composed of rules in $\Mlltc_3$. Then, to prove termination it suffices to prove that the number $n$ of alternations is finite.

For any $\phi\in \Mllc$, if $\phi_{Cross}$ is the number of crossing splits in $\phi$, any alternate rewriting path starting from $\phi$ counts at most $|\phi|_{\{\mathfrak Big\}}$ $B$-deactivations and $\phi_{Cross}$ untangle sequences. In fact, no rule in $\Mlltc_3$ generates new $B$-gates either crossing splits. This is underlined by the correspondence between equivalence relation over derivations $\simeq_D=\sim_{\tilde D}$  and rules permutations over derivations which do not change the structure of tree branching.
\end{proof}
\end{theorem}


We extend the Theorem \ref{corrMLLc} to proof diagrams in $\Mllc$. This leads the linear complexity of the test of sequentializability for proof diagrams in $\Mllc$.

\begin{theorem}[Multiplicative proof diagram correspondence]\label{corrMllc}
$$\vdash_{\MLLc} \Gamma \Leftrightarrow \exists \phi \in \Mllc \mbox{ such that } \phi: \pro \frr L,\Gamma, R.$$
\begin{proof}
The left-to-right implication immediate follows by Theorem \ref{corrMLLc}. For the proof of  right-to-left implicationwe have to also consider the cases when it occurs a $2$-cells in $\mathfrak Big$.
We observe that a proof diagram $\phi: \pro \frr L,\Gamma, R$ contains a gate of type $B \in \mathfrak Big$ iff  there is a subdiagram $\phi'\subseteq \phi$ of the form
$$\phi'=(\id_{L, \Gamma'}, g_\alpha, \id_{\Gamma'', R})\circ (\phi_2' ,\phi_1') \circ B \circ (\phi_1 ,\phi_2)$$
with $g_\alpha$ gate of type $\alpha\in \{Cut, \otimes\}$.
Then, during the sequentialization procedure, whenever a gate of type $\otimes$ or $Cut$ occurs, we consider the following cases:
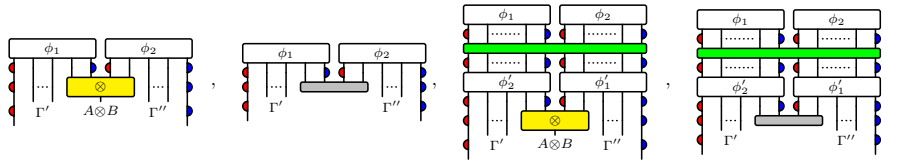
\begin{center}\resizebox{12cm}{!}{
$$\twocell{(prem1 *0 prem2)*1 (L *0 2 *0 1 *0 R *0 L *0 1 *0 2 *0 R)*1 (L *0 d *0 tenc *0 d *0 R) *1 (L *0 pitGam1 *0 pitAtenB *0 pitGam2 *0 R)} \; , \quad
\twocell{(prem1 *0 prem2) *1 (L *0 2 *0 1 *0 R *0 L *0 1 *0 2 *0 R)*1 (L *0 d *0 cutc *0 d *0 R) *1 (L *0 pitGam1  *0 pitGam2 *0 R)}\;, 
\quad
{\raisebox{-38.75pt}{\begin{tikzpicture} \begin{scope} [ x = 10pt, y = 10pt, join = round, cap = round, thick, black, solid, -] \draw  ; \draw (0.00,0.00)--(0.00,-0.50) (9.00,0.00)--(9.00,-0.50) ; \draw [rounded corners = 1pt, fill = white] (-0.25,6.50) rectangle (4.25,7.50) ; \node at (2.00,7.00) {$\scriptstyle \phi_1$} ; \draw [rounded corners = 1pt, fill = white] (4.75,6.50) rectangle (9.25,7.50) ; \node at (7.00,7.00) {$\scriptstyle \phi_2$} ; \draw (0.00,6.50)--(0.00,6.25) (1.00,6.50)--(1.00,6.00) (3.00,6.50)--(3.00,6.00) (4.00,6.50)--(4.00,6.25) (5.00,6.50)--(5.00,6.25) (6.00,6.50)--(6.00,6.00)  (8.00,6.50)--(8.00,6.00) (9.00,6.50)--(9.00,6.25) ; \draw (0.00,6.25)--(0.00,6.00) ; \draw (0.00,5.75)--(0.00,6.00) ; \draw [fill = red] (0.00,6.25) arc (90:270:0.25) ; \draw (0.00,5.75)--(0.00,6.25) ; \draw [dash pattern = on 0.25pt off 2pt] (1.25,6.00)--(2.75,6.00) ;  \draw (4.00,6.25)--(4.00,6.00) ; \draw (4.00,5.75)--(4.00,6.00) ; \draw [fill = blue] (4.00,5.75) arc (-90:90:0.25) ; \draw (4.00,5.75)--(4.00,6.25) ; \draw (5.00,6.25)--(5.00,6.00) ; \draw (5.00,5.75)--(5.00,6.00) ; \draw [fill = red] (5.00,6.25) arc (90:270:0.25) ; \draw (5.00,5.75)--(5.00,6.25) ;  \draw [dash pattern = on 0.25pt off 2pt] (6.25,6.00)--(7.75,6.00) ; \draw (9.00,6.25)--(9.00,6.00) ; \draw (9.00,5.75)--(9.00,6.00) ; \draw [fill = blue] (9.00,5.75) arc (-90:90:0.25) ; \draw (9.00,5.75)--(9.00,6.25) ; \draw (0.00,5.75)--(0.00,5.50) (1.00,6.00)--(1.00,5.50) (3.00,6.00)--(3.00,5.50) (4.00,5.75)--(4.00,5.50) (5.00,5.75)--(5.00,5.50) (6.00,6.00)--(6.00,5.50) (8.00,6.00)--(8.00,5.50) (9.00,5.75)--(9.00,5.50) ; \draw [rounded corners = 1pt, fill = green] (-0.25,5.00) rectangle (9.25,5.50) ; \draw (0.00,5.00)--(0.00,4.75) (1.00,5.00)--(1.00,4.50) (3.00,5.00)--(3.00,4.50) (4.00,5.00)--(4.00,4.75) (5.00,5.00)--(5.00,4.75) (6.00,5.00)--(6.00,4.50) (8.00,5.00)--(8.00,4.50) (9.00,5.00)--(9.00,4.75) ; \draw (0.00,4.75)--(0.00,4.50) ; \draw (0.00,4.25)--(0.00,4.50) ; \draw [fill = red] (0.00,4.75) arc (90:270:0.25) ; \draw (0.00,4.25)--(0.00,4.75) ; \draw [dash pattern = on 0.25pt off 2pt] (1.25,4.50)--(2.75,4.50) ;  \draw (4.00,4.75)--(4.00,4.50) ; \draw (4.00,4.25)--(4.00,4.50) ; \draw [fill = blue] (4.00,4.25) arc (-90:90:0.25) ; \draw (4.00,4.25)--(4.00,4.75) ; \draw (5.00,4.75)--(5.00,4.50) ; \draw (5.00,4.25)--(5.00,4.50) ; \draw [fill = red] (5.00,4.75) arc (90:270:0.25) ; \draw (5.00,4.25)--(5.00,4.75) ;  \draw [dash pattern = on 0.25pt off 2pt] (6.25,4.50)--(7.75,4.50) ; \draw (9.00,4.75)--(9.00,4.50) ; \draw (9.00,4.25)--(9.00,4.50) ; \draw [fill = blue] (9.00,4.25) arc (-90:90:0.25) ; \draw (9.00,4.25)--(9.00,4.75) ; \draw (0.00,4.25)--(0.00,4.00) (1.00,4.50)--(1.00,4.00)  (3.00,4.50)--(3.00,4.00) (4.00,4.25)--(4.00,4.00) (5.00,4.25)--(5.00,4.00) (6.00,4.50)--(6.00,4.00)  (8.00,4.50)--(8.00,4.00) (9.00,4.25)--(9.00,4.00) ; \draw [rounded corners = 1pt, fill = white] (-0.25,3.00) rectangle (4.25,4.00) ; \node at (2.00,3.50) {$\scriptstyle \phi_2'$} ; \draw [rounded corners = 1pt, fill = white] (4.75,3.00) rectangle (9.25,4.00) ; \node at (7.00,3.50) {$\scriptstyle \phi_1'$} ; \draw (0.00,3.00)--(0.00,2.75) (1.00,3.00)--(1.00,2.50) (2.00,3.00)--(2.00,2.50) (3.00,3.00)--(3.00,2.50) (4.00,3.00)--(4.00,2.75) (5.00,3.00)--(5.00,2.75) (6.00,3.00)--(6.00,2.50) (7.00,3.00)--(7.00,2.50) (8.00,3.00)--(8.00,2.50) (9.00,3.00)--(9.00,2.75) ; \draw (0.00,2.75)--(0.00,2.50) ; \draw (0.00,2.25)--(0.00,2.50) ; \draw [fill = red] (0.00,2.75) arc (90:270:0.25) ; \draw (0.00,2.25)--(0.00,2.75) ;   \draw (4.00,2.75)--(4.00,2.50) ; \draw (4.00,2.25)--(4.00,2.50) ; \draw [fill = blue] (4.00,2.25) arc (-90:90:0.25) ; \draw (4.00,2.25)--(4.00,2.75) ; \draw (5.00,2.75)--(5.00,2.50) ; \draw (5.00,2.25)--(5.00,2.50) ; \draw [fill = red] (5.00,2.75) arc (90:270:0.25) ; \draw (5.00,2.25)--(5.00,2.75) ;   \draw (9.00,2.75)--(9.00,2.50) ; \draw (9.00,2.25)--(9.00,2.50) ; \draw [fill = blue] (9.00,2.25) arc (-90:90:0.25) ; \draw (9.00,2.25)--(9.00,2.75) ; \draw (0.00,2.25)--(0.00,1.75) (1.00,2.50)--(1.00,1.50) (2.00,2.50)--(2.00,1.50) (3.00,2.50)--(3.00,2.00) (4.00,2.25)--(4.00,2.00) (5.00,2.25)--(5.00,2.00) (6.00,2.50)--(6.00,2.00) (7.00,2.50)--(7.00,1.50) (8.00,2.50)--(8.00,1.50) (9.00,2.25)--(9.00,1.75) ; \draw (0.00,1.75)--(0.00,1.50) ; \draw (0.00,1.25)--(0.00,1.50) ; \draw [fill = red] (0.00,1.75) arc (90:270:0.25) ; \draw (0.00,1.25)--(0.00,1.75) ; \draw [dash pattern = on 0.25pt off 2pt] (1.25,1.50)--(1.75,1.50) ; \draw [rounded corners = 1pt, fill = yellow] (2.75,1.00) rectangle (6.25,2.00) ; \node at (4.50,1.50) {$\scriptstyle \otimes$} ; \draw [dash pattern = on 0.25pt off 2pt] (7.25,1.50)--(7.75,1.50) ; \draw (9.00,1.75)--(9.00,1.50) ; \draw (9.00,1.25)--(9.00,1.50) ; \draw [fill = blue] (9.00,1.25) arc (-90:90:0.25) ; \draw (9.00,1.25)--(9.00,1.75) ; \draw (0.00,1.25)--(0.00,0.50) (1.00,1.50)--(1.00,0.75) (2.00,1.50)--(2.00,0.75) (4.50,1.00)--(4.50,0.75) (7.00,1.50)--(7.00,0.75) (8.00,1.50)--(8.00,0.75) (9.00,1.25)--(9.00,0.50) ; \draw (0.00,0.50)--(0.00,0.25) ; \draw (0.00,0.00)--(0.00,0.25) ; \draw [fill = red] (0.00,0.50) arc (90:270:0.25) ; \draw (0.00,0.00)--(0.00,0.50) ; \node at (1.50,0.25) {$\scriptstyle \Gamma'$} ; \node at (4.50,0.25) {$\scriptstyle A\otimes{B}$} ; \node at (7.50,0.25) {$\scriptstyle \Gamma''$} ; \draw (9.00,0.50)--(9.00,0.25) ; \draw (9.00,0.00)--(9.00,0.25) ; \draw [fill = blue] (9.00,0.00) arc (-90:90:0.25) ; \draw (9.00,0.00)--(9.00,0.50) ; \end{scope} \end{tikzpicture}}}
 \; , 
\quad
{\raisebox{-36.25pt}{\begin{tikzpicture} \begin{scope} [ x = 10pt, y = 10pt, join = round, cap = round, thick, black, solid, -] \draw  ; \draw (0.00,0.00)--(0.00,-0.50) (9.00,0.00)--(9.00,-0.50) ; \draw [rounded corners = 1pt, fill = white] (-0.25,6.00) rectangle (4.25,7.00) ; \node at (2.00,6.50) {$\scriptstyle \phi_1$} ; \draw [rounded corners = 1pt, fill = white] (4.75,6.00) rectangle (9.25,7.00) ; \node at (7.00,6.50) {$\scriptstyle \phi_2$} ; \draw (0.00,6.00)--(0.00,5.75) (1.00,6.00)--(1.00,5.50)  (3.00,6.00)--(3.00,5.50) (4.00,6.00)--(4.00,5.75) (5.00,6.00)--(5.00,5.75) (6.00,6.00)--(6.00,5.50)  (8.00,6.00)--(8.00,5.50) (9.00,6.00)--(9.00,5.75) ; \draw (0.00,5.75)--(0.00,5.50) ; \draw (0.00,5.25)--(0.00,5.50) ; \draw [fill = red] (0.00,5.75) arc (90:270:0.25) ; \draw (0.00,5.25)--(0.00,5.75) ; \draw [dash pattern = on 0.25pt off 2pt] (1.25,5.50)--(2.75,5.50) ;  \draw (4.00,5.75)--(4.00,5.50) ; \draw (4.00,5.25)--(4.00,5.50) ; \draw [fill = blue] (4.00,5.25) arc (-90:90:0.25) ; \draw (4.00,5.25)--(4.00,5.75) ; \draw (5.00,5.75)--(5.00,5.50) ; \draw (5.00,5.25)--(5.00,5.50) ; \draw [fill = red] (5.00,5.75) arc (90:270:0.25) ; \draw (5.00,5.25)--(5.00,5.75) ;  \draw [dash pattern = on 0.25pt off 2pt] (6.25,5.50)--(7.75,5.50) ; \draw (9.00,5.75)--(9.00,5.50) ; \draw (9.00,5.25)--(9.00,5.50) ; \draw [fill = blue] (9.00,5.25) arc (-90:90:0.25) ; \draw (9.00,5.25)--(9.00,5.75) ; \draw (0.00,5.25)--(0.00,5.00) (1.00,5.50)--(1.00,5.00) (3.00,5.50)--(3.00,5.00) (4.00,5.25)--(4.00,5.00) (5.00,5.25)--(5.00,5.00) (6.00,5.50)--(6.00,5.00)  (8.00,5.50)--(8.00,5.00) (9.00,5.25)--(9.00,5.00) ; \draw [rounded corners = 1pt, fill = green] (-0.25,4.50) rectangle (9.25,5.00) ; \draw (0.00,4.50)--(0.00,4.25) (1.00,4.50)--(1.00,4.00)  (3.00,4.50)--(3.00,4.00) (4.00,4.50)--(4.00,4.25) (5.00,4.50)--(5.00,4.25) (6.00,4.50)--(6.00,4.00)  (8.00,4.50)--(8.00,4.00) (9.00,4.50)--(9.00,4.25) ; \draw (0.00,4.25)--(0.00,4.00) ; \draw (0.00,3.75)--(0.00,4.00) ; \draw [fill = red] (0.00,4.25) arc (90:270:0.25) ; \draw (0.00,3.75)--(0.00,4.25) ; \draw [dash pattern = on 0.25pt off 2pt] (1.25,4.00)--(2.75,4.00) ;  \draw (4.00,4.25)--(4.00,4.00) ; \draw (4.00,3.75)--(4.00,4.00) ; \draw [fill = blue] (4.00,3.75) arc (-90:90:0.25) ; \draw (4.00,3.75)--(4.00,4.25) ; \draw (5.00,4.25)--(5.00,4.00) ; \draw (5.00,3.75)--(5.00,4.00) ; \draw [fill = red] (5.00,4.25) arc (90:270:0.25) ; \draw (5.00,3.75)--(5.00,4.25) ;  \draw [dash pattern = on 0.25pt off 2pt] (6.25,4.00)--(7.75,4.00) ; \draw (9.00,4.25)--(9.00,4.00) ; \draw (9.00,3.75)--(9.00,4.00) ; \draw [fill = blue] (9.00,3.75) arc (-90:90:0.25) ; \draw (9.00,3.75)--(9.00,4.25) ; \draw (0.00,3.75)--(0.00,3.50) (1.00,4.00)--(1.00,3.50) (3.00,4.00)--(3.00,3.50) (4.00,3.75)--(4.00,3.50) (5.00,3.75)--(5.00,3.50) (6.00,4.00)--(6.00,3.50) (8.00,4.00)--(8.00,3.50) (9.00,3.75)--(9.00,3.50) ; \draw [rounded corners = 1pt, fill = white] (-0.25,2.50) rectangle (4.25,3.50) ; \node at (2.00,3.00) {$\scriptstyle \phi_2'$} ; \draw [rounded corners = 1pt, fill = white] (4.75,2.50) rectangle (9.25,3.50) ; \node at (7.00,3.00) {$\scriptstyle \phi_1'$} ; \draw (0.00,2.50)--(0.00,2.25) (1.00,2.50)--(1.00,2.00) (2.00,2.50)--(2.00,2.00) (3.00,2.50)--(3.00,2.00) (4.00,2.50)--(4.00,2.25) (5.00,2.50)--(5.00,2.25) (6.00,2.50)--(6.00,2.00) (7.00,2.50)--(7.00,2.00) (8.00,2.50)--(8.00,2.00) (9.00,2.50)--(9.00,2.25) ; \draw (0.00,2.25)--(0.00,2.00) ; \draw (0.00,1.75)--(0.00,2.00) ; \draw [fill = red] (0.00,2.25) arc (90:270:0.25) ; \draw (0.00,1.75)--(0.00,2.25) ;  \draw (4.00,2.25)--(4.00,2.00) ; \draw (4.00,1.75)--(4.00,2.00) ; \draw [fill = blue] (4.00,1.75) arc (-90:90:0.25) ; \draw (4.00,1.75)--(4.00,2.25) ; \draw (5.00,2.25)--(5.00,2.00) ; \draw (5.00,1.75)--(5.00,2.00) ; \draw [fill = red] (5.00,2.25) arc (90:270:0.25) ; \draw (5.00,1.75)--(5.00,2.25) ;  \draw (9.00,2.25)--(9.00,2.00) ; \draw (9.00,1.75)--(9.00,2.00) ; \draw [fill = blue] (9.00,1.75) arc (-90:90:0.25) ; \draw (9.00,1.75)--(9.00,2.25) ; \draw (0.00,1.75)--(0.00,1.50) (1.00,2.00)--(1.00,1.25) (2.00,2.00)--(2.00,1.25) (3.00,2.00)--(3.00,1.50) (4.00,1.75)--(4.00,1.50) (5.00,1.75)--(5.00,1.50) (6.00,2.00)--(6.00,1.50) (7.00,2.00)--(7.00,1.25) (8.00,2.00)--(8.00,1.25) (9.00,1.75)--(9.00,1.50) ; \draw (0.00,1.50)--(0.00,1.25) ; \draw (0.00,1.00)--(0.00,1.25) ; \draw [fill = red] (0.00,1.50) arc (90:270:0.25) ; \draw (0.00,1.00)--(0.00,1.50) ; \draw [dash pattern = on 0.25pt off 2pt] (1.25,1.25)--(1.75,1.25) ; \draw [rounded corners = 1pt, fill = lightgray] (2.75,1.00) rectangle (6.25,1.50) ; \draw [dash pattern = on 0.25pt off 2pt] (7.25,1.25)--(7.75,1.25) ; \draw (9.00,1.50)--(9.00,1.25) ; \draw (9.00,1.00)--(9.00,1.25) ; \draw [fill = blue] (9.00,1.00) arc (-90:90:0.25) ; \draw (9.00,1.00)--(9.00,1.50) ; \draw (0.00,1.00)--(0.00,0.50) (1.00,1.25)--(1.00,0.75) (2.00,1.25)--(2.00,0.75) (7.00,1.25)--(7.00,0.75) (8.00,1.25)--(8.00,0.75) (9.00,1.00)--(9.00,0.50) ; \draw (0.00,0.50)--(0.00,0.25) ; \draw (0.00,0.00)--(0.00,0.25) ; \draw [fill = red] (0.00,0.50) arc (90:270:0.25) ; \draw (0.00,0.00)--(0.00,0.50) ; \node at (1.50,0.25) {$\scriptstyle \Gamma'$} ; \node at (7.50,0.25) {$\scriptstyle \Gamma''$} ; \draw (9.00,0.50)--(9.00,0.25) ; \draw (9.00,0.00)--(9.00,0.25) ; \draw [fill = blue] (9.00,0.00) arc (-90:90:0.25) ; \draw (9.00,0.00)--(9.00,0.50) ; \end{scope} \end{tikzpicture}}}
\; .$$}\end{center}
the first two cases are handled by the same strategy of Theorem\ref{corrMLLc}. The sequentialization procedure for the two new cases follows the intuition behind $B$-gates as proof diagram branchings twisting: $(\id_{L, \Gamma'}, g_\alpha, \id_{\Gamma'', R})\circ (\phi_2', \phi_1')\circ B \circ (\phi_1 ,\phi_2): \pro \frr L, \Gamma'', R, L ,\Gamma', R$ is sequentializable iff $\phi_1'\circ \phi_1: \pro \frr L, \Gamma',  R $ and $ \phi_2'\circ \phi_2: \pro \frr L, \Gamma'', R $ are.
\end{proof}
\end{theorem}

\begin{oss}\label{sigMll}
The signature $\Mlltc_2$ suffice to represent $\MLLc$ derivations, that is, $B$-gates are not needed in order to represent proofs and that the quotient $\catgen {\Mllc}$ equate more of these proof diagrams  than $\catgen{\Mlltc}$.
\end{oss}

Let consider the equivalence relation $\sim_D$ over derivations of $\MLLc$ sequent calculus defined as follows:
$$d'(\Gamma)\sim_{ D} d''(\Gamma) \mbox{ iff } \exists \phi_{d'(\Gamma)},\phi_{d''(\Gamma)'}\in \Mlltc \mbox{ such that  } [\phi_{d'(\Gamma)}]_{\Mllc}=[\phi_{d''(\Gamma)}]_{\Mllc}.$$
where $\phi_{d(\Gamma)}\in \Mllc$ is a diagrammatic representation of a $\MLLc$ derivation $d(\Gamma)$.
In other words, $d'(\Gamma)\sim_{ D} d''(\Gamma)$ whenever they can be represented by two proof diagrams which are equivalent modulo $\Mllc_3$. 

The standard proof equivalence of $\MLLc$ sequents is faithfully represented by $\sim_D$:

\begin{theorem}[Proof diagram representation]\label{proofRep}
Two derivations are equivalent modulo $\sim$ if and only if they are represented  by two equivalent proof diagrams with respect of $\catgen{\Mllc}$. That is:
$$d(\Gamma)\sim d'(\Gamma) \Leftrightarrow  d(\Gamma)\sim_D d'(\Gamma)$$
\begin{proof}
Given two derivation $d(\Gamma),d'(\Gamma)$ in $\MLLc$ sequent calculus, $d(\Gamma)\sim d'(\Gamma)$ iff there is a sequence of rules permutations from $d(\Gamma)$ to $d'(\Gamma)$. As remarked in Section \ref{secQ}, $\sim_{\tilde D}$ capture all rules permutations which do not affect the branching of a derivation tree and $\simeq_D\subset \sim_D$. 

This implies that even if we consider derivations up to rules permutations, it is possible to well-define the following function which associate to a derivation an equivalence class of proof diagrams in $\Mllc$:
$$
\xymatrix@C=.5em@R=.5em{
[ - ]_{\Mllc}: &
\{ \MLLc  \mbox{ derivations} \}  &
\fr & 
\{  \mbox{ morphisms in } \catgen{\Mllc}\}
\\
~& 
d(\Gamma)  &
\fr &
[\phi_{d(\Gamma)}]_{\Mllc}
}
$$

Moreover, in a diagrammatic representation of a derivation $\d(\Gamma)$, untangle sequences and their inverses permute pairs of proof diagram branches which correspond to the represented derivation branches. This means that $\sim_D$ captures all rules permutations missed by $\simeq_D$, then that $\sim=\sim_D$
\end{proof}
\end{theorem}

We define the following polygraph:

\begin{definition}[Polygraph of $\MLLc$ semantics]
The \emph{polygraph of multiplicative linear logic semantics} $\Sem {\MLLc}$ is given by extending the polygraph $\Mllc$ with the following the sets of $3$-cells  $\Sem{3}= \Mllc_3 \cup \Sem{\MLLc}^{Cut}$ where $\Sem {\MLLc}^{Cut}=\Mll_{Cut}\cup \Mllc_{Cut}$ is given by the following sets of $3$-cells:
\begin{itemize}
\item $\Mll_{Cut}$ is made of the following $3$-cells:
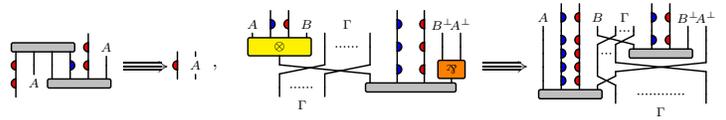
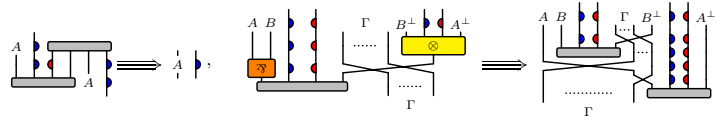
\begin{center}\resizebox{10cm}{!}{
$$
\xymatrix{ \twocell{(axc *0 L *0 topA) *1 (L *0 2 *0 R*0 L *0 1) *1 (L*0 pitA *0 cutc ) } \ar@3[r] & \twocell{L*0 midA}},\quad
\xymatrix{ \twocell{( topA *0 R *0 L *0 topB *0 topGam *0 R *0 L *0 topBb *0 topAb)*1(tenc *0 d *0 R *0 L *0 2 )*1 (ln *0 R *0 L *0 par ) *1 (d*0 cutc )*1 pitGam }
 \ar@3[r] &
\twocell{ (topA *0 R *0 L *0 topB *0 topGam *0 R *0 L *0 topBb *0 topAb)  *1 (4 *0 d *0 4) *1(1 *0 R *0 L *0 ln *0 R *0 L *0 2) *1 (1 *0 R *0 L *0 d *0 cutc *0 1) *1  (1 *0 R *0 L *0 1 *0 s ) *1 (1 *0 R *0 L *0 s *0 1 )  *1 (cutc *0 d) *1 pitGam}},
$$
}\end{center}
\begin{center}\resizebox{10cm}{!}{
$$
\xymatrix{ \twocell{(topA *0 R *0 axc) *1 (1*0 R *0 L *0 2 *0 R) *1 (cutc *0 pitA *0 R) } \ar@3[r] & \twocell{ midA *0 R}},
\quad
\xymatrix{ \twocell{(topA *0 topB *0 R *0 L *0 topGam *0 topBb *0 R *0 L *0 topAb)*1( 2*0  R *0 L *0 d  *0 tenc ) *1 (par *0 R *0 L *0 rn) *1 (cutc *0 d) *1 pitGam}
\ar@3[r] & 
\twocell{ (topA *0 topB *0 R *0 L *0 topGam *0 topBb *0 R *0 L *0 topAb) *1 (4 *0 d *0 4)*1 (2 *0 R *0 L *0 rn *0 R *0 L *0 1) *1  (1 *0 cutc *0 d *0 R *0 L *0 1) *1  (s *0 1 *0 R *0 L *0 1) *1 (1 *0 s *0 R *0 L *0 1)*1 (d *0 cutc)*1 pitGam}},
$$}\end{center}
for all $A,B \in \fmllc$, $\Gamma \in \smllc$;

\item $\Mllc_{Cut}$ is made of the following $3$-cells:
$$\xymatrix{ \twocell{(bot *0 R *0 vc) *1 (1 *0 R *0 L *0 1 *0 R) *1 ( cutc *0 R ) } \ar@3[r] & \twocell{R}}, \quad
\xymatrix{ \twocell{(vc *0 L*0  bot)*1 (L*0 1 *0 R*0 L *0 1) *1 (L*0 cutc) } \ar@3[r] & \twocell{L}};$$
\end{itemize}
\end{definition}

\begin{theorem}[Termination in $\Sem{\MLLc}$]\label{terSem}
The polygraph $\Sem{\MLLc}$ is terminating.
\begin{proof}
Any rewriting path in $\Sem{3}$ is a sequence of rewriting paths in $\Mllc$ and rewriting rules in $\Sem{\MLLc}^{Cut}$ occurrences. If the number these latter is finite in any rewriting path, we conclude by Theorem \ref{TerMllu} that there are no infinite rewriting paths in $\Sem{\MLLc}$.

We define the \emph{degree}  $\delta(g)=\|A\|$ of $Cut_A$-gates  $g\in\phi$ as the number of occurrences of $\parr$ and $\otimes$ symbols in the formula $A$. We define a weight $w(\phi)$ of a proof diagram $\phi\in \Mllc$ depending on the degrees of all its $Cut$-gates:
 $$w(\phi)=\sum^{g\in \phi}_{ g:Cut} 3^{\delta(g)}$$ 
We observe that $w(\phi)=w(\psi)$ whenever $\xymatrix@C=1em{\phi \ar@3[r]_{\Mllc_3} &\psi }$ since $\phi$ and $\psi$ have the same occurrences of $Cut$-gates. 

However, $w(\phi)>w(\psi)$  whenever $\xymatrix@C=1em{ \phi \ar@3[r]_{\Sem 3} &\psi} $. In fact, $\phi $ has an extra $Cut $-gates with respect to the one of $\psi$ or else in $\phi$ there is a $Cut_{A\otimes B}$-gate or a $Cut_{A\parr B}$-gate which is  replaced in $\psi$ by one $Cut_{A}$-gate and one $Cut_B$-gate. The inequality holds because for any $A,B\in \fmllc$ we have $3^{A\otimes B}=3^{A\parr B}=3^{\|A\|+\|B\|+1}> 3^{\|A\|}+3^{\|B\|}$, 

This concludes the proof since any rewriting path in $\Sem 3$ there is a finite number of occurrence rewriting rules in $\Sem{\MLLc}^{Cut}$

\end{proof}
\end{theorem}

Consequently, we have a \emph{cut-elimination} Theorem for sequentializable proof diagrams in $\Sem{\MLLc}$

\begin{theorem}[Cut-elimination]
An irreducible proof diagram $\phi\in \Sem{\MLLc}$ which represent a derivation contains no $Cut$-gates.
\begin{proof}
Proposition \ref{Bigera} assures that a $\Mllc_3$-irreducible proof diagram $\phi\in \Sem{\MLLc}$ of type $\phi:\pro \frr L,\Gamma,R$ contains no $B$-gates and, by  Theorem \ref{terSem}, neither crossing splits. Since twisting relations moves $\parr$ and $\bot$ gates downward in a proof diagram $\phi$, if a $Cut_A$-gate  occurs in $\phi$ then it  has to belong in a subdiagram with shape the source one of the rules in $\Sem 3$, thus $\phi$ is reducible.
\end{proof}
\end{theorem}

However, the twisting relations generates a wide family of critical pairs in the rewritings of $\Mlltc$, $\Mllc$ and $\Sem{\MLLc}$. Some of these critical peaks are not solvable. This leads the following:

\begin{proposition}[$\Sem{\MLLc}$ confluence]
The polygraph $\Sem{\MLLc}$  is not confluent.
\begin{proof}
In $\Sem{\MLLc}$ (but also in $\Mlltc$ and $\Mllc$) the following critical peak is not confluent:
$$
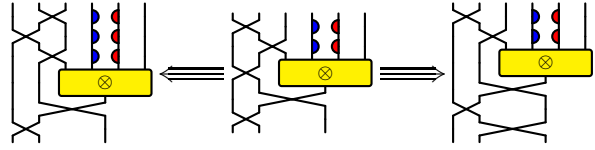
\xymatrix{ 
\twocell{(1 *0 s *0 R *0 L *0 1) *1 (s *0 1 *0 R *0 L *0 1) *1 (1 *0 s *0 R *0 L *0 1) *1 (2 *0 tenc) *1 (1 *0 s) *1 (s *0 1)} &  
\twocell{(s *0 1 *0 R *0 L *0 1) *1 (1 *0 s *0 R *0 L *0 1) *1 (s *0 tenc) *1 (1 *0 s) *1 (s *0 1)} \ar@3[r] \ar@3[l]&
\twocell{(s *0 1 *0 R *0 L *0 1) *1 (1 *0 s *0 R *0 L *0 1) *1 (2 *0 tenc) *1 (1 *0 s) *1 (s *0 1)*1 (1 *0 s)} 
}$$
This rules out a confluence for this polygraph.
\end{proof}
\end{proposition}

Since the signature of $\Sem {\MLLc}$ is the same of $\MLLc$,  we naturally extend the Theorem \ref{corrMllc}:

\begin{theorem}[Multiplicative linear logic correspondence]
$$\vdash_{\MLLc} \Gamma \Leftrightarrow \exists \phi \in \Sem {\MLLc} \mbox{ such that } \phi: \pro \frr L,\Gamma, R.$$
\end{theorem}

This correspondence, together with Theorem \ref{proofRep} ensures the well-definition of the following function:
\begin{definition}[Denotational semantics of proof diagrams]
For any $\MLLc$ derivation $d(\Gamma)$ we associate an equivalence classes of proof diagrams corresponding to a morphism of the category $\catgen{\Sem {\MLLc}}$ as follows:
$$
\xymatrix @C=.5em@R=.5em{
[ - ]_D: &
\{ \MLLc  \mbox{ derivations} \}  &
\fr & 
\{  \mbox{ morphisms in } \catgen{\Sem {\MLLc}}\}
\\
~& 
d(\Gamma)  &
\fr &
[{d(\Gamma)}]_D= [\phi_{d(\Gamma)}]_{\Sem {\MLLc}}
}
$$
where $\phi_{d(\Gamma)}$ is an arbitrary representation of $d(\Gamma)$.
\end{definition}

\begin{theorem}[Proof diagram semantics]
$[ - ]_D$ is a denotational semantics for $\MLLc$ sequent calculus.
\begin{proof}
We define the following  equivalence relation $\approx_D$ over $\MLLc$ derivations:
$$d'(\Gamma)\approx_D d''(\Gamma) \mbox{ iff } [d'(\Gamma)]_D= [d''(\Gamma)]_D$$

We remark the existence of a one-to-one correspondence between rewriting rules in $\Sem {\MLLc}^{Cut}$ and cut-elimination steps. Moreover,  if we denote by $\leftrightarrow^*_{Cut}$ the equivalence relation induced over equivalence classes in $\catgen{\Mllc}$ by the rewriting rules in $\Sem {\MLLc}^{Cut}$, we have that $\catgen{\Sem{\MLLc}}=\frac {\catgen{\Mllc}}{\leftrightarrow^*_{Cut}}$. This implies the following properties:

\begin{enumerate}
\item if $d(\Gamma)\fr_{Cut} \hat d(\Gamma)$, then $d(\Gamma)\approx_D \hat d(\Gamma)$: each cut-elimination step over the diagrammatic representation of $\phi_d(\Gamma)$ is replicated by a rewriting rule in $\Sem {\MLLc}^{Cut}$ eventually preceded by a rewriting path $\frrr^*_{\Mllc}$ in $\Mllc_3$.
\item $\approx_D$ is non-degenerated, i.e.~one can find a formula with at least two non-equivalent proofs: it suffice to take any formula $A\in \fmllc$ which exhibits two non-equivalent (with respect of $\sim$) cut-free derivations $d(A)$ and $d'(A)$, then trivially $d(A)\not \approx_D d'(A)$;
\item $\approx_D$ is a congruence, i.e.~if $d(\Delta)\approx d'(\Delta)$ and we obtain $d(\Gamma)$ and $d'(\Gamma)$ by applying the same inference rule to $d(\Delta)$ and $d'(\Delta)$ , then $d(\Gamma)\approx d(\Gamma)'$: it follow by the compatibility of rewriting with the sequential and parallel diagram compositions.
\end{enumerate}

We remark that $[-]_D$ is coherent with the involutivity of negation. In fact, the invariance of  diagram inputs and outputs with respect to rewriting impose the equivalence $A^{\bot \bot}=A$:
$$Ax_A=
\xymatrix{ \twocell{ axcA *1 (L *0  s *0 R) *1 (L *0  s *0 R) *1 (pitV  *0 pitA *0 pitAb*0 pitV) } \ar@3[r] &  \twocell{ axcAb  *1 (L *0  s *0 R) *1 (pitV  *0 pitA *0 pitAb*0 pitV) } \ar@3[r] & \twocell{ axcAbb *1 (L *0 2 *0 R) *1 (pitV  *0 pitA *0 pitAb*0 pitV) }}
=Ax_{A^{ \bot\bot }}$$
Similarly, De Morgan's laws follow by the definition of $Cut$-gates (see Remark \ref{CutDef}). By means of example, consider the equivalence of $A \parr B=(B^\bot \otimes A^\bot)^\bot$:
$$\xymatrix{ \twocell{ (L *0 d *0 R) *1 (phi4 *0 axcBbtenAb) *1 (L *0  par *0 R *0 L *0  2 *0 R) *1 (L *0 cutcAparB *0 1 *0 R)*1 (pitV *0  pitAbtenBbb *0 pitV)} \ar@3[r] &  \twocell{  (L *0 d *0 R) *1  phi4  *1 (L *0 par *0 R) *1 (L *0 1 *0 R) *1 (pitV *0  pitAbtenBbb *0 pitV) }}$$

From these properties we deduce that $[-]_D$ defines a denotational semantics for $\MLLc$ sequent calculus by means of equivalence classes of proof diagrams.
\end{proof}
\end{theorem}


\section{Conclusion}

In this paper we have presented the syntax of \emph{proof diagrams}, a particular class of string diagrams suitable for interpreting linear logic proof derivations. Even if proof diagrams syntax  reminds the intuitive $2$-dimensional representations of proof nets, their strings have a more  rigid structure with respect to proof net wirings. This allows for the definition of some \emph{control strings} and a consequent linear-time sequentializability test. Indeed,  we can test the possibility to interpret a proof diagrams as a $\MLLc$ derivation in linear time by checking the type of its inputs and outputs only.

Furthermore, the syntax of proof diagrams induce an equivalence relation over the syntax of $\MLLc$ sequent calculus derivations.  We here summarize some different equivalence relations over derivation we obtain by different rewriting systems over proof diagrams:

\begin{itemize}
\item an equivalence relation which captures all permutations of $\parr$ and $\bot$ rules but not some permutations involving $Cut$ and $\otimes$  which also permute derivation branches order. This equivalence induced by proof diagram syntax turns out to be invariant under a given set of diagram rewriting rules we call \emph{twisting relations};

\item an equivalence relation which captures all permutations of inference rules and turns out to be equivalent to the standard proof equivalence we always use to consider in sequent calculus;

\item an equivalence relation which both captures rules permutations and the cut-elimination. 

\end{itemize}

In the conclusive Theorem of this paper we define a  denotational semantics for $\MLLc$ sequent calculus by means of equivalence classes of proof diagrams. Moreover, we show that this diagram equivalence is defined by means of a terminating (but not confluent) rewriting.
%


\section*{Acknowledgements}
I would like to thank Michele Alberti, Marianna Girlando, Giulio Guerrieri, Paolo Pistone and Lionel Vaux for the fruitful exchanges during the redaction of this work. A special acknowledgment to Yves Guiraud  who wrote (and upgraded for the scope) the latex package for string diagrams representations \cite{catex} employed in the present manuscript.

{\small
\bibliography{Bibtes}{}  

\bibliographystyle{plain}
}
\end{document}